\documentclass[a4paper,11pt]{article}
\usepackage{jcappub} 
\usepackage{graphics}
\usepackage{color}
\usepackage[table,xcdraw,dvipsnames]{xcolor}
\usepackage{physics}
\usepackage{appendix}
\usepackage{slashed}
\usepackage[htt]{hyphenat}
\usepackage{cancel}
\usepackage{multirow}
\usepackage{mathtools}
\usepackage[T1]{fontenc}
\usepackage{lmodern}
\def\abs#1{\left| #1\right|}

\definecolor{blue-violet}{rgb}{0.7, 0.2, 0.8}

\title{\boldmath Gravitational Waves from Dimension-6 Assisted Peccei–Quinn Phase Transitions}

\author[a]{Nico Benincasa} 
\author[b,c]{and Kristjan Müürsepp}
\affiliation[a]{Université Claude Bernard Lyon 1, CNRS/IN2P3, IP2I UMR 5822, 4 rue Enrico Fermi, F-69100 Villeurbanne, France}
\affiliation[b]{INFN, Laboratori Nazionali di Frascati, C.P. 13, 00044 Frascati, Italy}
\affiliation[c]{Laboratory of High-Energy and Computational Physics, NICPB, R\"avala pst 10,  10143 Tallinn, Estonia}

\abstract{We consider a simple model-independent extension of the minimal KSVZ axion model, augmenting the Peccei–Quinn effective potential with a dimension-6 operator invariant under the Peccei–Quinn symmetry. We show that this operator can trigger a first-order phase transition and study its cosmological consequences. We map out the parameter space, consistent with present observational constraints, in which the phase transition produces a gravitational wave signal observable at current and future interferometers, and we investigate the resulting dark matter phenomenology. Finally, we present specific ultraviolet completions that generate the dimension-6 operator, each leading to distinct phenomenological signatures, which motivates our model-independent approach.}

\begin{document}
\maketitle
\flushbottom

\section{Introduction}

In view of the stringent bounds on the neutron electric dipole moment, the strong CP problem remains one of the outstanding motivations in our search for physics beyond the Standard Model (SM). The Peccei--Quinn (PQ) \cite{Peccei:1977hh, PhysRevLett.40.223, PhysRevLett.40.279} solution to the strong CP problem is achieved by introducing a new complex scalar field $\Phi$, charged under a global $U(1)_{\rm PQ}$ symmetry. Once the radial mode of $\Phi$ obtains a non-zero vacuum expectation value (vev) $\expval{\Phi} = f_a/\sqrt{2}$, the PQ symmetry is spontaneously broken, with the angular mode $a(x)$ corresponding to the Goldstone boson of the broken symmetry. In order to solve the strong CP problem, one then only needs to generate an effective coupling of $a(x)$ with the gluon field strength tensor,\footnote{Although not strictly needed, in specific ultraviolet (UV) realisations, couplings of the axion field $a(x)$ to the fermions or the photon are often also introduced; see e.g. \cite{DiLuzio:2020wdo} for more details.}
\begin{equation}
   \mathcal{L} \supset \frac{\alpha_s}{8\pi}\frac{a(x)}{f_a} G_{\mu \nu} \Tilde{G}^{\mu \nu},
   \label{eq:AxionCoupling}
\end{equation}
where $\alpha_s$ denotes the fine structure constant of $SU(3)_c$. Provided that the potential for $a(x)$ admits a global minimum at $\expval{a(x)} = 0$, this coupling allows the CP-violating term to be dynamically relaxed to zero. While the operator in
Eq.~\eqref{eq:AxionCoupling} is a total derivative, and hence leaves the shift
symmetry $a \to a + \text{constant}$ intact to all orders in perturbation theory,
non-perturbative QCD effects break it explicitly and generate a potential for
$a(x)$. Its curvature at the minimum fixes the mass of the QCD axion,
$m_a^2 f_a^2 = \chi_{\rm QCD}$, with $\chi_{\rm QCD}^{1/4} \simeq 75.5$ MeV the topological susceptibility of QCD, making $a(x)$ the pseudo-Nambu-Goldstone (pNGB) boson of the
broken $U(1)_{\rm PQ}$ symmetry~\cite{GrillidiCortona:2015jxo}.

The effective coupling appearing in Eq.~\eqref{eq:AxionCoupling} can be generated in several different UV models, among which the most common realisations are the Kim--Shifman--Vainshtein--Zakharov (KSVZ) \cite{PhysRevLett.43.103, SHIFMAN1980493} and Dine--Fischler--Srednicki--Zhitnitsky (DFSZ) models \cite{Zhitnitsky:1980tq, DINE1981199}. Although both of these require the introduction of a complex scalar singlet, they differ by the additional degrees of freedom that must be included beyond those of the SM. In particular, the minimal KSVZ realisation requires at least one additional coloured fermion, while the DFSZ model introduces an additional scalar electroweak doublet. In this work, we will exclusively focus on the minimal KSVZ model extended with a non-renormalisable dimension-6 operator, which we will describe in more detail in Section \ref{sec4:Analytics}.

In addition to solving the strong CP problem, the QCD axion is also a perfectly valid candidate for dark matter (DM), whose relic abundance is typically produced through the misalignment mechanism~\cite{Abbott:1982af,Dine:1982ah, Preskill:1982cy}, with an additional component from topological effects in the post-inflationary scenario. In this case, the energy density corresponding to DM can be identified with the energy stored in coherent oscillations of the axion field, once the axion mass exceeds the Hubble rate during the evolution of the Universe. Assuming that the velocity of the axion field during the onset of oscillations is small,\footnote{For the opposite case, see recent works on the \textit{kinetic misalignment mechanism} \cite{Co:2019jts, Co:2020dya, Eroncel:2024rpe, Morgante:2025lav}.} the DM abundance is only determined by two parameters: the value of the misalignment angle at the onset of oscillations $\theta_{i}\equiv a(t_{\rm osc})/f_a$ and the axion decay constant $f_a$, that is related to the PQ symmetry breaking vev.\footnote{Typically $f_a = v_a/N_{\rm DW}$, where $v_a$ denotes the vev of the PQ field, and $N_{\rm DW}$ the integer domain wall number. In this work we consider the minimal scenario with $N_{\rm DW} = 1$, in which case $f_a$ directly equals the vev of the PQ field.}

The exact region of parameter space that reproduces the observed amount of DM then depends on whether the PQ symmetry was broken before or after inflation. If the PQ symmetry breaks before inflation, and is not restored afterwards, the Hubble patch corresponding to a particular value of $\theta_i$ is inflated exponentially, encompassing the whole volume within our present horizon, and thus $\theta_i$ is a free parameter that can assume any value in the interval $[-\pi, \pi]$. Instead, if the PQ symmetry is broken after inflation, the axion field can take different values on different Hubble patches that were originally out of causal contact, but which at later times re-entered within the present-day horizon.
The $\theta_i$ governing the DM abundance in this case is obtained by taking the average over the values of $\theta_i$ across the different patches, whereby including anharmonic effects (see \cite{DiLuzio:2020wdo, GrillidiCortona:2015jxo} for more details), one obtains $  \sqrt{\expval{\theta_i^2}} \simeq 2.155$. The observed amount of DM can then be approximated by \cite{DiLuzio:2020wdo} 
\begin{equation}
    \Omega^a_{\rm DM}h^2 = 0.12 \left( \frac{f_a}{2 \times 10^{11} {\rm GeV}} \right)^{7/6}, \qquad \text{(neglecting topological defects)},
    \label{eq:OmegaDM}
\end{equation}
where in this simple estimate, we have neglected the contribution to DM from topological defects.

Due to the relative simplicity and theoretical motivation of axion models, experimental efforts attempting to probe, or rule out, QCD axions, and more generally axion-like particles, have attracted an ever-increasing interest from the community. To that end, laboratory experiments such as haloscopes, helioscopes and light-shining-through-wall experiments; precise measurements of astrophysical processes such as neutron star, white dwarf, or supernova cooling; as well as cosmological non-observation of axion miniclusters or isocurvature fluctuations all provide a useful handle to narrow down the available parameter space (see \cite{Marsh:2015xka,DiLuzio:2020wdo,Choi:2020rgn,Caputo:2024oqc} for more details on the different experiments and \cite{AxionLimits} for useful summary plots). In this work, we are mainly constrained by the lower bound on $f_a$, obtained from observations of supernova cooling, which in the simple KSVZ scenario is given by \cite{Caputo:2024oqc}
\begin{equation}
   f_a \gtrsim 4 \times 10^8\, { \rm GeV}.
\end{equation}

If the breaking of the global $U(1)_{\rm PQ}$ symmetry proceeds through a first-order phase transition (FOPT), one could have an additional experimental probe for the QCD axion through the observation of gravitational waves (GWs)~\cite{DelleRose:2019pgi, Yang:2024npd, Ahmadvand:2021vxs, Chiang:2020aui, Ghoshal:2020vud, VonHarling:2019rgb, Caprini:2024ofd,Turbang:2020,Ringwald:2020vei}.
In the minimal version of the KSVZ model, due to the absence of a thermally induced barrier near the false vacuum, the phase transition (PT) is almost always either second-order or a crossover~\cite{DelleRose:2019pgi, Ballesteros:2016xej}. However, this may be changed by going beyond the minimal scenario either by extending the field content, by making the theory scale-invariant, or by considering higher-dimensional operators.

In this work, we consider an extension of the KSVZ model, with the addition of a $U(1)_{\rm PQ}$ - preserving dimension-6 operator to the scalar potential. The same mechanism has long been studied for the electroweak transition, where a dimension-6 $\abs{H}^6/\Lambda^2$ operator can render the SM crossover strongly first order \cite{Grojean:2004xa, Delaunay:2007wb,Chala:2018ari}; here we apply it to the PQ sector.\footnote{It has also been argued~\cite{Postma:2020toi} that, since strong electroweak PT requires $v/\Lambda \gtrsim \order{0.1}$, one should include additional non-renormalisable operators beyond dimension-6. In this first exploratory analysis for the PQ transition, we will not consider such extensions, leaving them for future studies.}  To the best of our knowledge, PTs from KSVZ-type models with non-renormalisable operators have only been studied in a Master's thesis \cite{Turbang:2020} for QCD axion with a PQ breaking scale $f_a \lesssim 10^{9}\, {\rm GeV}$ and for the clockwork axion \cite{Chiang:2020aui} for lower PQ breaking scales $f_a \lesssim 10^{6}\, {\rm  GeV}$. Compared to previous works, we will focus on the parameter space $10^9\,{\rm GeV} < f_a < 10^{12}\,{\rm GeV}$, which brackets the value
$f_a \simeq 2\times10^{11}\,{\rm GeV}$ at which the post-inflationary axion in our benchmark estimate accounts for the full observed DM abundance, and extends into the regions
where it is under- and overproduced. In particular, we will identify observational prospects for the current
LIGO--Virgo--KAGRA network, as well as for future
detectors such as the Einstein Telescope (ET)
and Cosmic Explorer (CE).

In contrast to previous studies, we will make use of the finite counterterms to keep the zero-temperature minimum fixed at its tree-level value, and explain the necessary conditions in order to have two minima separated by a potential barrier, with the symmetry-breaking vacuum corresponding to the global minimum. As another novelty, we will not only focus on the thermal $O(3)$-symmetric tunneling solution, but also consider the possibility of having $O(4)$ quantum tunneling, which is made possible by the potential having a tree-level barrier even at zero temperature. This allows us to quantitatively explore scenarios in which the PT from the unbroken minimum to the broken one does not complete even at very low temperatures, and thus delineate regions of parameter space in which the PQ solution cannot be realised. We furthermore identify non-standard thermal histories, such as symmetry non-restoration (SNR), that can arise due to the negative scalar quartic coupling, and the previously unexplored possibility of having multi-step PQ transitions even in the case of a single scalar degree of freedom. Moreover, compared to previous works \cite{Turbang:2020}, we will expand the list of UV models that can give rise to the dimension-6 operator, thus motivating the use of effective field theories (EFTs) to trigger strong PQ transitions. 

Since we work in the post-inflationary regime, the breaking of $U(1)_{\rm PQ}$
is necessarily accompanied by the formation of a cosmic string network. These
are global strings, which lose energy predominantly through the emission of
axions rather than gravitational radiation \cite{Davis:1986xc, Vilenkin:1984ib}; the associated stochastic
background is therefore suppressed with respect to that of local
strings of comparable tension, and is most likely negligible over the range of $f_a$ considered here~\cite{Chang:2019mza, Chang:2021afa,Figueroa:2023zhu}. The GW signal we compute is thus dominated by
the PT itself. The axion emission, on the other hand, is not negligible
for the relic abundance, and is expected to contribute at a level comparable to
the misalignment contribution of Eq.~\eqref{eq:OmegaDM}~\cite{Gorghetto:2018myk, Gorghetto:2020qws,Buschmann:2019icd, Buschmann:2021sdq, Klaer:2017ond, Hiramatsu:2012gg}. A
quantitative treatment of the network requires dedicated lattice simulations
and lies beyond the scope of this work.

The rest of this paper is organised as follows. In Section~\ref{sec:KSZV}, we recall the minimal KSVZ model as a possible UV completion for addressing the strong CP problem. Section~\ref{sec3:PT_background} offers a general review of PT dynamics and the key quantities to be computed in the following sections. The results of the previous two sections will then be used in Section~\ref{sec4:Analytics}, which will focus on analytic estimates for the PT dynamics for a minimal KSVZ model extended by a dimension-6 operator. After obtaining some analytical understanding, we study the PT dynamics and the resulting GW signal numerically in Section~\ref{sec5:Results_and_Pheno}, and link the conclusions with DM phenomenology. Finally, we outline concrete UV models, motivated by DM and neutrino physics, that could give rise to the effective dimension-6 operator in Section~\ref{sec6:UVcompletions}, before presenting our conclusions in Section~\ref{sec7:concl}.

\section{Brief review of the KSVZ realisation}
\label{sec:KSZV}

The minimal KSVZ \cite{PhysRevLett.43.103, SHIFMAN1980493} model extends the SM with a new complex scalar $\Phi$, singlet under the SM group, and a fermion $\psi$ that transforms as a triplet under the $SU(3)_c$ group, a singlet under $SU(2)_L$, and which we take to have zero hypercharge under $U(1)_Y$. Under the new global $U(1)_{\rm PQ}$ symmetry, $\Phi$ and the chiral components of $\psi$ transform as
\begin{equation}
    \Phi \rightarrow e^{iq \alpha_{\rm PQ}} \Phi,\quad\psi_L \rightarrow e^{i \alpha_{\rm PQ} q/2}\psi_L, \quad \psi_R \rightarrow e^{-i \alpha_{\rm PQ} q/2}\psi_R,
    \label{eq:transformations}
\end{equation}
with $\alpha_{\rm PQ}$ denoting the constant transformation parameter of $U(1)_{\rm PQ}$, while $q$ fixes the overall normalisation of the PQ charges. The SM fields remain neutral under $U(1)_{\rm PQ}$. In what follows, without loss of generality, we will choose $q=1$.

Taking all of those new components into account, the SM Lagrangian can thus be extended with the additional terms:
\begin{equation}
\mathcal{L}_{\rm KSVZ} \supset \abs{\partial_\mu \Phi}^2
 + \Bar{\psi} i \slashed{D} \psi
 - \left( y_F \Phi \Bar{\psi}_L \psi_R + {\rm h.c.} \right)
 - V(\abs{\Phi}^2),
 \label{eq:KSVZ lagrangian}
 \end{equation}
where the scalar potential depends only on $\abs{\Phi}^2$ in order to preserve the global $U(1)_{\rm PQ}$ symmetry. In parts of the parameter space where $V(\abs{\Phi}^2)$ is minimised for non-zero values of $\abs{\Phi}^2$, we can parametrise the complex scalar field as\footnote{As mentioned in the Introduction we consider the minimal model with $N_{\rm DW} = 1$, and so $f_a$ can be identified with the $U(1)_{\rm PQ}$ symmetry-breaking scale $v_a$.}
\begin{equation}
    \Phi(x) = \frac{\phi(x)}{\sqrt{2}} e^{i a(x)/f_a}, \quad {\rm with} \quad \expval{\phi} = f_a.
    \label{eq:Phi_after_SSb}
\end{equation}
After integrating out the radial mode, which has a large mass $\sim \order{f_a}$, the low-energy Yukawa Lagrangian can be written as 
\begin{equation}
\mathcal{L}_{\rm KSVZ} \supset
 - m_\psi\, \Bar{\psi}\, e^{i a(x)\gamma_5/f_a}\, \psi\,,
 \qquad m_\psi = \frac{y_F f_a}{\sqrt{2}}.
\end{equation}
We can then perform a space-time dependent chiral field redefinition on the fermions, $\psi \rightarrow e^{-ia(x) \gamma_5/(2 f_a)}\psi$ to remove the phase containing $a(x)$, picking up derivative axion couplings to fermions, as well as an anomalous contribution from the non-invariance of the fermion path-integral measure~\cite{Fujikawa:1979ay}:
\begin{equation}
    \delta \mathcal{L}_{\rm KSVZ} \supset \frac{g_s^2}{32\pi^2} \frac{a(x)}{f_a} G_{\mu\nu} \Tilde{G}^{\mu\nu},
\end{equation}
where $g_s = \sqrt{4\pi \alpha_s}$ denotes the $SU(3)_c$ gauge coupling. This term has exactly the form of Eq.~\eqref{eq:AxionCoupling} that is needed to solve the strong CP problem. The potential $V(\abs{\Phi}^2)$ appearing in Eq.~\eqref{eq:KSVZ lagrangian} can be arbitrarily chosen, as long as it admits a global minimum at $f_a$.\footnote{One could of course also imagine scenarios where $f_a$ corresponds to the local minimum, with sufficiently long lifetime, similar to the Higgs vacuum metastability. However, we will not consider such exotic scenarios in this work.} We will specify its precise form in the next section.

\section{Review of phase transitions and the key quantities to be computed}

\label{sec3:PT_background}

Cosmological PTs typically occur when a global or a gauge symmetry of a theory is spontaneously broken. Of particular interest are FOPTs, whereby the order parameter of the theory, in this case the vev of a scalar field, undergoes a discontinuous change as the Universe evolves in time. Since thermal effects typically restore the symmetry at high temperatures (early times) \cite{Weinberg:1974hy}, we are particularly interested in scenarios wherein a scalar field develops a non-zero vev at some finite temperature, thus spontaneously breaking a symmetry under which it is charged.\footnote{In Section~\ref{sec:snr}, we will also briefly mention the opposite scenario, wherein the symmetry is not restored at high temperatures. However, we will not investigate the observational signatures of these scenarios here, leaving detailed studies for future work.} As we will explain in more detail below, FOPTs proceed via the nucleation of spherically symmetric true vacuum (symmetry-broken) bubbles in the false vacuum background. Subsequently, the collision of those bubbles and their interactions with the ambient plasma break spherical symmetry and thus give rise to the production of GWs. It is the possibility of observing GWs from FOPTs at the present and future observatories that has recently motivated a huge activity in the study of FOPTs in a variety of BSM models \cite{Breitbach:2018ddu, Figueroa:2023zhu, vonHarling:2017yew,  Madge:2023dxc, Fairbairn:2019xog, Addazi:2023jvg,  Bringmann:2023opz, Carena:2019une, Xue:2021gyq, Salvio:2023ynn,  Megias:2018sxv, Ghosh:2023aum, Wu:2023hsa, Ellis:2022lft,  Azatov:2022tii, Agashe:2019lhy, Romero:2021kby, An:2023jxf, DiBari:2021dri, Brandenburg:2021tmp, Greljo:2019xan, Ertas:2021xeh, Agashe:2020lfz, DelleRose:2019pgi,  VonHarling:2019rgb, Ahmadvand:2021vxs, Harigaya:2019shz,  Baldes:2023cih, Blasi:2023rqi, Athron:2023rfq, Cataldi:2024pgt, Benincasa:2022elt, Kumar:2021ffi, Conaci:2024tlc,  Kobakhidze:2017mru, Cataldi:2025nac, Chun:2023ezg, Azatov:2021ifm, Azatov:2021irb, Baldes:2021vyz, Vaskonen:2016yiu,  Marzola:2017jzl, Marzo:2018nov, Lewicki:2021xku, Lewicki:2024ghw, Lewicki:2024sfw, Bringmann:2026xcx, Balan:2025uke, Bringmann:2023iuz, Benincasa:2026dhg, Benincasa:2025tdr, Benincasa:2023vyp,Alanne:2020jwx, Kannike:2019wsn, Kannike:2019mzk, Jinno:2022fom, Giudice:2024tcp, Inomata:2024rkt, Schwaller:2015tja,Morgante:2022zvc,Pascoli:2026tuu, Levi:2022bzt, Kierkla:2022odc, Kierkla:2023von, Prokopec:2018tnq, Croon:2020cgk, Guo:2020grp, Dror:2019syi,Croon:2018erz, Croon:2018kqn, Balazs:2016tbi, Guo:2021qcq, Croon:2018new, Addazi:2016fbj, Pasechnik:2023hwv, Reichert:2021cvs, Huang:2020bbe, Caprini:2009fx, Grojean:2006bp, Grojean:2004xa, Jinno:2016knw, Hashino:2018wee,Espinosa:2011ax}. In what follows, we will offer a brief overview of the key quantities for the study of FOPTs and the associated GWs that are relevant for our chosen KSVZ model. For more general details on FOPTs and GWs, several excellent reviews are available \cite{Athron:2023xlk, Croon:2023zay, Hindmarsh:2020hop, Mazumdar:2018dfl, Quiros:1999jp}.

\subsection{Effective potential}
\label{sec:quantum_corrections}

The symmetry-breaking dynamics during the evolution of the Universe is dictated by the study of the phase structure of the theory as a function of temperature, which can be obtained by minimising the effective potential \cite{Dolan:1973qd, Coleman:1973jx}

\begin{equation}
    V_{\rm eff} = V_{\rm tree}(\phi) + V_{\rm CW}(\phi) + V_{\rm CT}(\phi) + V_{T}(\phi,T),
\end{equation}
where $V_{\rm tree}$ denotes the tree-level potential, $V_{\rm CW}$ denotes the Coleman-Weinberg potential, arising from zero-temperature 1-loop corrections, $V_{\rm CT}$ denotes the counterterms keeping the vev and the mass of the scalar undergoing PT unchanged by loop corrections at zero temperature and $V_T$ denotes additional loop corrections arising due to the ambient thermal bath. In our scenario of interest, the tree-level potential is given by\footnote{Notice that we omit the dimension-6 corrections to the kinetic terms that are inevitably present in many UV models and that can have sizeable effects on the PT dynamics. More details about our underlying assumptions can be found in Appendix \ref{sec:Appendix}.}
\begin{equation}
    V_{\rm tree}(\Phi(x)) = m^2 \abs{\Phi(x)}^2 + \lambda \abs{\Phi(x)}^4 +  \frac{C_1}{\Lambda^2}\abs{\Phi(x)}^6.
\label{eq:Vtree_beforeSSb}
\end{equation}
As described in more detail in Appendix \ref{sec:Appendix}, provided that $C_1 > 0$, we can redefine the couplings in Eq.~\eqref{eq:Vtree_beforeSSb} in such a way as to absorb the value of $C_1$ into $1/\Lambda^2$, thus setting $C_1 = 1$.
Parametrising the PQ field after symmetry breaking as in Eq.~\eqref{eq:Phi_after_SSb} we obtain the following potential for the radial mode\footnote{From now on, we leave the spacetime dependence of the fields implicit for the sake of readability.}
\begin{equation}
    V_{\rm tree}(\phi) = \frac{m^2}{2} \phi^2 +\frac{\lambda}{4} \phi^4 + \frac{1}{8 \Lambda^2} \phi^6.
\label{eq:treelevelpot:main}
\end{equation}

Notice that in order to have a barrier between the symmetric and broken minima even at tree-level, a necessary but not sufficient condition is $\lambda < 0$, whereas the stability of the potential is protected by the positive Wilson coefficient $C_1 > 0$ of the non-renormalisable term. The exact values of couplings for which the barrier and the minima away from the origin exist are examined in the next section.

The 1-loop Coleman-Weinberg potential calculated with dimensional regularisation, in the $\overline{\rm MS}$-scheme is given by\footnote{In the effective potential, $\phi$ denotes the generic constant background-field configuration $\langle\phi\rangle$, not necessarily evaluated at its minimising value $f_a$.}~\cite{Coleman:1973jx}
\begin{equation}
    V_{\rm CW}(\phi) = \sum_i n_i\frac{M_i^4(\phi)}{64 \pi^2} \left[ \ln \left( \frac{M_i^2(\phi)}{\mu^2} \right) - \frac{3}{2} \right],
    \label{eq:VCW_1loop}
\end{equation}
where the summation index $i$ runs over the scalar and fermion degrees of freedom, with $n_\phi=n_a=1$ and $n_\psi=-12$. The field-dependent masses $M_i^2$ can be computed by differentiating the potential twice with respect to the appropriate field $i = a,\phi,\psi$ and are given by
\begin{align}
    M_\phi^2 (\phi) &= m^2 + 3\lambda \phi^2 + \frac{15}{4}\frac{\phi^4}{\Lambda^2},\\
    M_a^2(\phi) &= m^2 + \lambda\phi^2 + \frac{3}{4}\frac{\phi^4}{\Lambda^2},\\
    M_\psi^2(\phi) &=  \frac{1}{2} y_F^2\phi^2.
\end{align}
To make sure that the vev of $\phi$, computed from the tree-level potential, and the curvature at $\expval{\phi} = f_a$ do not get shifted from their tree-level values at the 1-loop level, we use the so-called \textit{on-shell-like} scheme, adding finite counterterms to the effective potential: 
\begin{equation}
    V_{\rm CT} =\delta m^2 \phi^2 + \delta \lambda \phi^4.
\end{equation}
These finite terms are determined by\footnote{We use this scheme to make sure that at zero temperature the symmetry-breaking minimum can be easily computed by tree-level expressions and that it remains minimum even after including loop effects.}
\begin{equation}
    \frac{\partial (V_{\rm CW} + V_{\rm CT}) }{\partial \phi}\Bigg\vert_{\phi=f_a} = 0, \quad \frac{\partial^2 (V_{\rm CW} + V_{\rm CT}) }{\partial \phi^2}\Bigg\vert_{\phi=f_a} = 0,
\end{equation}
with their analytic form given by
\begin{equation}
    \delta m^2= \frac{-3\partial_\phi V_\text{CW} + f_a \partial_\phi^2 V_\text{CW}}{4 f_a}\Bigg\vert_{\phi=f_a},\quad    \delta \lambda = \frac{\partial_\phi V_\text{CW} - f_a\partial_\phi^2 V_\text{CW}}{8f_a^3}\Bigg\vert_{\phi=f_a}.
    \label{eq:deltam2_and_deltalam}
\end{equation}
Note that $\partial_\phi^2 V_\text{CW}$ yields terms like $(\partial_\phi M_a^2)^2\ln (M_a^2/\mu^2)$. Since $M_a^2\vert_{\phi = f_a} = 0$ by the tree-level minimisation condition, while $\partial_\phi M_a^2 \vert_{\phi = f_a} \neq 0$, computing the curvature of $V_\text{CW}$ at $\phi=f_a$ leads to an infrared divergence (IR) due to the $\ln M_a^2$ term. In order to cure that issue, we follow the prescription of~\cite{Cline:1996mga, Cline:2011mm} and introduce the following replacement when computing $\partial_\phi^2 V_\text{CW}$: $\ln M_a^2(f_a)\rightarrow \ln M_\phi^2(f_a)$.\footnote{The IR-divergence arising from evaluating the Goldstone boson masses at the minimum is known in the literature as the \textit{Goldstone boson catastrophe} \cite{Martin:2013gka, Martin:2014bca, Elias-Miro:2014pca}. While our approach of regulating the IR-divergence by $M_\phi(f_a)$ constitutes a valid first approximation, more formally one should resum an infinite set of IR-divergent diagrams contributing to the Goldstone boson catastrophe to obtain an IR-finite result \cite{Martin:2014bca, Croon:2020cgk}. However, we expect that our results would not be significantly altered by this more formal procedure.}

We fix the renormalisation scale $\mu = \Lambda$, and so all couplings of the theory should be understood as having been initialised at that scale. Since we mostly focus on scenarios for which $\order{10^{-2}}\lesssim m/\Lambda,f_a/ \Lambda \lesssim \order{1}$, the problem of large logs does not appear.\footnote{Near the origin where $\phi \rightarrow 0$, we have $V_{\rm CW} \rightarrow \frac{m^4}{32\pi^2} \left( \ln \left[ \frac{m^2}{\Lambda^2}\right] -\frac{3}{2} \right)$ and there is no problem of large logs, as long as $\order{10^{-2}} \lesssim m/\Lambda \lesssim \order{1}$.} However, to quantify the remaining uncertainty from the choice of the renormalisation scale, renormalisation group (RG)-improved methods \cite{Ford:1992mv,Bando:1992wy, Manohar:2020nzp, Kannike:2025ykx}, or better yet 3-dimensional EFT techniques \cite{Ginsparg:1980ef, Appelquist:1981vg} would be needed and are left for future studies.

\subsection{Thermal corrections}
\label{sec:thermal_corrections}
The 1-loop thermal potential is given by \cite{Dolan:1973qd}
\begin{equation}
    V_{T}(\phi, T) = \frac{T^4}{2\pi^2} \left[ n_\phi J_B\left( \frac{M_\phi^2(\phi)}{T^2} \right) +  n_a J_B\left( \frac{M_a^2(\phi)}{T^2} \right) + n_\psi J_F\left( \frac{M_\psi^2(\phi)}{T^2} \right)  \right],
\end{equation}
with 
\begin{equation}
    J_{B/F}(y^2) = \int_{0}^{\infty} {\rm d}k\, k^2 \ln \left[ 1 \mp e^{- \sqrt{k^2 + y^2}} \right].
\end{equation}
For $M_{\phi, a, \psi}(\phi) \ll T$ one can use the high-temperature expansion of the $J_{B,F}$ functions, which gives $V_T^{\rm high-T} = V_T^F + V_T^B$, with
\begin{equation}
\label{eq:VTF}
    V_{T}^F \simeq \sum_F n_F \left(\frac{7 \pi^2}{720}T^4 - \frac{M_F^2 T^2}{48}  - \frac{M_F^4}{64 \pi^2} \ln \left[ \frac{M_F^2}{T^2} \right] + \frac{M_F^4}{64 \pi^2} \ln a_F \right) + \order{\frac{M_F^6}{T^2}}, 
\end{equation}
for $F = \psi$,
and
\begin{equation}
\label{eq:VTB}
    V_{T}^B \simeq \sum_B n_B \left( -\frac{\pi^2}{90}T^4 + \frac{M_B^2 T^2}{24} - \frac{M_B^3 T}{12 \pi} - \frac{M_B^4}{64 \pi^2} \ln \left[ \frac{M_B^2}{T^2} \right] + \frac{M_B^4}{64 \pi^2} \ln  a_B   \right) + \order{\frac{M_B^6}{T^2}},
\end{equation}
for $B = \phi, a$, with $a_B=16\pi^2 \exp(3/2-2\gamma_E)$, $a_F=\pi^2 \exp(3/2-2\gamma_E)$ and $\gamma_E\simeq 0.5772$ the Euler–Mascheroni constant, and where the field-independent terms yield the radiation pressure, up to a minus sign:
\begin{equation}
\label{eq:rad_pressure}
    p_\text{rad}=\frac{\pi^2T^4}{90}\left(\sum_Bn_B-\frac{7}{8}\sum_Fn_F\right)=\frac{\pi^2g_*}{90}T^4,
\end{equation}
with $g_*$ the effective number of relativistic degrees of freedom.\footnote{Here we define $g_\ast$ only in terms of the fields that are relevant for our PT analysis, i.e $\phi,\psi,a$. However, when computing GW spectra, we also include the SM degrees of freedom, meaning $g_\ast$ appearing in the following sections should be understood as being defined by $g_\ast=\left(\sum_Bn_B-\frac{7}{8}\sum_Fn_F\right) + 106.75 $.}

In addition, we will include the effect of \textit{daisy resummation}, which is needed to resum the IR-divergent bosonic contributions to the scalar potential at high temperatures. In particular, we will focus on the \textit{Parwani} method \cite{Parwani:1991gq}, whereby we replace
\begin{equation}
    M_\phi^2(\phi) \rightarrow M_\phi^2(\phi) + \Delta_{T}, \quad M_a^2(\phi) \rightarrow M_a^2(\phi) + \Delta_{T},
\end{equation}
inside $V_{\rm CW}, V_{T}$,\footnote{Note that we do not include the Parwani masses inside $V_{\rm CT}$ since this term is meant to only cancel the loop corrections to the vev at \textit{zero} temperature.} with the leading-order thermal mass $\Delta_{T}$ obtained by retaining only the $T^2$ terms in $V_T^{B,F}$ given by\footnote{We neglect subleading thermal self-energy contributions, including terms proportional to $mT$ arising from the bosonic cubic terms, as is conventional when using leading-order thermal masses in the Parwani prescription.}
\begin{equation}
    \Delta_{T} \simeq \frac{\partial^2 V^{\text{high-T},T^2}_{T}}{\partial \phi^2}\Bigg\vert_{\phi=0} \simeq  \frac{T^2}{24}\frac{\partial^2 }{\partial \phi^2}\left[n_\phi M^2_\phi(\phi)+n_a M^2_a(\phi)-\frac{n_\psi}{2} M^2_\psi(\phi)\right]= \left(\frac{ y_F^2}{4} + \frac{\lambda}{3} \right) T^2.
    \label{eq:ThermalMasses}
\end{equation}

\subsection{Tunneling and the bounce solution}

FOPTs proceed via quantum or thermal tunneling  from a false vacuum $\phi_f$ corresponding to a field configuration with higher potential energy, to a deeper minimum --- the true vacuum $\phi_t$ \cite{Coleman:1977py, Callan:1977pt, Linde:1981zj}. At leading order, the process can be described by a semiclassical approximation
\begin{equation}
    \Gamma(T) = A(T)e^{-S[\phi_b,T]},
\end{equation}
where $\Gamma(T)$ is the bubble nucleation rate --- also known as the decay rate of the false vacuum.

The functional $S[\phi,T]$ appearing in the exponent is the action corresponding to the saddle point (instanton) solution to the Euclidean equation of motion in 3+1 dimensions which, assuming spatial spherical symmetry, takes the form
\begin{equation}
    S[ \phi, T]  = 4\pi \int_{-\frac{1}{2T}}^{\frac{1}{2T}} d \tau \int_{0}^\infty d r\,r^2\left[ \frac{1}{2} \left( \frac{\partial \phi}{\partial \tau}\right)^2 + \frac{1}{2} \left( \frac{\partial \phi}{\partial r}  \right)^2 + V_\text{eff}(\phi,T) - V_\text{eff}(\phi_f, T) \right],
    \label{eq:general action}
\end{equation}
where $r$ denotes the radial coordinate and $\tau = -i t$ is the Euclidean time compactified along a circle of circumference $1/T$. The instanton solution or the bounce $\phi_b$ along which Eq.~\eqref{eq:general action} is evaluated can be found by solving the Euclidean equation of motion 
\begin{equation}
    \frac{\partial^2 \phi}{\partial \tau^2} + \frac{\partial^2 \phi}{\partial r^2} + \frac{2}{r}\frac{\partial \phi}{\partial r} = \frac{\partial V_{\rm eff}(\phi,T)}{\partial \phi},
    \label{eq:bounce_PDE}
\end{equation}
with the boundary conditions
\begin{equation}
\frac{\partial \phi}{\partial \tau}\bigg\rvert_{\tau = 0, \pm \frac{1}{2T}} = 0, \quad \frac{\partial \phi}{\partial r}\bigg\rvert_{r = 0} = 0, \quad \lim_{r \rightarrow \infty} \phi(r) = \phi_f.
\end{equation}

It is useful to characterise the bounce by a single length scale. Following~\cite{Bigazzi:2020phm} we define the
bubble radius $R_b$ as the radial coordinate at which the field along the
bounce configuration sits midway between the two minima,
\begin{equation}
 \phi(R_b) \equiv \frac{\phi_t + \phi_f}{2} \simeq \frac{f_a}{2},
 \label{eq:Rbdef}
\end{equation}
where we neglected the temperature
dependence of the minima, $\phi_t \simeq f_a$ and $\phi_f \simeq 0$. This is
the convention adopted throughout, both in the prefactor of
Eq.~\eqref{eq:AO4} below and in the percolation analysis of
Section~\ref{sec:O4}. Comparing $T$ with $R_b^{-1}$ one can then identify the
low-temperature ($T\ll R_b^{-1}$) and high-temperature ($T\gg R_b^{-1}$)
regimes, in which the partial differential equation reduces to an ordinary
differential equation with $O(4)$ and $O(3)$ symmetry respectively.

Explicitly, the $d$-dimensional action $S_d$, with $d$ the dimension of the Euclidean space, is given by
\begin{equation}
    S_d(T) = \frac{2 \pi^{d/2}}{\Gamma_B(d/2)} \int_{0}^{\infty} d\Tilde{r} \Tilde{r}^{d-1} \left[ \frac{1}{2}\left( \frac{d \phi}{d \Tilde{r}} \right)^2 + V_{\rm eff}(\phi,T) - V_{\rm eff}(\phi_f,T) \right],
\end{equation}
where the $\Gamma_B$ function (not to be confused with the nucleation rate) is given by
\begin{equation}
    \Gamma_B(z) = \int_0^\infty t^{z-1}e^{-t} dt,
\end{equation}
and where the bounce equation with $O(d)$ symmetry is given by 
\begin{equation}
    \frac{d^2 \phi}{d \tilde{r}^2} + \frac{d-1}{\Tilde{r}} \frac{d \phi}{d \Tilde{r}} = \frac{\partial V_{\rm eff}(\phi,T)}{\partial \phi} ,
\end{equation}
with $\Tilde{r} = r$, for $T \gg R_b^{-1}$ ($d$ = 3) and $\Tilde{r} = \sqrt{r^2+\tau^2}$ for $T \ll R_b^{-1}$ ($d$ = 4). Between these two extremes, one should solve Eq.~\eqref{eq:bounce_PDE} numerically. However, as was shown in \cite{Kobakhidze:2017mru} in the context of the electroweak transition, smoothly interpolating between the $O(3)$ and $O(4)$ actions is typically a very good approximation across the full temperature range. In practice, we follow the common approach of connecting the two regimes by choosing the minimum action at each $T$:
\begin{equation}
    S[\phi_b,T] = \min \left( S_4(T), \frac{S_3(T)}{T} \right).
\end{equation}
In particular, at sufficiently low temperature $S_3/T$ diverges and the $O(4)$ branch always
dominates; this is the regime relevant for Section~\ref{sec:O4}.

Finally, the prefactor $A(T)$ of the exponent describes the fluctuations around the saddle point solution. To formally determine $A(T)$, one should compute the fluctuation determinant from the Euclidean path integral around the bounce solution. However, for our purposes we simply resort to dimensional analysis to infer
\begin{equation}
    A(T) = \frac{1}{R_b^4} \left(\frac{S_4(T)}{2\pi}\right)^2 \quad \text{for} \quad S[\phi_b,T] = S_4(T),
    \label{eq:AO4}
\end{equation}
and
\begin{equation}
    A(T) = T^4\left( \frac{S_3(T)}{2\pi T} \right)^{3/2} \quad \text{for} \quad S[\phi_b,T] = \frac{S_3(T)}{T}.
\end{equation}
Note that in the limit in which thermal corrections to the effective potential
are negligible, $T \ll m$, the $O(4)$ action and the prefactor of
Eq.~\eqref{eq:AO4} become temperature independent. This observation underlies
the analysis of Section~\ref{sec:O4}, where it permits the percolation
integral to be evaluated in closed form.

\subsection{Phase-transition milestone temperatures}
\label{sec:PT_temp}
In this section, we will elaborate on the PT dynamics during the thermal history of the Universe. We will first focus on the simplest scenario, where at early times the PQ symmetry is restored due to sizeable thermal corrections to the quadratic term. The more involved scenarios with possible SNR are left for Section~\ref{sec:snr}. 

In this scenario, whilst the symmetry is restored, the potential has a single minimum $\phi_f = 0$. As the Universe cools down, the significance of the thermal corrections diminishes, and another minimum away from $\phi_f = 0$ can appear due to the negative quartic term in the tree-level potential. As the temperature decreases further, the new minimum, with $\phi_t \neq 0$, will be deeper than the original one. The time at which the potential energies for the true and false vacuum are equal defines the critical time (and the corresponding critical temperature obtained from the time-temperature relation):
\begin{equation}
    V_{\rm eff}(\phi_t, T_c) = V_{\rm eff}(0,T_c).
\end{equation}
For $T < T_c$, true vacuum bubbles can start to form. More precisely, the \textit{onset of nucleation} is defined as the time $t_n$ at which the Universe contains one bubble per Hubble volume
\begin{equation}
    N(t) = \frac{4\pi}{3}\int_{t_c}^{t_n} dt\frac{\Gamma(t)P_f(t)}{H(t)^3} = \frac{4\pi}{3}\int_{T_n}^{T_c} \frac{{\rm d} T}{T} \frac{\Gamma(T)P_f(T)}{H(T)^4} = 1,
\end{equation}
with $H(T)$ the Hubble rate, and where we used the time-temperature relation in the radiation domination regime:
\begin{equation}
\label{eq:time_temp}
    \frac{d T}{d t} = - H(T)T.
\end{equation}

In this work, in the initial larger scan (to be described in Section \ref{sec5:Results_and_Pheno}) we also use a simpler heuristic by requiring that the bubbles nucleate at least as fast as the Universe expands, defining the nucleation temperature $T_n$ as
\begin{equation}
    \frac{\Gamma(T_n)}{H^4(T_n)} = 1.
    \label{eq: Simpler_nuc}
\end{equation}
Assuming that the bubbles nucleate during radiation domination, we have 
\begin{equation}
    H(T) \simeq 1.66 \sqrt{g_{\ast}} \frac{T^2}{M_{\rm Pl}}, 
\end{equation}
with $\sqrt{8\pi^3/90}\simeq1.66$ and $M_\text{Pl}$ the Planck mass, and so we can infer from Eq.~\eqref{eq: Simpler_nuc}
\begin{equation}
    \frac{S_3(T_n)}{T_n} \simeq 4 \ln \left[ \frac{1}{1.66 \sqrt{g_{\ast}}} \frac{M_{\rm Pl}}{T_n} \right], \quad \text{and} \quad     S_4(T_n) \simeq 4 \ln \left[ \frac{1}{1.66 \sqrt{g_{\ast}}} \frac{M_{\rm Pl}}{R_b T_n^2} \right].
\end{equation}
In the literature, the nucleation temperature is often used as the milestone temperature at which the thermodynamic parameters determining the strength of the FOPT signal are evaluated. However, the power spectrum of the GWs associated with FOPT is produced at the completion of the transition once the different PT bubbles collide and disrupt the motion of the surrounding plasma. Since for strong supercooling, the PT can last long after the onset of nucleation, it is appropriate to associate the completion of the transition with percolation instead of nucleation.\footnote{Especially since in some cases the PT can proceed even when the nucleation criterion is never satisfied \cite{Athron:2023xlk}.} The percolation temperature is defined as the temperature at which a connected cluster of bubbles forms in a Hubble volume; the cosmological PT usually completes soon after percolation.
In the context of early Universe PTs, this happens when the probability of being in the true vacuum exceeds the critical value $P_t(T_p) \simeq 0.29$ \cite{10.1063/1.1338506, LI2020112815, LIN2018299}, which implies
\begin{equation}
    P_f(T_p) = 1-P_t(T_p) = e^{-I(T_p)} \simeq 0.71 \implies I(T_p) \simeq 0.34,
\end{equation}
wherein \cite{PhysRevD.23.876, PhysRevLett.44.631}
\begin{equation}
    I(T) = \frac{4 \pi}{3} \int_T^{T_c} dT' \frac{\Gamma(T')}{H(T') T^{'}} 
    \frac{1}{3c_s^2(T')} \frac{a^3(T')}{a^3(T)} \left( \int_T^{T'} {\rm d} \tilde{T} \frac{v_w}{3 \tilde{T} H(\tilde{T})c_s^2(\Tilde{T})} \frac{a(T)}{a(\tilde{T})} \right)^3,
\end{equation}
and where $v_w$ denotes the bubble wall speed.\footnote{For more details on the exact value of $v_w$ see Section \ref{sec3.6:hydro}.} Notice the explicit appearance of $c_s$, the speed of sound in the false vacuum plasma, that was pointed out in \cite{Matuszak:2026xsz}:
\begin{equation}
    c^2_{s} \equiv \frac{dp}{d\rho}=\frac{dp/dT}{d\rho/dT} = \frac{\partial V_\text{eff}}{\partial T} \left(T\frac{\partial^2 V_\text{eff}}{\partial T^2}\right)^{-1}\Bigg\vert_{\phi = \phi_f},
\end{equation}
which arises from the time-temperature relation in the false vacuum:
\begin{equation}
    \frac{dT}{dt} = -3 H T c_s^2,
\end{equation}
generalising the relation Eq.~\eqref{eq:time_temp}.
For very strongly supercooled PTs, the energy density of the Universe may include a non-negligible or even dominant component of vacuum energy $\Delta V$. In this case the Hubble rate should be generalised to~\footnote{More generally, as emphasized in~\cite{Matuszak:2026xsz} $H$ should be defined in terms of the volume-averaged energy density $\bar{\rho}$ --- which is introduced below --- as $H^2 = 8\pi \bar{\rho}/(3 M_{\rm Pl}^2)$. This is the formula that we also use in our refined scan with the \texttt{TransitionListener} package, as discussed in Section~\ref{sec5:Results_and_Pheno}. However, for an analytical understanding Eq.~\eqref{eq:Hubble_vac+rad} is a sufficient approximation.}
\begin{equation}
    H^2(T) = \frac{8\pi}{3 M_{\rm Pl}^2} \left( \frac{g_{\ast} \pi^2}{30} T^4 +\Delta V \right).
    \label{eq:Hubble_vac+rad}
\end{equation}
When vacuum energy dominates the energy budget, the Universe undergoes a period of thermal inflation in which it is rapidly expanding. Therefore, to make sure that percolation actually occurs and the PT completes, it is necessary to ensure that the fraction of space in the false vacuum decreases faster than the Universe expands. A sufficient condition is given by \cite{PhysRevD.46.2384, Ellis:2018mja}
\begin{equation}
    H(T) \left( 3 + 3 T c_s^2 \frac{d }{dT}I(T) \right) < 0.
    \label{eq:falsevacuum_decrease}
\end{equation}
 As shown in \cite{Ellis:2018mja}, this requirement can impose significant constraints on the duration of supercooling and set a lower bound on the percolation temperature.
Typically, it is assumed that if this condition holds at percolation, the decrease of the false vacuum volume fraction is fast enough such that the PT can complete. However, in this work, mirroring the approach of \cite{Matuszak:2026xsz}, we directly check whether there is a temperature $T_{\rm end} < T_p$ below which the false vacuum fraction asymptotically approaches zero, as by definition a PT completes when $P_t(T) \rightarrow 1$. Concretely, we only keep parameter points that satisfy $P_t(T_{\rm end}) = 0.99$ for some $T_{\rm end} < T_p$.

Finally, a part of the vacuum energy released during a FOPT is converted into thermal energy of the ambient plasma, reheating the Universe. Since the two phases can in general be at different temperatures during the transition, the resulting reheating temperature must be obtained by explicitly tracking the energy density $\rho_t$ of the true-vacuum phase. Assuming local energy conservation and negligible heat transport ahead of the bubble wall, the volume-averaged energy density $\bar\rho = P_t \rho_t + (1-P_t)\rho_f$ obeys the continuity equation $\dot{\bar\rho} = -3H(\bar\rho+\bar p)$, while the false-vacuum energy density redshifts as a decoupled fluid,
\begin{equation}
\dot\rho_f = -3H\left(\rho_f + p_f\right) \, .
\end{equation}
Combining these two relations yields an evolution equation for the true-vacuum energy density~\cite{Matuszak:2026xsz}:
\begin{equation}
\dot\rho_t = -3H\left(\rho_t + p_t\right) + \frac{\dot P_t}{P_t}\left(\rho_f - \rho_t\right) \, ,
\label{eq:rho_t}
\end{equation}
where the first term describes the usual adiabatic redshift of a perfect fluid, while the second term captures the injection of latent heat into the true vacuum as the transition proceeds, sourced by the growth of the true-vacuum fraction $P_t$. Integrating Eq.~\eqref{eq:rho_t} from an initial time at which $\rho_t = \rho_f$ (before bubble nucleation becomes efficient) up to the percolation time $t_\text{p}$, and inverting the relation $\rho_t(T_t)$ obtained from the effective potential, gives the true-vacuum temperature $T_t$ as a function of time. Recalling that the PT completes shortly after percolation, we define the reheating temperature as the temperature in the true vacuum at percolation time~\cite{Matuszak:2026xsz}
\begin{equation}
T_\text{reh} \equiv T_t(t_\text{p}) \, .
\label{eq:ourTreh}
\end{equation}
Finally, we note that in the literature, the reheating temperature is often also approximated by using energy conservation:
\begin{equation}
    T^{\rm app}_{\rm reh} = T_p (1+\alpha)^{1/4}.
\end{equation}
We have explicitly verified that for all parameter points considered in our numerical analysis, $T_{\rm reh}^{\rm app}$ computed with this approximation differs from $T_{\rm reh}$ computed with Eq.~\eqref{eq:ourTreh} at most at the level of a few per cent.

\subsection{Thermodynamic parameters for the gravitational-wave signal}

The spectral features of the GW signal are determined by thermodynamic parameters $\alpha$ and $\beta/H$ characterising the energy released and the (dimensionless) inverse duration of the PT respectively. The PT strength $\alpha$ is given by~\cite{Giese:2020rtr, Giese:2020znk}
\begin{equation}
    \alpha=\frac{\bar{\theta}_f(T_*)-\bar{\theta}_t(T_*)}{3w_f(T_*)},\qquad \bar{\theta}_{t/f}(T)=\left(\rho_{t/f}(T)-\frac{p_{t/f}(T)}{c^2_{s,t/f}(T)}\right),
\end{equation}
with the thermodynamic quantities defined in terms of the effective potential at the minima $\phi_t,\phi_f$ as
\begin{equation}
    p_{t/f}=-V_\text{eff}(\phi_{t/f}),\quad \rho_{t/f}=T \frac{\partial p_{t/f}}{\partial T} -p_{t/f},\quad w_{f}=p_{f}+\rho_{f}=T \frac{\partial p_{f}}{\partial T},
\end{equation}
where $\bar{\theta}$ is the pseudo-trace of the energy-momentum tensor of the plasma, $\rho$ is the energy density, $p$ is the pressure, $c_s$ is the sound speed of the plasma, $w$ is the enthalpy and where $T_\ast$ represents a reference temperature relevant for the production of the GW, which can be identified with the percolation temperature $T_p$, defined in the previous subsection. For weak FOPT, where $\alpha \lesssim \order{1}$, one can approximate $T_p \simeq T_n$. Note that for a relativistic fluid, we have $c_s^2=1/3$. Then the PT strength takes the form commonly used in the literature:
\begin{equation}
    \alpha = \frac{4\Delta V_\text{eff}-T\Delta\frac{\partial V_\text{eff}}{\partial T}}{3w_f}=\frac{\Delta V_\text{eff}-\frac{T}{4}\Delta\frac{\partial V_\text{eff}}{\partial T}}{\rho_f^\text{rad}},
\end{equation}
where $\Delta$ represents the difference between quantities evaluated in the false and true vacuum respectively and where we have used $p_f=\rho_f/3$, since all the particles relevant for the analysis of the PT are relativistic in the false vacuum $\phi_f=0$. Note that the usual formula where we take $w_f=4/3\rho_f^\text{rad}$ is not always valid. For instance, in the scenario of a multi-step PT or SNR, which we will discuss in Section \ref{sec:snr}, the false vacuum may not be located at $\phi_f = 0$, and so $V_{\rm eff} (\phi_f, T) \neq -p_f^{\rm rad}$. Thus, 
\begin{equation}
    w_f=T \frac{\partial p_f^\text{rad}}{\partial T} = \frac{4}{3}\frac{\pi^2g_*}{30}T^4=\frac{4}{3}\rho_f^\text{rad}
\end{equation}
should only be used when $\phi_f = 0$, i.e., when $V_\text{eff}(0,T)=-p_f^\text{rad}$ as in Eq.~\eqref{eq:rad_pressure}.
Therefore, when computing the PT strength $\alpha$, the enthalpy should always be derived via the derivative of the effective potential, not simply by replacing it with $4/3\rho_f^\text{rad}$.

For an exponential bubble nucleation rate, the inverse duration of the PT, normalised to the Hubble rate, is given by \cite{Matuszak:2026xsz}:
\begin{equation}
    \frac{\beta}{H(T_{p})} = \frac{\max(v_{w},c_s)}{R_p H(T_p)} \left(\frac{8\pi}{P_t(T_p)}\right)^{1/3}, 
    \label{eq:betaH_gen}
\end{equation}
with the mean bubble separation at percolation given by\footnote{In this work, whenever we refer to temperature or sound speed, these quantities are understood to be evaluated in the \textit{false vacuum}.}
\begin{equation}
    R_p = \left[\int_{T_p}^{T_c} dT' \frac{\Gamma(T')}{T' H(T')} \frac{e^{-I(T')}}{3\, c_{s}^2(T')} \left(\frac{a(T')}{a(T_p)} \right)^3 \right]^{-1/3}.
    \label{eq:meanbubblesep}
\end{equation}

For faster transitions that typically include smaller bubbles and weaker GW signals, one can use the simpler approximation
\begin{equation}
    \frac{\beta}{H(T_{\ast})} = T_\ast \frac{d (S_3/T)}{d T}\bigg\rvert_{T=T_\ast}.
\end{equation}
However, since we are mainly interested in strong FOPTs, we will always employ the more general formula Eq.~\eqref{eq:betaH_gen} with $R_p$ computed from Eq.~\eqref{eq:meanbubblesep}.

\subsection{Hydrodynamic parameters for the gravitational-wave signal}
\label{sec3.6:hydro}

The efficiency with which the vacuum energy released during the PT is converted into each of the three GW sources  introduced in the next subsection depends crucially on the dynamics of the bubble wall itself. Determining the bubble wall velocity $v_w$ from first principles requires solving the coupled Boltzmann equations for all species interacting with the wall, which is computationally involved. We instead rely on the local thermal equilibrium (LTE) approximation, which provides an upper bound on the wall velocity, $v_w\leq v_w^\text{LTE}$~\cite{Ai:2023see}. Since the GW amplitude grows with $v_w$, as we will see in Section~\ref{sec37:GWSignal} this amounts to an optimistic estimate of the potential GW signal.

A more subtle question concerns whether the bubble walls reach a terminal velocity before colliding with their neighbours, or whether they keep accelerating and become ultra-relativistic. The transition between these two regimes is controlled by the balance between the pressure difference driving the wall and the  frictional pressure exerted on it by the plasma. At leading order, this friction is due to the difference in the effective masses of the particles across the wall, and in the ultra-relativistic limit, produces a pressure independent of the wall Lorentz factor $\gamma_w$, the so-called Bödeker–Moore friction~\cite{Bodeker:2009qy}
\begin{equation}
P_{\rm LO} = \frac{T^2}{24}\sum_i c_i n_i \Delta m_i^2 \equiv \frac{3}{4}w_f\alpha_\infty,
\end{equation}
where the sum runs over all particles $i$ that acquire a mass difference $\Delta m_i^2$ across the wall and $n_i$ counts their internal degrees of freedom. The coefficient $c_i$ arises from the difference between Bose-Einstein and Fermi-Dirac statistics: $c_i = 1$ for bosons and $c_i = -1/2$ for fermions~\footnote{This follows from our convention of including the minus sign in the fermionic degrees of freedom, i.e $n_f = -12$.}. 

Whenever $\alpha > \alpha_\infty$, the leading-order friction alone is unable to stop the wall, and it would keep accelerating indefinitely if no further friction sources were present: resulting in the so-called runaway regime. A next-to-leading-order effect, sourced by the soft radiation of gauge bosons from the accelerating wall, provides an additional friction pressure, growing linearly with $\gamma_w=(1-v_w^2)^{-1/2}$~\cite{Bodeker:2017cim}. However, since our model does not gauge the PQ symmetry, there are no gauge bosons whose mass would depend on the background field $\phi$, and so the next-to-leading-order friction is absent for our scenario.

The fraction of the released vacuum energy  retained as kinetic energy in the bubble walls at the time of collision, rather than transferred to the surrounding plasma, defines the collision efficiency factor, which, given the above discussion, is expressed as
\begin{equation}
\kappa_\text{col} =
\begin{cases}
1-\dfrac{\alpha_\infty}{\alpha} & \text{for } \alpha >\alpha_\infty\\
0 & \text{for } \alpha \leq\alpha_\infty
\end{cases}\,.
\end{equation}

The bulk of the released vacuum energy is instead transferred to the surrounding plasma, where it sources sound waves once the bubbles collide. Denoting by $v(\xi)$ and $w(\xi)$ the self-similar velocity and enthalpy profiles of the expanding bubble as a function of the similarity coordinate $\xi=r/t$, the fraction of vacuum energy converted into bulk fluid motion is~\cite{Espinosa:2010hh}
\begin{equation}
\kappa_\text{sw} = \frac{4}{\alpha w_f v_w^3}\int d\xi~ \xi^2 \frac{v^2w}{1-v^2}.
\end{equation}
In practice, instead of directly evaluating $\kappa_{\rm sw}$, we use the fit functions provided in~\cite{Espinosa:2010hh} in our phenomenological analysis in Section \ref{sec5:Results_and_Pheno}.

Finally, the decay of sound waves after bubble collisions seeds magnetohydrodynamic (MHD) turbulence in the plasma, whose efficiency is parametrised as a fixed fraction of the sound-wave efficiency~\cite{Caprini:2024hue}
\begin{equation}
\kappa_\text{turb} = \epsilon\kappa_\text{sw}, \quad \epsilon = 0.1.
\end{equation}

These three efficiency factors $\kappa_\text{col}$, $\kappa_\text{sw}$, $\kappa_\text{turb}$, together with the PT strength $\alpha$, the mean bubble separation $R_p$, the percolation and reheating temperatures $T_p$ and $T_\text{reh}$ provide the full set of thermodynamic and hydrodynamic inputs needed to construct the power spectrum of the stochastic GW background introduced in the following subsection.

\subsection{Gravitational-wave spectra}

\label{sec37:GWSignal}

The GW spectrum from a FOPT is governed by three distinct sources: bubble collision, overlapping of sound waves and MHD turbulence. Describing the different spectra quantitatively is a formidable task, requiring extensive numerical simulations. Thus, for the present study, we employ the state-of-the-art  numerical fits recommended by the LISA Cosmology working group, which can be applied to a wide variety of models~\cite{Caprini:2024hue, Matuszak:2026xsz}. The total contribution to the stochastic GW background redshifted to today is given by
\begin{equation}
h^2\Omega_{\text{GW},0}=\mathcal{R}\left( h^2\Omega_{\text{col}} + h^2\Omega_{\text{sw}} + h^2\Omega_{\text{turb}}\right),
\end{equation}
where the redshift factor of the GW amplitude is given by
\begin{equation}
    \mathcal{R}=\Omega_{\gamma} \left( \frac{g_{s,0}}{g_{s,\rm reh}} \right)^{4/3}  \frac{g_{\ast, \rm reh}}{g_{\ast, \rm 0}} ,
\end{equation}
with $\Omega_\gamma = 2.473 \times 10^{-5}/h^2$ the photon energy density in radiation and $g_{s,0} = 3.91$, $g_{\ast,0} = 2$ the effective degrees of freedom contributing, respectively, to the entropy and energy density today. The same quantities with the subscript `reh', are instead evaluated at reheating, after the PT has transferred its energy to the relativistic degrees of freedom.

The contribution from the collisions of bubble walls is given by
\begin{equation}
    h^2\Omega_{\text{col}}(f)= h^2
A_\text{col}K_\text{col}^2\left[R_pH(T_p)\right]^2 S_\text{col}(f),\quad  A_\text{col}\simeq 3 \times 10^{-3}, 
\end{equation}
with the kinetic energy fraction $K_\text{col}$ and the spectral shape $S_\text{col}(f)$ given by
\begin{equation}
  K_\text{col}=\kappa_\text{col}\frac{\alpha}{1+\alpha},\quad S_\text{col}(f) = \left(\frac{f}{f_\text{col}}\right)^{2.4}\left[\frac{1}{2}+\frac{1}{2}\left(\frac{f}{f_\text{col}}\right)^{1.2}\right]^{-4},\quad f_\text{col}\simeq 0.487 \frac{H_{*,0}}{R_pH(T_p)},
\end{equation}
where the Hubble parameter $H_*$ at the time of GW generation, redshifted to today is given by
\begin{equation}
    H_{*,0} = \frac{a_*}{a_0}H_*\simeq 1.644\times 10^{-5}~\text{Hz}\left(\frac{T_\text{reh}}{100~\text{GeV}}\right)\left(\frac{g_{*,\text{reh}}}{100}\right)^{1/2}\left(\frac{100}{g_{s,\text{reh}}}\right)^{1/3}.
\end{equation}
The contribution from the bulk motion of the plasma (the so-called sound wave contribution) is given by
\begin{equation}
    h^2\Omega_{\text{sw}}(f)= h^2
A_\text{sw}K_\text{sw}^2\left[R_pH(T_p)\right] \Upsilon_\text{sw} S_\text{sw}(f), \quad A_\text{sw}\simeq 0.11,
\end{equation}
with the kinetic energy fraction $K_\text{sw}$ and the spectral shape $S_\text{sw}(f)$ given by
\begin{equation}
   K_\text{sw}= 0.6\kappa_\text{sw}\frac{\alpha}{1+\alpha},\quad S_\text{sw}(f)=N \left(\frac{f}{f_\text{sw,1}}\right)^3\left[1+\left(\frac{f}{f_\text{sw,1}}\right)^2\right]^{-1}\left[1+\left(\frac{f}{f_\text{sw,2}}\right)^4\right]^{-1},
\end{equation}
where $N$ is derived from the normalisation  condition
\begin{equation}
    \int_{-\infty}^\infty S_\text{sw}(f)~d\ln f=1\Longrightarrow N=\frac{4}{\pi}\frac{f_\text{sw,1}\left(f_\text{sw,1}^4+f_\text{sw,2}^4\right)}{f_\text{sw,2}^3\left(\sqrt{2}f_\text{sw,1}^2-2f_\text{sw,1}f_\text{sw,2}+\sqrt{2}f_\text{sw,2}^2\right)},
\end{equation}
with
\begin{equation}
    f_\text{sw,1}\simeq 0.2\frac{H_{*,0}}{R_pH(T_p)},\quad f_\text{sw,2}\simeq \frac{0.5}{\Delta_w}\frac{H_{*,0}}{R_pH(T_p)},\quad \Delta_w=\frac{\vert v_w-c_s\vert}{\max(v_w,c_s)}.
\end{equation}

The suppression factor $\Upsilon_\text{sw}$ in a radiation-dominated universe is given by~\cite{Guo:2020grp}
\begin{equation}
\Upsilon_\text{sw}=1-\frac{1}{\sqrt{2\tau_\text{sh}H(T_p) + 1}},\quad\tau_\text{sh}\equiv \frac{2R_p}{\sqrt{3K_\text{sw}}},
\end{equation}
which is a generalisation of $\Upsilon_\text{sw} = \min(\tau_\text{sh}H(T_p),1)$ used in~\cite{Caprini:2024hue}, and which reduces to $\tau_\text{sh}H(T_p)$ in the regime $\tau_\text{sh}H(T_p)\ll 1$. Lastly, the contribution from turbulence is given by
\begin{equation}
    h^2\Omega_{\text{turb}}(f)= h^2
A_\text{turb}K_\text{turb}^2\left[R_pH(T_p)\right]^3 S_\text{turb}(f),\quad A_\text{turb}\simeq 0.255,
\end{equation}
with the kinetic energy fraction $K_\text{turb}$ and the spectral shape $S_\text{turb}(f)$ given by
\begin{equation}
    K_\text{turb}=0.6\kappa_\text{turb}\frac{\alpha}{1+\alpha},\quad \kappa_\text{turb}=\epsilon\kappa_\text{sw}, \quad\epsilon=0.1,
\end{equation}
\begin{equation}
    S_\text{turb}(f) =\left(\frac{f}{f_\text{turb,1}}\right)^3\left[1+\left(\frac{f}{f_\text{turb,2}}\right)^{2.15}\right]^{-7.9}\begin{cases}
    \ln^2\left(1+\frac{H_{*,0}}{2\pi f_\text{turb,3}}\right) & \text{for $f\leq f_\text{turb,3}$}\\
    \ln^2\left(1+\frac{H_{*,0}}{2\pi f}\right) & \text{for $f> f_\text{turb,3}$}
    \end{cases},
\end{equation}
and
\begin{equation}
    f_\text{turb,1}=H_{*,0},\quad f_\text{turb,2}=2.2\frac{H_{*,0}}{R_pH(T_p)},\quad f_\text{turb,3}=\frac{\sqrt{3K_\text{turb}}}{4}\frac{H_{*,0}}{R_pH(T_p)}.
\end{equation}

In the SM, the effective number of relativistic species is 
parametrised in terms of the effective number of neutrino species, which
corresponds to  $N_{\nu,\text{eff}}^\text{SM} = 3.044$. The value from the combined observation of the cosmic microwave background (CMB), baryon acoustic oscillation and Big Bang nucleosynthesis (BBN), is measured to be $N_{\nu ,\text{eff}}= 2.99 \pm 0.07$~\cite{Goldstein:2026iuu}.  Therefore, one can define the discrepancy between the theory and the observation as $\Delta N_\text{eff} \equiv N_{\nu ,\text{eff}} - N_{\nu ,\text{eff}}^\text{SM} = -0.054 \pm 0.137$ within the 95\% joint confidence level. The upper limit on $\Delta N_\text{eff}$ is then $\Delta N_\text{eff} < 0.083$. In the limit where all the extra contribution to $N_{\nu ,\text{eff}}$ comes from GW, $\Delta N_\text{eff}^\text{GW}$, the latter is then constrained as
\begin{equation}
   \Delta N_\text{eff}^\text{GW}=\frac{8}{7}\left(\frac{11}{4}\right)^{4/3}\frac{1}{h^2\Omega_\gamma} \int_{f_\text{BBN}}^{f_\text{max}}d\ln f~h^2\Omega_\text{GW,0}(f)<0.083 \quad(1.96\sigma ~\text{bound})   
\end{equation}
\begin{equation}
\label{eq:neff}
   \Longleftrightarrow ~\int_{f_\text{BBN}}^{f_\text{max}}d\ln f~h^2\Omega_\text{GW,0}(f) \lesssim 4.67\times 10^{-7},
\end{equation}
where $f_\text{BBN}\simeq 1.5\times 10^{-11}$ Hz~\cite{Maggiore:2018sht} is the present-day frequency of the mode that entered the horizon at BBN.\footnote{Our results are insensitive to whether we consider BBN or CMB as the lower bound ($f_\text{CMB}\simeq 3\times 10^{-17}$ Hz~\cite{Maggiore:2018sht}), as the GW power spectrum in our model is suppressed at these frequencies.} The upper bound $f_\text{max}\simeq 10^{12}-10^{13}$ Hz comes from the constraint on the reheating temperature: the highest possible temperature of the Universe after inflation is given by $T_\text{reh}^\text{max}\simeq 6\times 10^{15}$ GeV~\cite{Servant:2023tua}. For a typical FOPT GW spectrum, the bound in Eq.~\eqref{eq:neff} can be interpreted as a bound on $h^2\Omega_\text{GW,0}(f)$ for the frequency range considered in Eq.~\eqref{eq:neff}, because the spectrum  is devoid of a very narrow peak~\cite{Maggiore:2018sht}.

\section{KSVZ with dimension-6 operator: analytical estimates}

\label{sec4:Analytics}

PTs are strongly supercooled in scenarios where a sufficiently high barrier between the false
and true minima persists down to very low temperatures, or even until $T=0$.
In this case, once the PT completes, the bubbles release a large amount of
vacuum energy, which is partially converted into a GW signal. In this work, in
order to maximise the potential detectability of GWs from a PQ transition, we
therefore consider only those scenarios in which the barrier stays intact
until $T=0$.

\subsection{Tree-level estimates}
\label{sec:tree-level}
The $U(1)_{\rm PQ}$ symmetry-breaking minimum of the tree-level potential in Eq.~\eqref{eq:treelevelpot:main} is given by 
\begin{equation}
\label{eq:vev}
    \expval{\phi} \equiv f_a = \sqrt{\frac{2}{3}} \sqrt{-\lambda \Lambda^2 + \Lambda \sqrt{\lambda^2 \Lambda^2 - 3m^2}},
\end{equation}
while the location of the peak of the barrier is given by 
\begin{equation}
        \phi_{\rm max} = \sqrt{\frac{2}{3}} \sqrt{-\lambda \Lambda^2 - \Lambda \sqrt{\lambda^2 \Lambda^2 - 3m^2}}.
\end{equation}

In particular, a tree-level barrier between the minimum at the origin (where the curvature satisfies $m^2>0$) and the minimum at a non-zero field value $f_a$ (where the curvature is given by $M_\phi^2(f_a)>0$) always exists as long as 
\begin{equation}
    \lambda < - \frac{\sqrt{3}m}{\Lambda}.
    \label{eq:existence of barrier}
\end{equation}
We also require the minimum that breaks the PQ symmetry to be the global one. Thus we need to ensure $V(0) > V(f_a)$, which gives
\begin{equation}
    \lambda < -\frac{2\,m}{\Lambda}.
    \label{eq:deeperminima}
\end{equation}

Finally, we estimate the maximum height of the tree-level barrier, with respect to the false vacuum. Expanding in $m/\Lambda$, we find
\begin{equation}
    V(\phi_\text{max}) \simeq -\frac{m^4}{4\lambda} + \order{\frac{m^6}{\Lambda^2}} \lesssim \frac{m^3 \Lambda}{8}
    \label{eq:Maxbarrier}
\end{equation}
where we used Eq.~\eqref{eq:deeperminima} in the last inequality. The barrier therefore grows as Eq.~\eqref{eq:deeperminima} approaches saturation. Consequently, Eq.~\eqref{eq:Maxbarrier} suggests that near $\lambda \lesssim -2m/\Lambda$ the bounce action stays large down to low temperatures, due to the presence of a high barrier, yielding a strong FOPT with significant amount of supercooling.

\subsection{1-loop analysis at zero temperature}
\label{sec:one-loop_constraint}
In the previous subsection we studied the existence of the minima separated by a potential barrier, taking only tree-level contributions into account. In this subsection we extend the analysis by the inclusion of 1-loop quantum corrections.

We begin by recalling that for $T=0$, the presence of a tree-level minimum at the origin, and at a non-zero field value $f_a$ requires $m^2>0$ and $\lambda<-\sqrt{3}m/\Lambda$ respectively. 
At one loop order, because of the finite counterterms, the physical mass remains unchanged, and so the curvature of the effective potential at $\phi = f_a$ is still given by $M_\phi^2(f_a)$: hence the minimum at $\phi = f_a$ still requires Eq.~\eqref{eq:existence of barrier} to be satisfied. However, to ensure the existence of a minimum at the origin, one-loop corrections change the condition $m^2>0$ to
\begin{equation}
    \xi_\text{upper}(m,\lambda,\Lambda,\mu)-\frac{3f_a^2}{16\pi^2}y_F^4>0,
\end{equation}
with
\begin{align}
    \xi_\text{upper}(m,\lambda,\Lambda,\mu)=~&\frac{234f_a^6+147\lambda\Lambda^2f_a^4+4\lambda^2\Lambda^4f_a^2-12m^2\lambda\Lambda^4}{64\pi^2\Lambda^4}\ln\left[\frac{m^2+3f_a^2\lambda+\frac{15f_a^4}{4\Lambda^2}}{\mu^2}\right]\nonumber\\
    & +\frac{45\lambda f_a^4+36\lambda^2\Lambda^2f_a^2+4m^2(16\pi^2-\lambda)\Lambda^2}{64\pi^2\Lambda^2} + \frac{m
    ^2\lambda}{4\pi^2}\ln\frac{m^2}{\mu^2}.
    \end{align}
This, then, imposes an upper bound on the Yukawa coupling $y_F$ at the scale $\mu$:
\begin{equation}
    y_F< y_F^\text{max}\equiv\left(\frac{16\pi^2\xi_\text{upper}(m,\lambda,\Lambda,\mu)}{3f_a^2}\right)^{1/4}.
    \label{eq:yFupper}
\end{equation}

Regarding the globality of the minimum at $\phi = f_a$, the condition that it is deeper than the minimum at the origin modifies the simple bound on $\lambda$ as given by Eq.~\eqref{eq:deeperminima} into a lower bound on $y_F$: 
\begin{equation}
    y_F > y_F^\text{min}\equiv\left(\frac{\xi_\text{lower}(m,\lambda,\Lambda, \mu)}{24\Lambda^4f_a^4}\right)^{1/4},
    \label{eq:yFlower}
\end{equation}
with 
\begin{align}
    \xi_\text{lower}(m,\lambda,\Lambda, \mu) =~&    99f_a^8+16\Lambda^2(8\pi^2+9\lambda)f_a^6+[16\lambda(16\pi^2+3\lambda)\Lambda^4-96m^2\Lambda^2]f_a^4 \nonumber\\ &+ 32m^2(16\pi^2-3\lambda)\Lambda^4 f_a^2
    -32m^4\Lambda^4\ln\frac{m^2}{\mu^2}\nonumber\\ &+ (711f_a^8+408\lambda\Lambda^2f_a^6+16\lambda^2\Lambda^4 f_a^4+16m^4\Lambda^4)\ln\left[\frac{m^2+3f_a^2\lambda+\frac{15f_a^4}{4\Lambda^2}}{\mu^2}\right].
\end{align}

\begin{figure}[!h]
    \centering
    \includegraphics[width=0.49\linewidth]{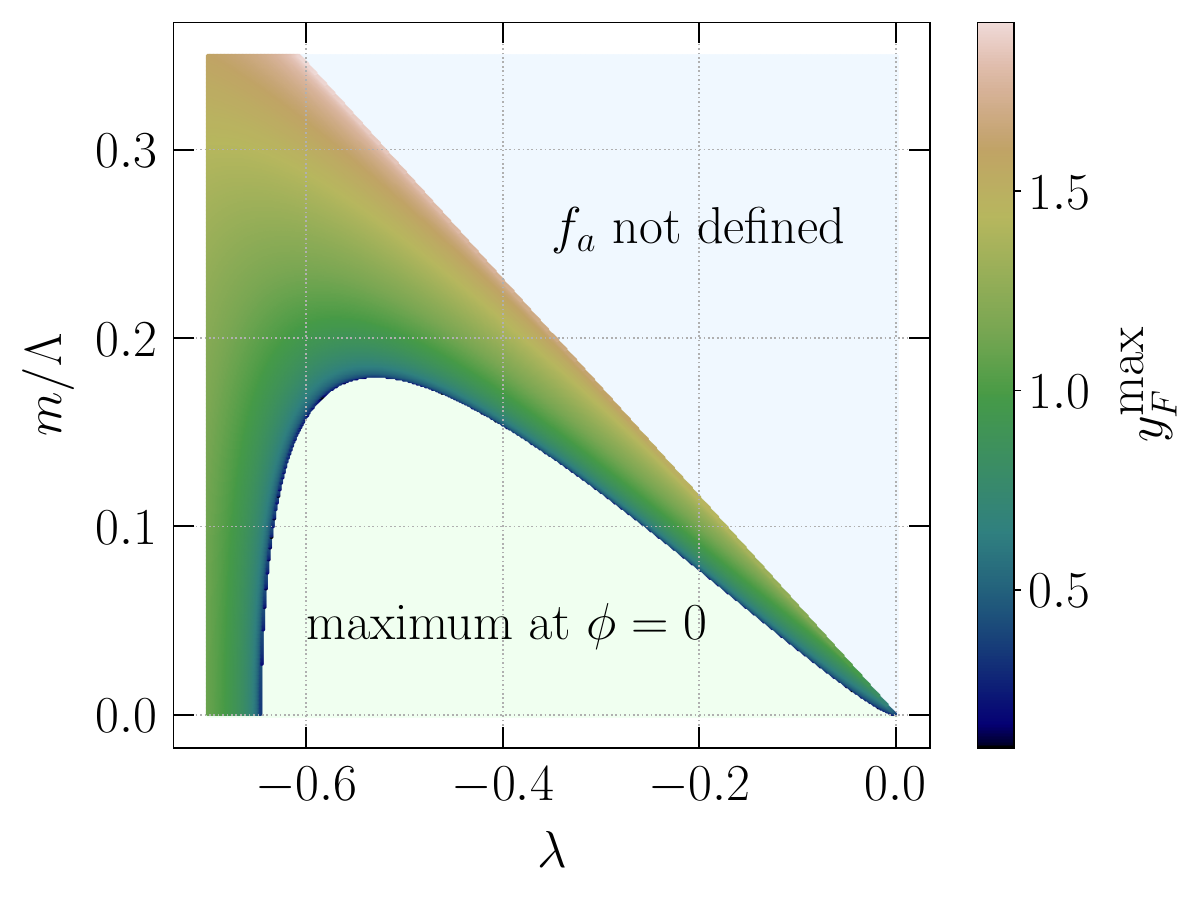}
\includegraphics[width=0.49\linewidth]{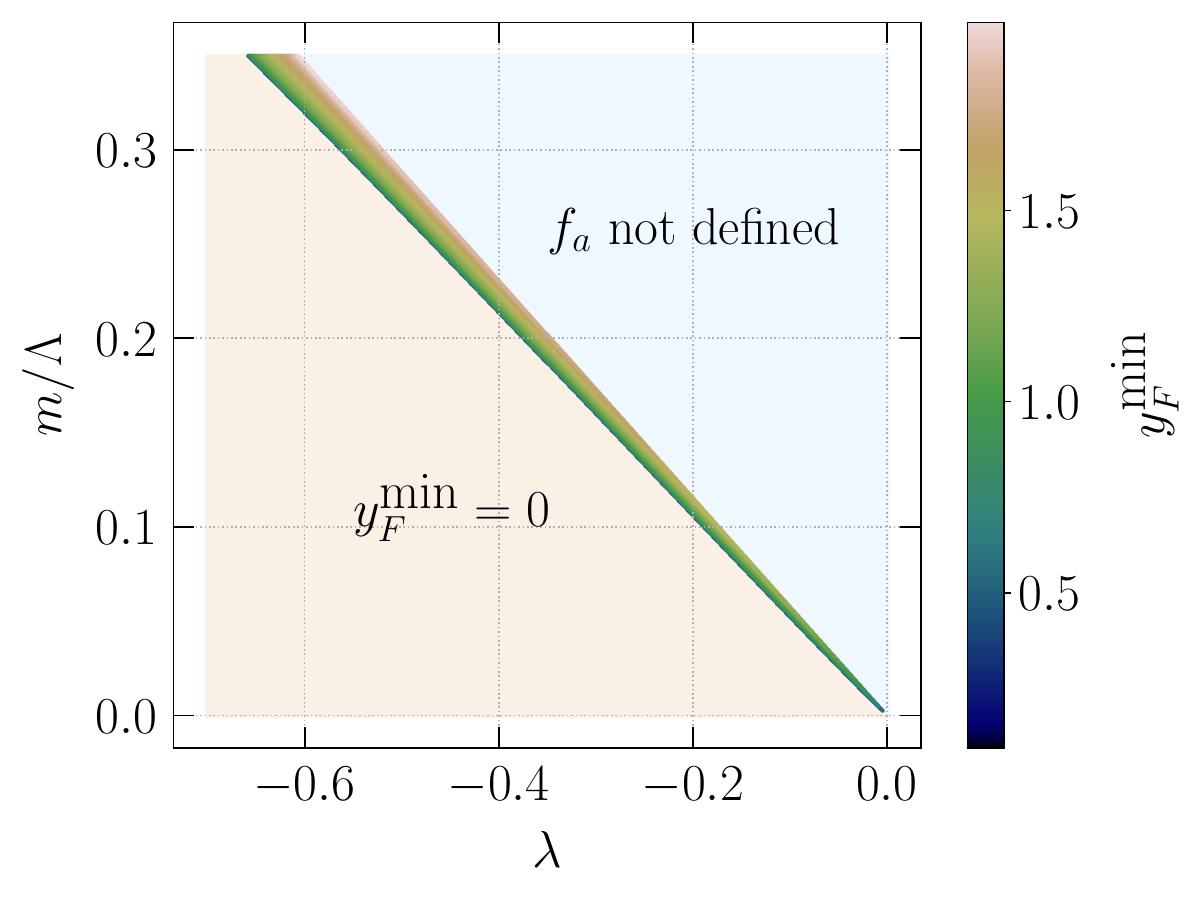}
    \caption{ The upper and lower bounds on the Yukawa coupling $y_F$, obtained respectively from the one-loop requirements at $T=0$ that the origin be a local minimum separated by a barrier from the minimum at $\phi = f_a$, and that the minimum at $f_a$ be the global one. In the beige region all positive values of $y_F$ are allowed. The blue region excludes points for which the vev $f_a$ does not exist, while in the green region the effective potential has a maximum at the origin.}
    \label{fig:yukawa_bound}
\end{figure}

These two bounds on the Yukawa coupling are depicted by the colour band in Figure~\ref{fig:yukawa_bound}. The green exclusion area in the left panel is due to the fact that the potential admits a single minimum --- at $\phi = f_a$, the extremum at the origin being a maximum. In this case, the upper bound $y_F^\text{max}$ is not defined. On the other hand, for $\xi_{\rm lower} < 0$, the extremum at $f_a$ is always the global minimum of the potential. This is the reason for denoting it by $y_F^{\rm min} = 0$ in the right panel of Figure~\ref{fig:yukawa_bound}, since in this case no lower bound on $y_F$ is imposed. In both panels, the blue exclusion area is due to Eq.~\eqref{eq:existence of barrier} not being satisfied, i.e. the vev $f_a$ is not defined.

Regarding perturbativity, in our numerical analysis we always set $\mu = \Lambda$, so the bounds in Figure~\ref{fig:yukawa_bound} are imposed at that scale. Since $\Lambda$ is the highest scale at which the EFT is defined, imposing perturbativity there ensures
that no UV Landau pole occurs within its domain of validity. In the
IR, $\lambda$ and $y_F$ are couplings of the low-energy theory only above
the thresholds at which $\phi$ and $\psi$ are integrated out; below these, the effective theory contains only the derivatively coupled axion. For the region of parameter space considered here, these thresholds (of the order $\sim f_a$) lie at most only one order of magnitude below $\Lambda$, such that the RG evolution of $\lambda$ and $y_F$ extends over a relatively short interval, thus preventing the presence of Landau poles in the IR.

\subsection{Finite temperature effects and non-standard thermal histories}
\label{sec:snr}

Previously, we always assumed that the PT starts when in addition to the minimum at the origin, another minimum with lower potential energy appears at non-zero field values. However, due to the negative quartic $\lambda < 0$, for some parameter values, the minimum at the origin may not exist at high temperature. In this case we will have a \textit{non-restoration} of the PQ symmetry (SNR), whereby the vev satisfies $\expval{\phi} \neq 0$ for $T \lesssim \Lambda$, beyond which one should specify the UV completion of the dimension-6 operator.

At high temperatures, using Eqs~\eqref{eq:VTF}, \eqref{eq:VTB}, the potential is well approximated by
\begin{equation}
    V_{\rm HT} = V_{\rm tree}(\phi) + \frac{T^2}{24}\left[M^2_\phi(\phi)+M^2_a(\phi)+6M^2_\psi(\phi)\right] =\frac{m^2_{\rm eff}(T)}{2} \phi^2 +\frac{\lambda_{\rm{eff}}(T)}{4} \phi^4 + \frac{1}{8 \Lambda^2} \phi^6,
\end{equation}
up to field-independent terms, with 
\begin{align}
& m^2_{\rm eff}(T)\equiv m^2 + \Delta_T=m^2+\left(\frac{y_F^2}{4} + \frac{\lambda}{3}\right) T^2,\\
&\lambda_{\rm eff}(T)\equiv \lambda + \frac{3}{4}\frac{T^2}{\Lambda^2},
\end{align}
where $\Delta_{T}$ is given by Eq.~\eqref{eq:ThermalMasses}. The curvature at the origin is then obtained as follows:
\begin{equation}
\label{eq:thermal_curvature}
    \frac{\partial^2 V_{\rm HT}}{\partial \phi^2}\bigg\rvert_{\phi = 0} = m^2+\left(\frac{y_F^2}{4} + \frac{\lambda}{3}\right) T^2.
\end{equation}
Thus, the PQ symmetry will not be restored at the highest temperatures accessible within the EFT if there is $T_{\rm SNR} \lesssim \Lambda$ such that
\begin{equation}
 T_{\rm SNR}^2 = -\frac{12m^2}{3y_F^2+ 4\lambda}\, .
\end{equation}
Given $\lambda < 0$, the origin of the field space is a local maximum for $T_{\rm SNR}<T\lesssim\Lambda$, while it becomes a local minimum for $T<T_{\rm SNR}$, separated by a barrier from the global minimum. The latter is located at non-zero field values, and can be approximated by\footnote{This expression relies on the leading high-temperature expansion and is
therefore valid as long as $M_{\phi,a,\psi}(\tilde f_a) < T$. It degrades once
$T$ approaches the scale of the field-dependent masses, where the neglected
cubic term and the zero-temperature one-loop corrections both become relevant.}
\begin{equation}
    \expval{\phi}\big\vert_{T \lesssim \Lambda} \equiv \tilde{f}_a(T) = \sqrt{\frac{2}{3}} \sqrt{- \lambda_{\rm eff}(T)\Lambda^2 + \Lambda \sqrt{\lambda_{\rm eff}^2(T) \Lambda^2 - 3 m_{\rm eff}^2(T)} }\,.
\end{equation}
Note that similarly to the zero-temperature tree-level analysis, where we derived Eq.~\eqref{eq:existence of barrier}, this minimum can only appear if
\begin{equation}
\lambda^2_{\rm eff}(T) > \frac{3m^2_{\rm eff}(T)}{\Lambda^2},
\end{equation}
which is automatically satisfied in the SNR scenario, where $m_{\rm eff}^2(T) < 0$.

\begin{figure}[!h]
    \centering
\includegraphics[width=0.49\linewidth]{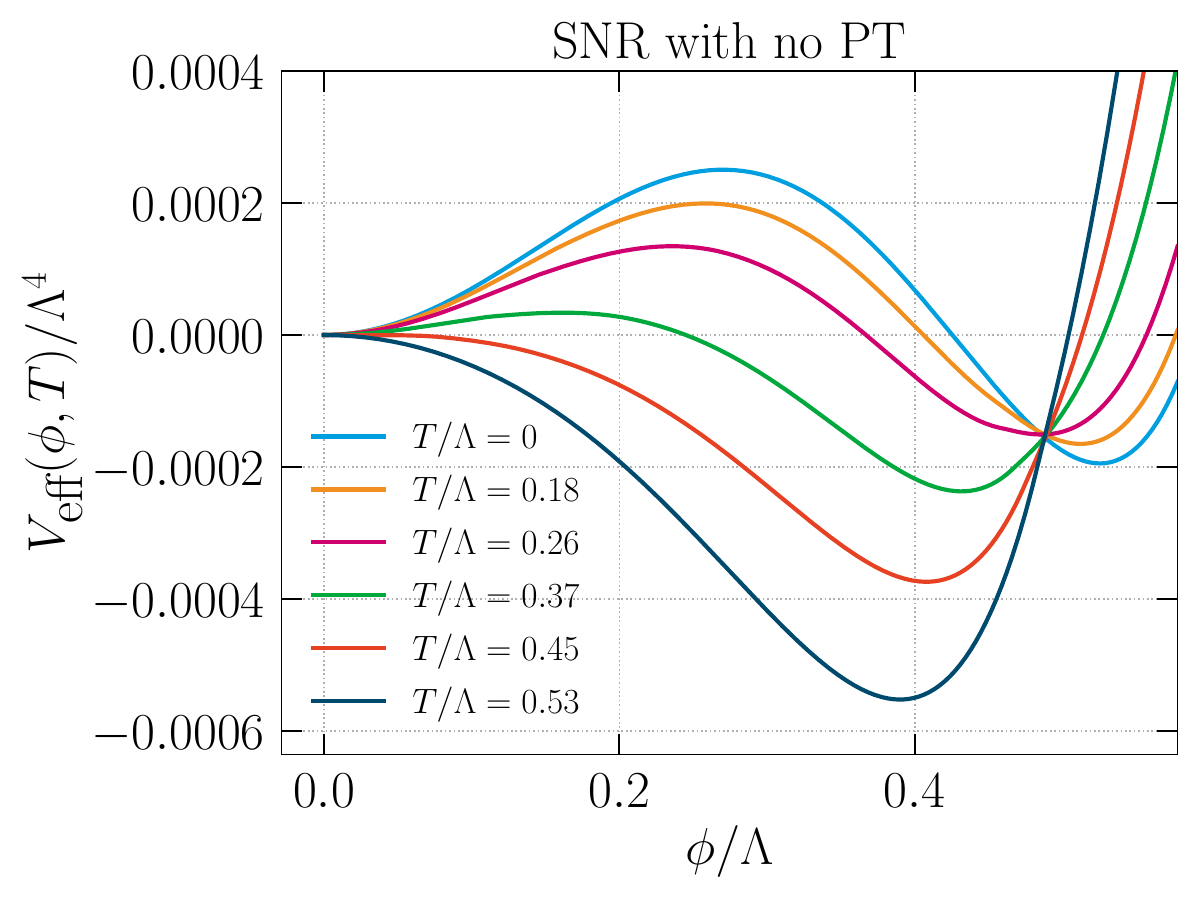}
\includegraphics[width=0.49\linewidth]{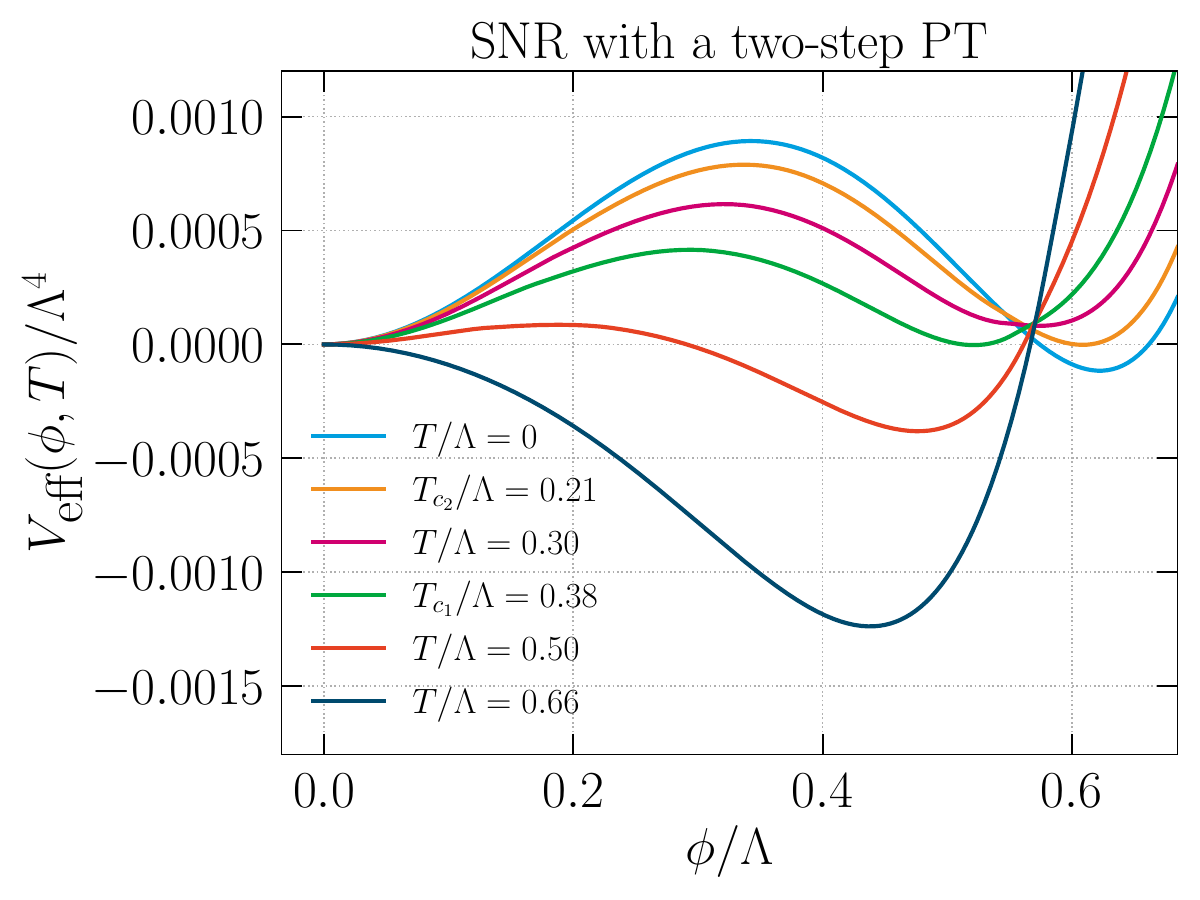}
    \caption{Evolution of the effective potential with the temperature in two SNR scenarios. In the left panel the benchmark point $m/\Lambda= 0.15,  \lambda= -0.29,  y_F= 0.17$, with $T_{\rm SNR}/\Lambda = 0.51$ leads to a scenario where there is no PT as the minimum at $f_a$, from which the Universe starts, is always the global one. In the right panel, the benchmark point $m/\Lambda = 0.22,\lambda=  -0.42, y_F =0.27$, with $T_{\rm SNR}/\Lambda = 0.64$ leads to a scenario in which the symmetry-broken phase is successively stable, metastable and stable again as the temperature decreases. This implies two critical temperatures $T_{c_1}$ and $T_{c_2}$, thus potentially a two-step PT, yielding a double-peak GW signal.}
    \label{fig:snr}
\end{figure}

Once the temperature is lowered, the maximum at the origin turns into a minimum, while the second minimum $\tilde{f}_a(T)$ typically stays intact. Then two distinct scenarios are possible:
\begin{enumerate}
    \item The minimum $\tilde{f}_a(T)$ is always deeper than the minimum at the origin. Then, the PQ field stays trapped at $\tilde{f}_a(T)$ and there is no FOPT, as shown in the left panel of Figure~\ref{fig:snr}.
    \item For some finite temperature interval in the range $ T\in[0, T_{\rm SNR}]$, the minimum at the origin is deeper than the minimum at $\tilde{f}_a(T)$, as shown in the right panel of Figure~\ref{fig:snr}.
\end{enumerate}
Regarding the latter possibility, we recall that in the previous Subsection we formulated analytical conditions on the parameter space ensuring that the symmetry-breaking minimum is always the global one at zero temperature. Thus, the second scenario entails the interesting possibility of a two-step PT, whereby one first tunnels from the symmetry broken to the symmetry unbroken phase at some intermediate $T< T_{c_1}$, and then back to the symmetry broken phase at lower temperatures $T< T_{c_2}<T_{c_1}$, where $T_{c_1}$ and $T_{c_2}$ are critical temperatures. This is illustrated in the right panel of Figure~\ref{fig:snr}. If both of these transitions are first order, the potentially observable distinctive multi-peaked GW signal could allow one to distinguish the signal from the KSVZ model with dim-6 operator from other KSVZ models that can yield FOPTs such as for instance those relying on scale invariance~\cite{DelleRose:2019pgi}.

\subsection{Zero-temperature quantum tunneling}
\label{sec:O4}

The potential barrier separating the true and false vacuum survives to $T=0$ by construction, so vacuum decay through an $O(4)$-symmetric bounce remains available for $T\ll R_b^{-1}$ when thermal nucleation is too slow to drive percolation. In this Subsection we show that such possibility can be assessed analytically, at the cost of a single bounce computation per parameter point.

For $T \ll R_b^{-1}$, and sufficiently low temperatures, the effective potential, and the $O(4)$ action $S_4$, obtained from Eq.~\eqref{eq:bounce_PDE} with $d=4$, carry no temperature dependence, and so neither does the nucleation rate
\begin{equation}
  \Gamma_0 = \frac{1}{R_b^4}\left(\frac{S_4}{2\pi}\right)^{2}e^{-S_4}\,.
  \label{eq:GammaO4}
\end{equation}
In the regime of extreme supercooling, where the temperature corrections to the effective potential can be neglected, the vacuum energy dominates the expansion, so that the Hubble rate is also constant,
\begin{equation}
  H^2_V = \frac{8\pi}{3M_{\rm Pl}^2}\,\Delta V_0\,,\qquad
  \Delta V _0\equiv V_{\rm eff}(0) - V_{\rm eff}(f_a)\,,
  \label{eq:HdS}
\end{equation}
and the Universe is de Sitter. Both ingredients of the percolation integral are then constant and it can be evaluated in closed form. Writing the exponent of the false-vacuum fraction $P_f=e^{-I_V(t)}$ as
\begin{equation}
  I_V(t)=\frac{4\pi}{3}\int_{t_V}^{t}\!\mathrm{d}t'\,\Gamma_0\,a^3(t')\,
        \left[\int_{t'}^{t}\frac{v_w\,\mathrm{d}\tilde t}{a(\tilde t)}\right]^{3},
\end{equation}
where $t_V$ marks the onset of vacuum energy domination, i.e $\Delta V_0 = \rho_{\rm rad}(t_V)$. Then, given the de Sitter geometry, $a(t)=e^{H_Vt}$, the individual scale factors $a(t')$ and $a(t)$ cancel out of the combination
\begin{align}
  a^3(t')\left[\int_{t'}^{t}\frac{v_w\,\mathrm{d}\tilde t}{a}\right]^{3}
  = e^{3H_Vt'}\left[\frac{v_w}{H_V}\left(e^{-H_Vt'}-e^{-H_Vt}\right)\right]^3  = \frac{v_w^3}{H_V^3}\left(1-e^{-H_V(t-t')}\right)^{3},
\end{align}
leaving a quantity that depends only on the elapsed time $t-t'$. Defining $x\equiv H_V(t-t_V)$, the integral is elementary:
\begin{equation}
  I_V(x)=\mathcal{N}\,F(x)\,,\qquad
  \mathcal{N}\equiv\frac{4\pi v_w^3\,\Gamma_0}{3H_V^4}\,,
  \label{eq:IdS}
\end{equation}
\begin{equation}
  F(x)=\int_0^x\!\left(1-e^{-u}\right)^3\mathrm{d}u
      = x-\frac{11}{6}+3e^{-x}-\frac32 e^{-2x}+\frac13 e^{-3x}\,,
  \label{eq:Fx}
\end{equation}
so that $dI_V/dt = H_V\mathcal{N}\left(1-e^{-x}\right)^3\to H_V\mathcal{N}$ after a few $e$-folds. 

Imposing the completion condition Eq.~\eqref{eq:falsevacuum_decrease}, $dI_V/dt>3H_V$, therefore reduces to the
Guth--Weinberg criterion~\cite{PhysRevD.23.876, Guth:1982pn}
\begin{equation}
  \mathcal{N}>3
  \qquad\Longleftrightarrow\qquad
  \frac{\Gamma_0}{H_V^4}>\frac{9}{4\pi v_w^3}\simeq 0.72\quad(v_w\to1)\,.
  \label{eq:GW}
\end{equation}
Using Eq.~\eqref{eq:GammaO4} this becomes an upper bound on the action,
\begin{equation}
  S_4 \;<\; S_4^{\rm crit}
  \;=\;  2\ln\!\frac{S_4^{\rm crit}}{2\pi} - 4\ln (R_bH_V)  + \ln\!\frac{4\pi v_w^3}{9}\,,
  \label{eq:S4crit}
\end{equation}
where, $R_b$ and $\Delta V_0$ can be expressed in units of the cut-off scale $\Lambda$:
\begin{equation}
  R_bH_V = R_b\Lambda
         \sqrt{\frac{8\pi}{3}\frac{\Delta V_0}{\Lambda^4}}\;
         \frac{\Lambda}{M_{\rm Pl}}\,.
  \label{eq:RbH}
\end{equation}

When Eq.~\eqref{eq:GW} is satisfied, the percolation temperature $T_p$ follows without further numerical work: we simply need to solve $I_V(x_p)=0.34$, i.e. $F(x_p)=0.34/\mathcal{N}$, for $x_p=H_V(t_p-t_V)$ and using $T\propto a^{-1}$ during vacuum domination to finally obtain $T_p$:
\begin{equation}
  T_p = T_V\,e^{-x_p}\,,\qquad
  T_V=\left(\frac{30\,\Delta V_0}{\pi^2 g_\ast}\right)^{1/4}.
\end{equation}

Note that $\mathcal{N}>3$, i.e $F(x_p)\lesssim 0.11$, implies that quantum tunneling can extend the supercooling only by $x_p\lesssim 1.1$ e-folds before completion fails altogether: the transition either percolates within about one Hubble time ($t_p-t_V \lesssim 1.1 H_V^{-1}$), after vacuum energy starts dominating, or never percolates at all.

\section{Results and phenomenology}

\label{sec5:Results_and_Pheno}

For numerical studies, we choose the cut-off scale at four different values $\Lambda \in [10^{10}, 10^{11}, 5\times 10^{11},10^{12}]$ GeV and vary the rest of the couplings, for each $\Lambda$, in the following range:
\begin{equation}
\label{eq:parameter_space}
    m\in [0, 35] \times\frac{\Lambda}{100},\quad \lambda\in [ -0.6, 0],\quad y_F\in [0.1, 2].
\end{equation}

\begin{figure}[!h]
    \centering
\includegraphics[width=0.49\linewidth]{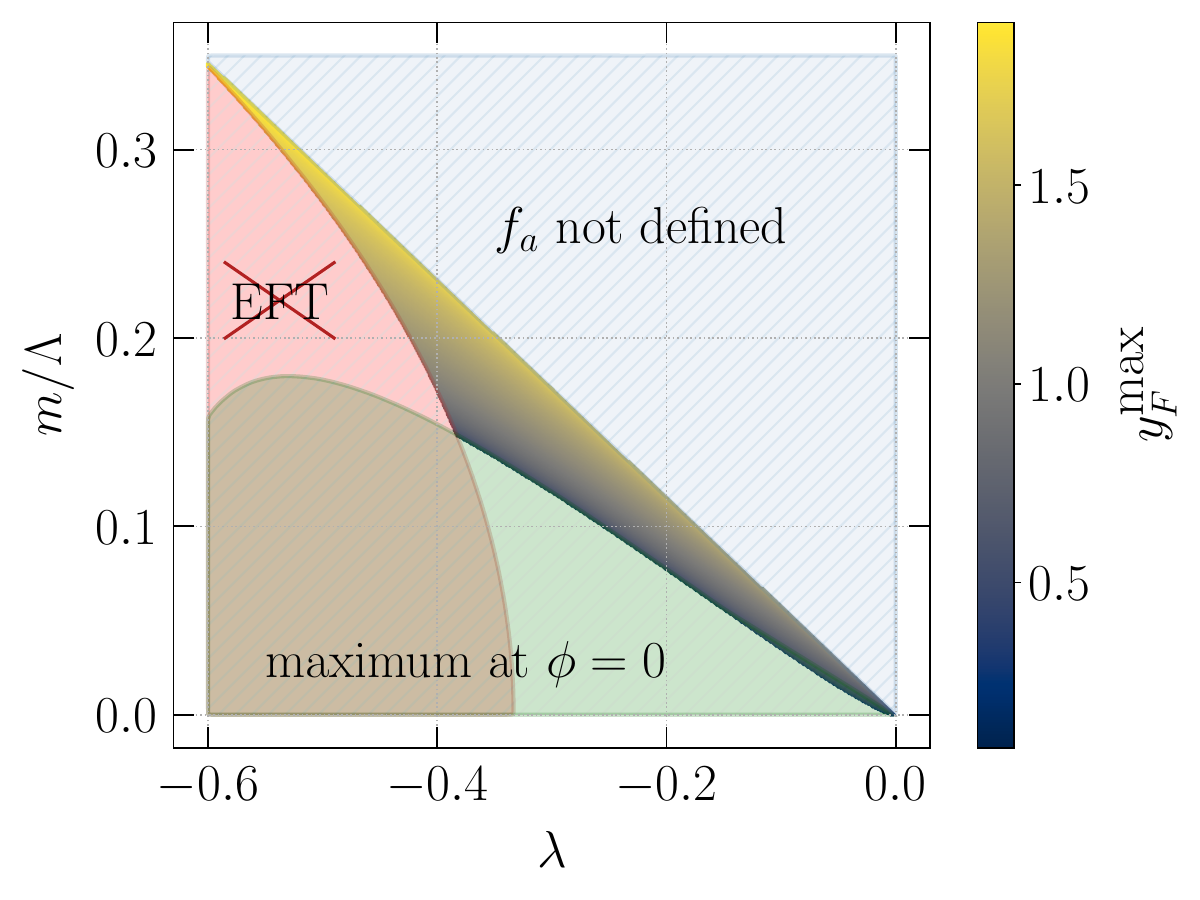}
\includegraphics[width=0.49\linewidth]{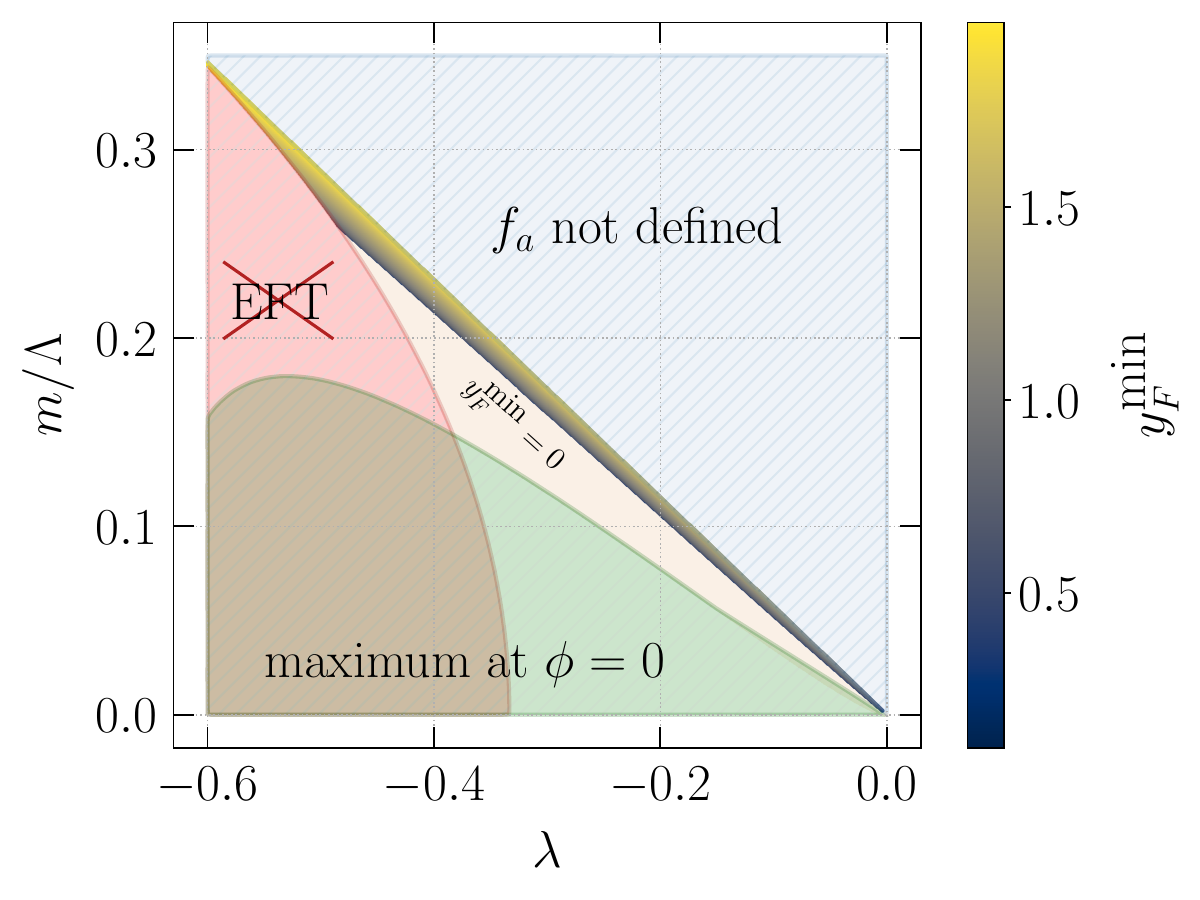}
    \caption{Allowed parameter space. It is surrounded by three exclusion regions: in blue  where the vev at $f_a$ is not found and thus the PQ symmetry is not spontaneously broken, in red where the EFT is beyond its validity region and in green where there is no minimum at the origin, thus violating our assumption on the presence of a barrier at $T=0$. This allowed region is further constrained by the maximum and minimum values that the Yukawa coupling $y_F$ can take. In the left panel the colour code shows the upper bound $y_F^\text{max}$ on $y_F$, obtained from the one-loop requirement of having a minimum at the origin at $T=0$. In the right panel the color code instead quantifies the lower bound $y_F^\text{min}$ imposed  in order for the minimum at $f_a$ to be the global one at one-loop order. It is shown with a beige colour when it is unconstrained, thus having $y_F^\text{min}=0$.}
    \label{fig:scan}
\end{figure}

The section of the  parameter space compatible with the constraints in Eqs.~\eqref{eq:existence of barrier}, \eqref{eq:yFupper}, \eqref{eq:yFlower} is shown in Figure~\ref{fig:scan}. The blue area excludes points for which $f_a$ is not defined for real-valued couplings $\lambda$ and $m^2$, which amounts to the condition in Eq. \eqref{eq:existence of barrier} not being satisfied. The red excluded area results from the EFT requirement that the vev $f_a$ should be smaller than and not too close to the scale $\Lambda$: $\Lambda > 3/2~f_a$, which, via Eq.~\eqref{eq:vev}, sets a lower bound for $\lambda$,\footnote{Combining this with our choice of parameter values given above, we also guarantee $M_{\phi,\psi}/\Lambda < 1$, which is required for consistency.}
\begin{equation}
    \lambda > - \left(\frac{1}{3}+\frac{9m^2}{4\Lambda^2}\right).
    \label{eq:EFT_consistency}
\end{equation}
Finally, because we are looking for strong FOPTs, we only study points for which the effective potential keeps a barrier at $T=0$. This imposes an upper bound for the allowed values of $y_F$, shown by the colour bar on the left panel of Figure \ref{fig:scan}. The green area excludes the region where the extremum at the origin is actually a maximum, thus violating our requirement of having two minima at $T=0$. In the right panel of Figure~\ref{fig:scan}, the beige region does not impose any lower bound on the Yukawa coupling, hence denoted by $y_F^\text{min}=0$. Finally, for the remaining region, the colour bar signifies the lower bounds for $y_F$.

\begin{figure}[!h]
    \centering
\includegraphics[width=0.99\linewidth]{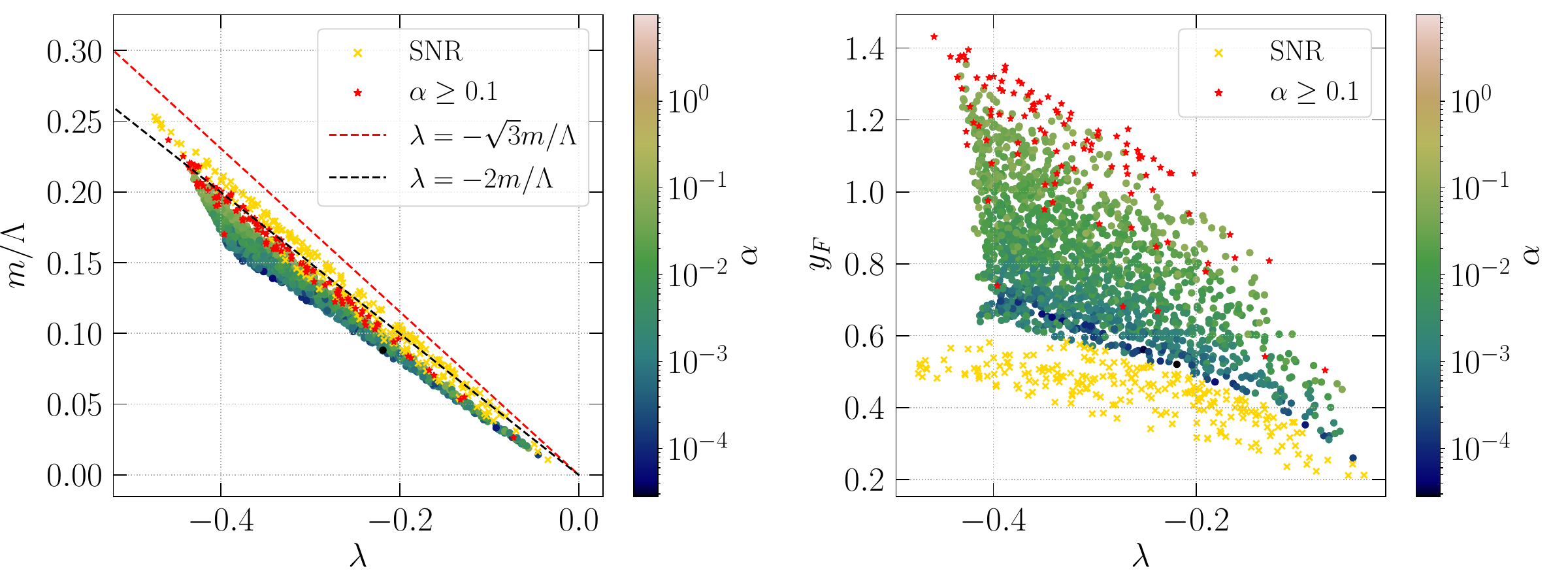}
    \caption{Projection of the parameter space that allows for FOPTs, for $\Lambda=10^{10}$ GeV. The tree-level condition for the existence of another minimum away from the origin is denoted by a red dashed line, while requiring this minimum to also be the global one at tree-level imposes a more restrictive constraint denoted by the black dashed line. Points with PT strength $\alpha\geq0.1$ are denoted by red stars, while points leading to SNR are represented by yellow crosses.}
    \label{fig:parameters}
\end{figure}

We begin our numerical studies by scanning the constrained parameter space for $\Lambda=10^{10}$ GeV, analysing PTs using the {\texttt{CosmoTransitions}} package~\cite{Wainwright:2011kj} up to the highest temperature compatible with the EFT approach, $T=\Lambda$. Those points that can give rise to the FOPT are shown in Figure~\ref{fig:parameters}, with the PT strength $\alpha$ indicated by a colour bar. From the left panel it is evident that the PT strength increases with the slope $\frac{\lambda}{m/\Lambda}$. The points with higher values for the PT strength parameter $\alpha\geq0.1$, denoted by red stars focus around the black dashed line defined as $\lambda=-2m/\Lambda$. This can be understood by recalling the discussion of Subsection~\ref{sec:tree-level}, where it was argued that points around $\lambda=-2m/\Lambda$ should result in the presence of a large barrier even at zero T, and thus be responsible for a strong FOPT. The PT strength of points leading to SNR, given by yellow crosses, is however bounded by $\alpha\lesssim 10^{-2}$. Those points result from the two-step PT scenario mentioned in Section~\ref{sec:snr}.

We notice that there are points between the red and black dashed lines. If we were considering the tree-level condition in Eq.~\eqref{eq:deeperminima} for the globality of the minimum at $f_a$, these points would be excluded. However, the 1-loop constraint derived in Section~\ref{sec:one-loop_constraint}, that we consider in our analysis, is actually less stringent than its tree-level version, in that it allows for points on either side of the $\lambda = - 2m/\Lambda$ diagonal. 

Finally, in the right panel, we can see that stronger PTs are obtained for larger $y_F$, while the SNR points are obtained for lower values of the Yukawa coupling. Although the fermionic Coleman--Weinberg contribution alone raises the potential at the broken minimum in the parameter region considered, the finite counterterms required by our on-shell-like renormalisation conditions compensate for this effect. In fact, using Eq.~\eqref{eq:VCW_1loop} and Eq.~\eqref{eq:deltam2_and_deltalam} we see that the combined fermionic Coleman--Weinberg and counterterm contribution to the vacuum-energy difference is
\begin{equation}
\frac{3 y_F^4 f_a^4}{128\pi^2}>0,
\end{equation}
which explains why increasing $y_F$ tends to increase $\Delta V$ and therefore strengthens the PT. On the other hand, the curvature of the effective potential around the origin, shown in Eq.~\eqref{eq:thermal_curvature} receives a positive contribution from $\psi$, and so the curvature can more easily be made negative for smaller values of $y_F^2$, resulting in an SNR.

One shortcoming of \texttt{CosmoTransitions} is that it is designed for fast (weak) electroweak PTs, considering the nucleation temperature as the temperature at which the PT occurs, and determining it by an approximate condition $S_3(T_n)/T_n=140$. However, as mentioned in Section~\ref{sec:PT_temp}, the relevant temperatures, in order to derive the spectrum of the stochastic GW background from FOPTs with non-negligible amount of supercooling, are the percolation and reheating temperatures. During the elaboration of this paper, a new Python package, \texttt{TransitionListener}~\cite{Matuszak:2026xsz}, based on \texttt{CosmoTransitions} functionalities, while improving on its precision and scope of potential applications, was released. Given the computational cost of applying the improved \texttt{TransitionListener} functionalities, we then performed a refined analysis only focusing on the promising points around the diagonal $\lambda=-2m/\Lambda$, as in the left panel of Figure~\ref{fig:parameters}. 

\begin{figure}[!h]
    \centering
\includegraphics[width=1\linewidth]{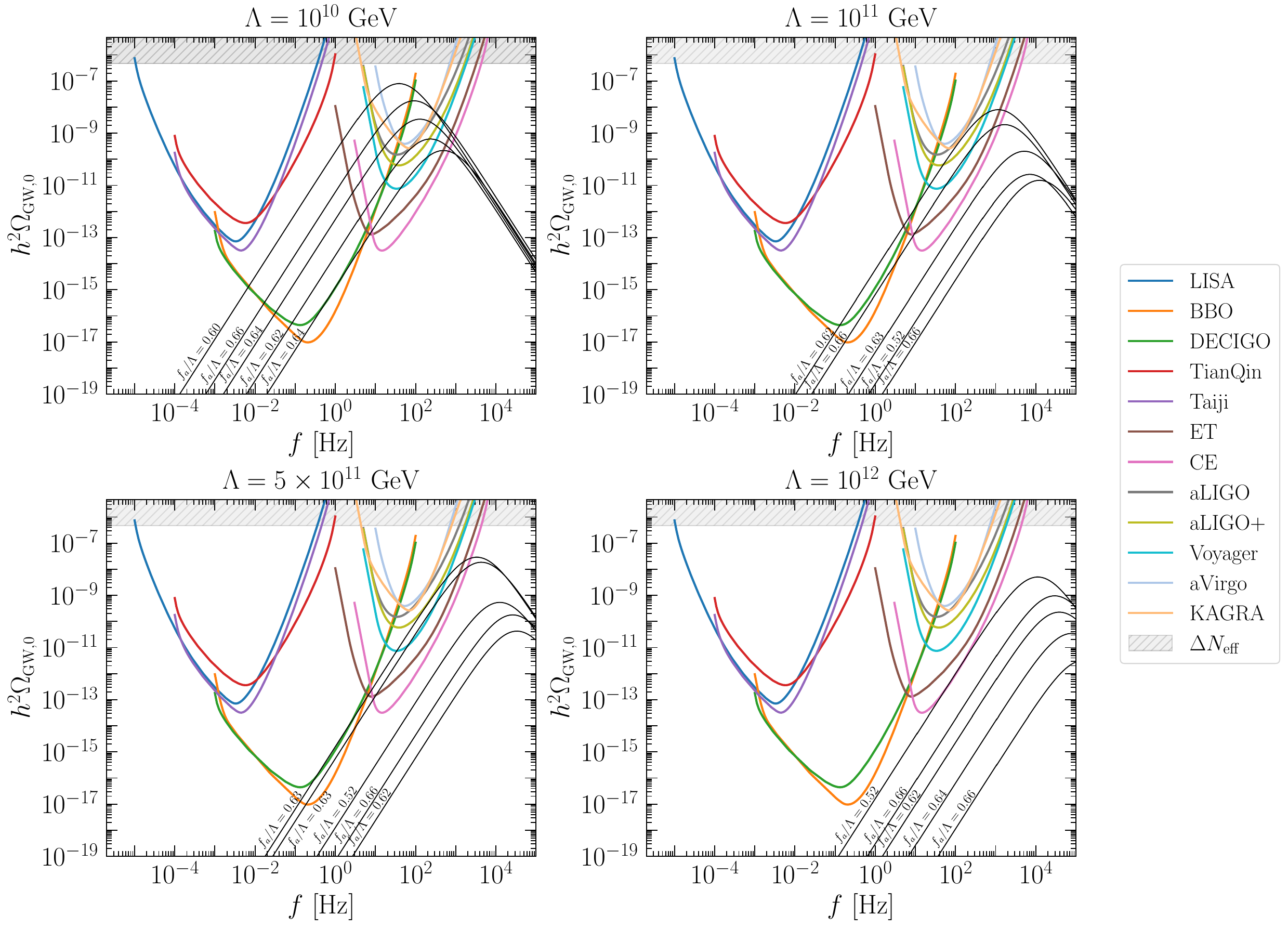}
    \caption{GW power spectrum $h^2\Omega_\text{GW,0}(f)$ as a function of the frequency $f$ for each value of $\Lambda \in [10^{10}, 10^{11}, 5\times 10^{11},10^{12}]$ GeV. In each panel we present the GW power spectrum for each benchmark point with a solid black line, to which we associate the corresponding value of the ratio $f_a/\Lambda$. Sensitivity curves of different GW interferometers are depicted by coloured solid lines and the exclusion area from the constraint on $\Delta N_\text{eff}$ is shown by a gray band.}
    \label{fig:gw_all_LAM}
\end{figure}

With the new package at hand, we scan the parameter space satisfying $\lambda\simeq -2m/\Lambda$ for each value of $\Lambda \in [10^{10}, 10^{11}, 5\times 10^{11},10^{12}]$ GeV and compute the GW power spectrum $h^2\Omega_\text{GW,0}$ for several benchmark points, which we display a selection of in Table~\ref{tab:BPs}. We present our results in Figure~\ref{fig:gw_all_LAM}, where, next to each power spectrum, we indicate the value of the corresponding ratio $f_a/\Lambda\simeq \sqrt{2m/\Lambda}$. In addition, we show the power-law-integrated sensitivity curves~\cite{Thrane:2013oya} of different GW interferometers, such as the space-based detectors LISA~\cite{LISA:2017pwj}, BBO~\cite{Corbin:2005ny}, DECIGO~\cite{Kawamura:2011zz}, TianQin~\cite{TianQin:2015yph}, Taiji~\cite{Hu:2017mde}, and the ground-based detectors ET~\cite{Punturo:2010zz}, CE~\cite{Reitze:2019iox}, Advanced LIGO~\cite{LIGOScientific:2014pky}, its upgrades Advanced LIGO Plus~\cite{Cooper:2022jfr} and LIGO Voyager~\cite{LIGO:2020xsf}, Advanced Virgo~\cite{VIRGO:2014yos}, KAGRA~\cite{Somiya:2011np}, assuming an observation time of 4 years and signal-to-noise ratio of 10. Furthermore, we display the constraint from Eq.~\eqref{eq:neff} from the contribution of GWs to $\Delta N_\text{eff}$, as a gray exclusion band.

\begin{table}[!h]
\centering
\begin{tabular}{cc}
\begin{tabular}{|p{3.03cm}|c|c|c|c|}
\hline
 & $m/\Lambda$ & $\lambda$ & $y_F$\\
 \hline
\multirow{5}{*}{\hspace{3.5mm}$\Lambda=10^{10}$ GeV} & 0.23 & -0.45 & 1.44 \\
\cline{2-4}
 & 0.21 & -0.42 & 1.36\\
\cline{2-4}
 & 0.17 & -0.36 & 1.04\\
\cline{2-4}
 & 0.17 & -0.35 & 1.20\\
\cline{2-4}
 & 0.18 & -0.39 & 1.04\\
\hline
\end{tabular} & \begin{tabular}{|c|c|c|c|}
\hline
 & $m/\Lambda$ & $\lambda$ & $y_F$ \\
 \hline
\multirow{5}{*}{$\Lambda=10^{11}$ GeV} & 0.19 & -0.38 &  1.25 \\
\cline{2-4}
 & 0.21 & -0.42 & 1.29 \\
\cline{2-4}
 & 0.19 & -0.42 & 1.06 \\
\cline{2-4}
 & 0.20 & -0.40 & 1.31 \\
\cline{2-4}
 & 0.12 & -0.25 & 1.02 \\
\hline
\end{tabular} \\[4.25em]
\begin{tabular}{|c|c|c|c|}
\hline
 & $m/\Lambda$ & $\lambda$ & $y_F$ \\
 \hline 
\multirow{5}{*}{$\Lambda=5\times 10^{11}$ GeV} & 0.21 & -0.42 & 1.29 \\
\cline{2-4}
 & 0.17 & -0.37 & 1.09 \\
\cline{2-4}
 & 0.20 & -0.40 & 1.31 \\
\cline{2-4}
 & 0.12 &-0.25 & 1.02 \\
\cline{2-4}
 & 0.19 & -0.39 & 1.22 \\
\hline
\end{tabular} & \begin{tabular}{|c|c|c|c|}
\hline
 & $m/\Lambda$ & $\lambda$ & $y_F$ \\
 \hline 
\multirow{5}{*}{$\Lambda=10^{12}$ GeV} & 0.21 & -0.42 & 1.29 \\
\cline{2-4}
 & 0.17 & -0.37 & 1.09 \\
\cline{2-4}
 & 0.18 & -0.39 & 1.13 \\
\cline{2-4}
 & 0.19 & -0.41 & 1.07 \\
\cline{2-4}
 & 0.12 & -0.25 & 1.02 \\
\hline
\end{tabular} \\
\end{tabular}
\caption{Different sets of five benchmark points, for each value of $\Lambda \in [10^{10}, 10^{11}, 5\times 10^{11},10^{12}]$ GeV, used to derive the GW power spectrum $h^2\Omega_\text{GW,0}(f)$ in Figure~\ref{fig:gw_all_LAM}.}
\label{tab:BPs}
\end{table}

We first directly see that it is easier to obtain a GW spectrum which reaches the sensitivity of the considered detectors for lower values of $\Lambda$. For instance, in the upper left panel, the signal could potentially be detected by the detectors aLIGO, aVirgo or KAGRA, which are currently in service. This is expected because the frequency at the peak of the GW signal typically scales like  $f^\text{peak}\sim 10^{-5}~\text{Hz}\times(T_\text{reh}/100~\text{GeV})\beta/H(T_p)$, with $T_\text{reh}\sim \Lambda/10$. Thus, for $\Lambda=10^{10}$ GeV, we obtain $f^\text{peak}\sim 10^2~\text{Hz}\times\beta/H(T_p)$, while for $\Lambda=10^{12}$ GeV, we obtain  $f^\text{peak}\sim 10^4~\text{Hz}\times\beta/H(T_p)$. It is therefore easier to find points giving a GW signal within the considered detector sensitivities for $\Lambda =10^{10}$ GeV, as landing in an observable frequency range only requires weakly supercooled PTs with $\beta/H(T_p)\sim \order{1}$. By contrast, points with $\Lambda=10^{12}$ GeV would require strongly supercooled PTs, with $\beta/H(T_p)\sim 10^{-2}$, in order to reach the detectors sensitivity, as the inverse time duration of a PT is inversely correlated to its strength. Therefore, these points will not be observable with our present detection capabilities, since causality requires, $R_p H(T_p) \lesssim 1$, which, assuming $v_w \simeq 1$ and using Eq.~\eqref{eq:betaH_gen} translates into the bound $\beta/H(T_p) \gtrsim 4.43$. Hence, for very high scales $\Lambda \geq 10^{12}$ GeV, the peak frequency cannot be shifted to observable values by increasing the duration of the PT, i.e. by decreasing $\beta/H(T_p)$, and such high scale could only be probed by future high frequency GW detectors.

\begin{figure}[!h]
    \centering
\includegraphics[width=0.8\linewidth]{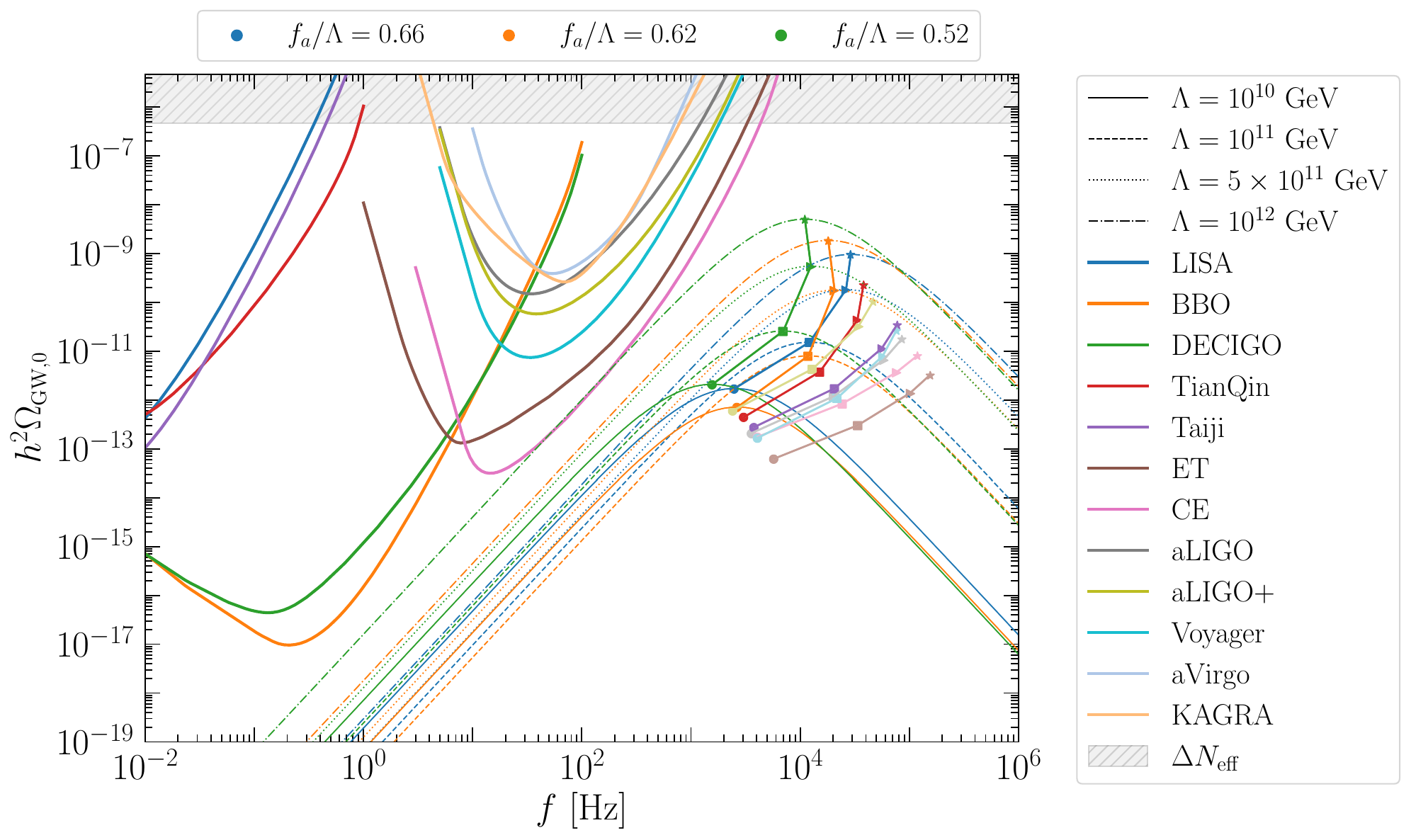}
    \caption{GW power spectrum $h^2\Omega_\text{GW,0}(f)$ as a function of the frequency $f$ for three benchmark points: $m/\Lambda = 0.21$, $\lambda = -0.42$, $y_F = 1.29$ (blue); $m/\Lambda = 0.17$, $\lambda = -0.37$, $y_F = 1.10$ (orange); $m/\Lambda = 0.12$, $\lambda = -0.25$, $y_F = 1.02$ (green). For each benchmark point we show the GW signal for $\Lambda = 10^{10}$ GeV (solid), $\Lambda = 10^{11}$ GeV (dashed), $\Lambda = 5\times 10^{11}$ GeV (dotted) and $\Lambda = 10^{12}$ GeV (dash-dotted), as well as the value of the ratio $f_a/\Lambda$. Moreover, we display the peak of each power spectrum with a dot $\bullet$ ($\Lambda = 10^{10}$ GeV), a square $\blacksquare$ ($\Lambda = 10^{11}$ GeV), a triangle $\blacktriangleright$ ($\Lambda = 5\times 10^{11}$ GeV) and a star $\boldsymbol{\star}$ ($\Lambda = 10^{12}$ GeV), for these three benchmark points, as well as for seven other ones. Sensitivity curves of different GW interferometers are depicted by coloured solid lines and the exclusion area from the constraint on $\Delta N_\text{eff}$ is shown by a gray band.}
    \label{fig:gw_common_BP}
\end{figure}

Among the analysed points, we find that ten of them give rise to a FOPT for each $\Lambda \in [10^{10}, 10^{11}, 5\times 10^{11},10^{12}]$ GeV. It is then interesting to look at how the value of the scale $\Lambda$ influences the GW signal for the same benchmark point. In Figure~\ref{fig:gw_common_BP}, we then present the GW spectrum for only three of these benchmark points, for each value of $\Lambda$, so that the plot does not look cumbersome. However, we also indicate the position of the peak $h^2\Omega_\text{GW,0}^\text{peak}$ of the GW signal for the ten benchmark points. We can then observe a trajectory $h^2\Omega_\text{GW,0}^\text{peak}$ as a function of $\Lambda$. It is apparent that increasing $\Lambda$ also increases $h^2\Omega_\text{GW,0}^\text{peak}$ and the corresponding frequency $f^\text{peak}$, thus shifting the peak of the signal to the upper-right direction. Nevertheless, for even higher values of $\Lambda$, the peak frequency $f^\text{peak}$ starts to decrease. We can see that it decreases even faster when the PT is stronger. Indeed, the curvature of the $h^2\Omega_\text{GW,0}^\text{peak}$ trajectory is large for stronger PTs (e.g.  green and orange ones), while it is small for weaker PTs (e.g pink and brown ones). Therefore, even though from Figure~\ref{fig:gw_all_LAM} it seems easier to find points leading to a GW signal in the sensitivity bands of the current GW detectors for smaller values of $\Lambda$, actually, for a point giving a too weak GW signal for $\Lambda=10^{10}$ GeV for instance, it is possible for this same point to reach the sensitivity of those detectors for a large enough $\Lambda$, provided that the PT completes.

The curvature of these $h^2\Omega_\text{GW,0}^\text{peak}$ trajectories can be explained as follows: the peak amplitude and frequency respectively scale as $h^2\Omega_\text{GW,0}^\text{peak}\sim [\beta/H(T_p)]^{-p}$, with $p\in[1,2,3]$ and $f^\text{peak}\sim T_\text{reh}\times\beta/H(T_p)$. For strongly curvy trajectories like the green or orange ones, the ratio between the vacuum energy $\Delta V$ and the percolation temperature $T_p$ strongly increases with $\Lambda$, as does $H(T_p)\sim \frac{T_p^2}{M_{\rm Pl}}\sqrt{1+\Delta V/T_p^4}$, thus implying the decrease of $\beta/H(T_p)$ with $\Lambda$. On the other hand the reheating temperature moderately increases with $\Lambda$. Therefore, in the interval $\Lambda\in[10^{10}, 10^{12}]$ GeV, $T_\text{reh}\times\beta/H(T_p)$ first increases, reaching a maximum, and then decreases. This is why the peak of the signal first moves to higher frequencies, then to lower frequencies, meanwhile always increasing in amplitude. For weakly curvy trajectories like the pink or brown ones, we still have $dT_\text{reh}/d\Lambda>0$ and $d^2T_\text{reh}/d\Lambda^2<0$ with the same magnitude. However, this time $\Delta V/T_p^4$ slowly increases with $\Lambda$, and thus $\beta/H(T_p)$ still decreases with $\Lambda$ but at smaller rate $d^2(\beta/H(T_p))/d\Lambda^2$. This implies that in the interval $\Lambda\in[10^{10}, 10^{12}]$ GeV, $T_\text{reh}\times\beta/H(T_p)$ increases but does not reach a maximum. Therefore, for these trajectories, the peak of the signal solely moves in the upper-right direction.

For the points of the scan for which the percolation temperature through thermal tunneling could not be determined, we have applied the analysis of Section~\ref{sec:O4}. Since the $O(4)$ action is temperature independent, and depends only on the ratios of the dimensionful parameters, this requires a single bounce solution per parameter point, regardless of whether we consider $\Lambda = 10^{10}$ GeV or higher. We find $S_4 \in [296, \infty)$, with $S_4$ diverging for parameter values that make the effective potential nearly degenerate at $T=0$, to be compared with the threshold of Eq.~\eqref{eq:S4crit}. The latter depends logarithmically on the ratio $\Lambda/M_{\rm Pl}$, with $M_{\rm Pl}$ introducing a fixed scale, as we saw in Eq.~\eqref{eq:RbH}. In particular $S^{\rm crit}_4$ evaluates to $S_4^{\rm crit} \in [88.3, 110]$ for $\Lambda = 10^{10} $ GeV, $ S^{\rm crit}_4 \in [79.1, 101.8]$ for $\Lambda = 10^{11}$ GeV, $ S^{\rm crit}_4 \in [72.6, 95.4]$ for $\Lambda = 5 \times 10^{11}$ GeV and $S^{\rm crit}_4 \in [69.9, 92.6]$ for $\Lambda = 10^{12}$ GeV. Not a single point satisfies the completion criterion Eq.~\eqref{eq:GW}: the closest exceeds $S_4^{\rm crit}$ by $\Delta S_4\simeq 186$, corresponding to $\Gamma_0/H_V^4$ falling short of satisfying the Guth-Weinberg criterion by approximately 94 orders of magnitude.

Overall, within the refined scan around $\lambda\simeq-2m/\Lambda$, we find
that all strong PQ FOPTs which successfully complete do so through thermal, and not quantum,
tunnelling. None of the points for which thermal percolation fails is rescued
by subsequent $O(4)$ vacuum tunnelling.
The scan spans the range $0.18 \lesssim f_a/\Lambda \lesssim 2/3$, with $3 \%$ of the scanned points resulting in a completed FOPT. Those points clustered  in a narrow band: $0.42\lesssim f_a/\Lambda\lesssim 2/3$, with the fraction of completing points falling monotonically from about $70\%$ near $ 0.6 \lesssim f_a/\Lambda \lesssim 2/3$ to $5 \%$ for $0.42 \lesssim f_a/\Lambda \lesssim 0.5$. Demanding a completed PQ FOPT thus strongly constrains the decay constant: it is restricted to $0.42\lesssim f_a/\Lambda\lesssim 2/3$, so that $f_a/\Lambda$ varies by only a factor $\simeq 1.6$. This is why the ratios $f_a/\Lambda$ quoted in Figures~\ref{fig:gw_all_LAM} and~\ref{fig:gw_common_BP} cluster so tightly.

Having fixed $f_a$ to within a narrow band below $f_a \simeq \tfrac23\Lambda$, the axion relic abundance ceases to be an independent quantity and becomes a prediction of the cut-off scale alone. The PQ symmetry is broken here after inflation, by a thermal transition at $T\sim\Lambda/10$, so that the misalignment angle is fixed by averaging over causally disconnected patches and Eq.~\eqref{eq:OmegaDM} applies. Combining it with $0.42\lesssim f_a/\Lambda\lesssim 2/3$ gives the results in Table~\ref{tab:DMabundance}.
\begin{table}[!h]
\centering
\begin{tabular}{ccc}
\hline
\noalign{\vskip 2pt}
$\Lambda$ [GeV] & $f_a$ [GeV] & $\Omega^a_{\rm DM} h^2$ \vspace{2pt}\\
\hline
\noalign{\vskip 2pt}
$10^{10}$        & $(4.2$--$6.7)\times10^{9}$  & $0.0013$--$0.0022$ \\
$10^{11}$        & $(4.2$--$6.7)\times10^{10}$ & $0.019$--$0.033$ \\
$5\times10^{11}$ & $(2.1$--$3.3)\times10^{11}$ & $0.13$--$0.22$ \\
$10^{12}$        & $(4.2$--$6.7)\times10^{11}$ & $0.29$--$0.49$ \\
\hline
\end{tabular}
\caption{Axion relic abundance for the four values of the cut-off scale $\Lambda$, using $0.42\lesssim f_a/\Lambda\lesssim 2/3$ from the requirement that the PQ transition completes via a strong FOPT. The observed value $\Omega^a_{\rm DM}h^2=0.12$ is reproduced for $\Lambda\simeq(3-5)\times10^{11}$~GeV.}
\label{tab:DMabundance}
\end{table}
The observed value $\Omega^a_{\rm DM}h^2=0.12$ is thus reproduced for $\Lambda\simeq(3-5)\times10^{11}$~GeV, below which the QCD axion accounts for only a fraction of the DM --- a few percent at $\Lambda=10^{10}$~GeV and $\sim20\%$ at $10^{11}$~GeV --- while above it the abundance is overproduced by up to a factor of four at $\Lambda=10^{12}$~GeV, so that the highest cut-off scales we consider are excluded on cosmological grounds for the minimal QCD axion scenario. We stress that this bound is conservative: in the post-inflationary scenario, the decay of the cosmic-string network contributes an additional, comparable population of axions, which would strengthen the overproduction constraint and lower the value of $\Lambda$ at which the observed abundance is saturated.

This sharpens the multimessenger picture. The GW signal is most readily observable at low $\Lambda$, where the peak frequency lies within the reach of the ground-based interferometers, when the axion is a subdominant component of the DM. In this case, other particles of the UV theory responsible for generating the dimension-6 operator can provide a viable DM candidate --- some examples will be discussed in Section \ref{sec6:UVcompletions}. Conversely, the scales at which the axion saturates the relic abundance push the signal towards frequencies which are beyond the reach of the present GW detectors, while remaining potentially accessible for future GW experiments. The region in which both a detectable stochastic background and complete axion DM abundance are achieved is confined to $\Lambda\simeq(3-5)\times10^{11}$~GeV. 

\section{UV completion for the effective operator}

\label{sec6:UVcompletions}

In this section, we present several renormalisable UV completions that generate the effective dimension-six operator. In Section~\ref{sec61.realsinglet}, we begin with the simplest possible UV completion, obtained by extending the PQ scalar sector with a heavy real scalar singlet. This proof-of-concept example demonstrates that the effective potential in Eq.~\eqref{eq:Vtree_beforeSSb} can arise from a renormalisable UV theory. Building on this construction, the following two subsections consider physically motivated scenarios inspired by neutrino physics and multi-axion models. Besides providing UV completions of the effective operator, these models also offer viable DM candidates, including in regions of parameter space where the conventional axion misalignment mechanism underproduces the observed DM abundance.

\subsection{Real scalar singlet}

\label{sec61.realsinglet}

First, as a proof of principle, we consider the minimal extension of the PQ scalar sector by a heavy real singlet. Our aim is not to construct a complete UV model, but rather to demonstrate that the tree-level potential in Eq.~\eqref{eq:treelevelpot:main} can arise from a renormalisable theory after integrating out heavy degrees of freedom.

The relevant parts of the UV Lagrangian including a new scalar singlet $S$ are given by
\begin{equation}
    \mathcal{L}_{\rm UV} \supset  \abs{\partial_{\mu} \Phi}^2 + \frac{1}{2}(\partial_{\mu}S)^2 -\mu_\Phi^2 \abs{\Phi}^2 -\frac{\mu_S^2}{2}S^2 - \frac{\kappa}{3} S^3 - \kappa_{\Phi S} \abs{\Phi}^2 S  - \lambda_\Phi \abs{\Phi}^4 -\frac{\lambda_S}{4}S^4 -\frac{\lambda_{\Phi S}}{2} \abs{\Phi}^2 S^2. 
\end{equation}
To match the UV theory to the EFT Lagrangian $\mathcal{L}_{\rm EFT}$ at tree-level we use the standard EFT result,
\begin{equation}
    \mathcal{L}_{\rm EFT}(\Phi)  = \mathcal{L}_{\rm UV}(\Phi, \hat{S}(\Phi)),
\end{equation}
where $\hat{S}$ can be found by solving its equation of motion in the background of $\Phi$.\footnote{For very clear explanations about this method and its theoretical background see \cite{Henning:2014wua}.} The equation of motion for S is given by
\begin{equation}
    \left(\Box  + \mu_S^2  + \lambda_{\Phi S} \abs{\Phi}^2 \right)S + \kappa_{\Phi S} \abs{\Phi}^2 + \kappa S^2 + \lambda_S S^3 = 0. 
    \label{eq:singlet_PQ_EOM}
\end{equation}
Organising the solution in powers of $\abs{\Phi}^2$, we then find that to leading order
\begin{equation}
        \hat{S} = - \frac{\kappa_{\Phi S} \abs{\Phi}^2}{\left(\Box  + \mu_S^2  + \lambda_{\Phi S} \abs{\Phi}^2 \right)} + \order{\abs{\Phi}^4}.
\end{equation}
Assuming that $\mu_S$ is larger than any other energy scale: $\Box, \abs{\Phi}^2 \ll \mu_S^2$, we can then expand this equation as,
\begin{equation}
    \hat{S} = - \frac{\kappa_{\Phi S}\abs{\Phi}^2}{\mu_S^2} + \frac{\kappa_{\Phi S}}{\mu_S^4}\left( \Box + \lambda_{\Phi S} \abs{\Phi}^2 \right)\abs{\Phi}^2 + \order{\mu_S^{-6}}.
\end{equation}
Plugging this back in the Lagrangian, we then obtain
\begin{equation}
    \mathcal{L}_{\rm EFT} \supset \abs{\partial_{\mu} \Phi}^2 -\Tilde{\mu}^2\abs{\Phi}^2 - \Tilde{\lambda} \abs{\Phi}^4 -  \frac{\Tilde{c}_{\Phi \Box}}{\Tilde{\Lambda}^2} \abs{\Phi}^2 \Box \abs{\Phi}^2 - \frac{\Tilde{c}_{\Phi \delta}}{\Tilde{\Lambda}^2} \abs{\Phi \partial_\mu \Phi }^2 - \frac{\Tilde{c}_{\Phi}}{\Tilde{\Lambda}^2} \abs{\Phi}^6.
    \label{eq:singlet_matched_lag}
    \end{equation}
    with
\begin{equation}
\begin{aligned}
    \Tilde{\mu}^2 &= \mu_{\Phi}^2, \quad \Tilde{\lambda} = \lambda_\Phi - \frac{1}{2}\frac{\kappa_{\Phi S}^2}{\mu_S^2}, \quad \Tilde{c}_{\Phi} = \frac{\lambda_{\Phi S} \kappa_{\Phi S}^2 }{2\mu_S^2} - \frac{\kappa \kappa_{\Phi S}^3}{3\mu_S^4}, \\   \Tilde{\Lambda} &= \mu_{S}, \quad \Tilde{c}_{\Phi \Box} = \frac{\kappa_{\Phi S}^2}{2 \mu_S^2} , \quad \Tilde{c}_{\Phi \delta} = 0.
    \label{eq:singlet_match_coeff}
\end{aligned}
\end{equation}
These parameters can then be mapped to the $m^2, \lambda, \Lambda$, which appear in Eq.~\eqref{eq:treelevelpot:main} by using the field redefinitions given in Appendix \ref{sec:Appendix}:
\begin{align}
\label{eq:singlet_finalmatching}
m^2 &= \mu_\Phi^2\,, \nonumber\\[4pt]
\lambda\; &= \lambda_\Phi - \frac{\kappa_{\Phi S}^2}{2\mu_S^2}
          - \frac{2}{3}\frac{\kappa_{\Phi S}^2\,\mu_\Phi^2}{\mu_S^4}\,, \\[4pt]
\frac{1}{\Lambda^2} &= \frac{\kappa_{\Phi S}^2}{6\,\mu_S^6}
  \Big[\left(3\lambda_{\Phi S}-8\lambda_\Phi\right)\mu_S^2
       + 2\kappa_{\Phi S}\left(2\kappa_{\Phi S}-\kappa\right)\Big]. \nonumber
\end{align}

\subsection{UV completion inspired by Type-I seesaw}
\label{sec62:seesaw}

Next, we construct a UV completion of Eq.~\eqref{eq:treelevelpot:main} motivated by neutrino physics. To begin with, we extend the minimal KSVZ model by introducing SM-singlet right-handed neutrinos\footnote{Here for simplicity we focus on one generation of right-handed neutrinos, while the discussion below can be straightforwardly generalised to multiple neutrino generations.} $N_R$ as in the Type-I seesaw mechanism \cite{Minkowski:1977sc,Yanagida:1979as}. The relevant part of the Lagrangian is
\begin{equation}
\mathcal{L}\supset
-\frac12 M_M \bar N_R^c N_R
-y_\nu \bar L \tilde H N_R
+{\rm h.c.},
\label{eq:TypeIseesaw_Lag}
\end{equation}
where $\tilde H=i\sigma_2H^\ast$. After electroweak symmetry breaking the Yukawa interaction generates the Dirac mass
\begin{equation}
m_D=\frac{y_\nu v_h}{\sqrt2},
\end{equation}
with $v_h$ denoting the Higgs vev.

The lepton-number-breaking Majorana mass can similarly originate from spontaneous
symmetry breaking. To that end, we promote $U(1)_L$ to an exact symmetry of the
Lagrangian --- which forbids the first term of Eq.~\eqref{eq:TypeIseesaw_Lag} --- and introduce a
complex SM-singlet scalar $\Phi_M$, neutral under $U(1)_{\rm PQ}$ but carrying lepton
number $L=-2$, with the Yukawa interaction
\begin{equation}
\mathcal{L}\supset
y_M \bar N_R^cN_R\Phi_M
+{\rm h.c.}
\label{eq:MajoranaYukawa}
\end{equation}
Since this operator conserves $U(1)_L$, lepton number is violated only after $\Phi_M$ develops a vev,
\begin{equation}
\Phi_M(x)=
\frac{\phi_M(x)+f_M}{\sqrt2}
e^{iJ(x)/f_M},
\label{eq:MajoronExpression}
\end{equation}
yielding the Majorana mass
\begin{equation}
M_M=\sqrt{2} y_Mf_M.
\label{eq:MajoranaMass}
\end{equation}
For $M_M\gg m_D$, the light neutrino mass is given by the usual seesaw relation
\begin{equation}
m_\nu\simeq\frac{m_D^2}{M_M}.
\end{equation}
Assuming Yukawa couplings of order unity, reproducing $m_\nu \simeq 0.1\,\mathrm{eV}$
requires $f_M \simeq 2 \times 10^{14}\,\mathrm{GeV}$. Since $f_M \propto y_\nu^2$,
moderately smaller Dirac Yukawas, $y_\nu \sim 10^{-1}$, lower $f_M$ closer to --- though still above --- the values of $f_a$ in the
conventional axion window,
\begin{equation}
    10^{9}\,\mathrm{GeV} < f_a < 10^{12}\,\mathrm{GeV}\,,
    \label{eq:axion-window}
\end{equation}
which corresponds to the range of $f_M$ values relevant for the Majoron DM scenario discussed below.

The relevant terms of the scalar potential preserving both $U(1)_{\rm PQ}$ and $U(1)_L$ are then given by
\begin{equation}
V(\Phi,\Phi_M) \supset
\mu_\Phi^2|\Phi|^2
+\mu_M^2|\Phi_M|^2
+\lambda_\Phi|\Phi|^4
+\lambda_M|\Phi_M|^4
+\lambda_p|\Phi|^2|\Phi_M|^2,
\label{eq:UVaxionneutrinpot}
\end{equation}
where, for simplicity, we neglect Higgs portal interactions, which are not relevant for the present analysis owing to the large hierarchy $v_h \ll f_a, f_M$. Writing $\Phi_M$ in terms of its radial mode as in Eq.~\eqref{eq:MajoronExpression}, and
fixing $\mu_M^2 = -\lambda_M f_M^2$ so that $f_M$ minimises the $\Phi_M$
potential in the limit $\Phi\to0$, the equation of motion for $\phi_M$ reads\footnote{With this definition $f_M$ is a Lagrangian parameter rather than the
exact vev: the portal coupling induces
$\langle\phi_M\rangle/f_M = -\lambda_p f_a^2/(4\lambda_M f_M^2)$, so that
$\langle|\Phi_M|\rangle = (f_M + \langle\phi_M\rangle)/\sqrt{2}$. The Majorana
mass in Eq.~\eqref{eq:MajoranaMass} and the Majoron decay constant are corrected by the same
relative amount, which is negligible in the regime
$f_a^2 \ll 4\lambda_M f_M^2$ where the effective description applies. Thus in what follows we identify $f_M$ with the Majoron vev.}
\begin{equation}
  \left(\Box + 2\lambda_M f_M^2 + \lambda_p|\Phi|^2\right)\phi_M
  + \lambda_p f_M |\Phi|^2
  + 3\lambda_M f_M \phi_M^2
  + \lambda_M \phi_M^3 = 0\,,
  \label{eq:phiM-eom}
\end{equation}
where $2\lambda_M f_M^2 = m_{\phi_M}^2$ is the tree-level mass of the radial
mode.

Consequently, integrating out $\phi_M$ proceeds identically to the singlet
scalar discussed in Section~\ref{sec61.realsinglet}: Eq.~\eqref{eq:phiM-eom} has
the same structure as Eq.~\eqref{eq:singlet_PQ_EOM} upon the identifications
\begin{equation}
  \mu_S^2 \to 2\lambda_M f_M^2\,,\quad
  \lambda_{\Phi S} \to \lambda_p\,,\quad
  \kappa_{\Phi S} \to \lambda_p f_M\,,\quad
  \kappa \to 3\lambda_M f_M\,,\quad
  \lambda_S \to \lambda_M\,,
  \label{eq:dictionary}
\end{equation}
supplemented by
\begin{equation}
  \lambda_\Phi \to \lambda_\Phi\,,
  \qquad
  \mu_\Phi^2 \to \mu_\Phi^2 + \tfrac12 \lambda_p f_M^2\,,
  \label{eq:dictionary-extra}
\end{equation}
the latter arising from the constant $ f_M^2/2$ piece in $\abs{\Phi_M}^2 = (\phi_M + f_M)^2/2$, which is invisible in the $\phi_M$ equation of motion. The
derivative expansion requires $|\Phi|^2 \ll 2\lambda_M f_M^2$, i.e.\
$f_a^2 \ll 4\lambda_M f_M^2$. Applying Eq.~\eqref{eq:singlet_match_coeff} then yields
\begin{equation}
  \tilde\mu^2 = \mu_\Phi^2 + \frac{\lambda_p f_M^2}{2}\,,\quad
  \tilde\lambda = \lambda_\Phi - \frac{\lambda_p^2}{4\lambda_M}\,,\quad
  \tilde c_{\Phi\Box} = \frac{\lambda_p^2}{4\lambda_M}\,,\quad
  \tilde c_{\Phi\delta} = 0\,,\quad
  \tilde\Lambda = \sqrt{2\lambda_M}\, f_M\,,
  \label{eq:seesaw-wilson}
\end{equation}
while the coefficient of the sextic operator vanishes identically,

\begin{equation}
  \tilde c_\Phi
  = \frac{\lambda_{\Phi S}\kappa_{\Phi S}^2}{2\mu_S^2}
  - \frac{\kappa\,\kappa_{\Phi S}^3}{3\mu_S^4}
  \;\longrightarrow\;
  \frac{\lambda_p^3}{4\lambda_M} - \frac{\lambda_p^3}{4\lambda_M} = 0\,.
  \label{eq:cPhi-vanishes}
\end{equation}

The sextic term of the canonically normalised potential is thus generated
\emph{entirely} by the field redefinition of Eq.~\eqref{eq:TildePhi_def}. With
$\tilde c_k = \tfrac14\tilde c_{\Phi\delta} - \tilde c_{\Phi\Box}
= -\lambda_p^2/(4\lambda_M)$, Eq.~\eqref{eq:redef:match} gives
\begin{align}
  m^2 &= \mu_\Phi^2 + \frac{\lambda_p f_M^2}{2}\,,
  \nonumber\\[4pt]
  \lambda &= \tilde\lambda
            - \frac{\lambda_p^2\, m^2}{6\lambda_M^2 f_M^2}\,,
  \label{eq:seesaw-final}\\[4pt]
  \frac{1}{\Lambda^2}
    &= -\frac{\lambda_p^2\,\tilde\lambda}{3\lambda_M^2 f_M^2}\,,
  \nonumber
\end{align}
with $\tilde\lambda = \lambda_\Phi - \lambda_p^2/(4\lambda_M)$. Note that
$\Lambda^2 > 0$ requires $\tilde\lambda < 0$ --- precisely the regime in
which the quartic is destabilised and the sextic operator is needed to
stabilise the potential. The  Type-I seesaw completion with spontaneously induced Majorana mass for the right-handed neutrinos, therefore reproduces
Eq.~\eqref{eq:treelevelpot:main} automatically, fixing $\Lambda$ in terms of
$\lambda_p$, $\lambda_M$ and $f_M$.

If the scalar potential contains no explicit source of $U(1)_L$ breaking, the angular field $J(x)$ is the massless Majoron \cite{Chikashige:1980ui}. Its interactions are highly suppressed\footnote{The coupling to photons is generated at the two-loop level and is thus even more suppressed.} \cite{Heeck:2019guh},
\begin{equation}
g_{\nu\nu J} \sim \frac{m_\nu}{f_M}\,,
\qquad
g_{ffJ} \sim \frac{1}{16\pi^2}\,\frac{m_f\, m_D^\dagger m_D}{v_h^2\, f_M}\,
\end{equation}
which, for the symmetry-breaking scales considered here,
$f_M\gtrsim10^{12}$ GeV, render the Majoron extremely weakly coupled to the SM. Consequently, the minimal singlet Majoron scenario is compatible with existing astrophysical and cosmological constraints from stellar cooling, supernova, and laboratory searches \cite{Brune:2018sab}.

Explicit $U(1)_L$-breaking effects promote the Majoron to a pNGB with mass $\mu_J$, which we treat as a free parameter. Such a particle provides a viable DM candidate \cite{Frigerio:2011in}, particularly in the region $f_a\lesssim10^{11}$ GeV where the axion misalignment alone underproduces the observed relic abundance.\footnote{See \cite{Akita:2026gzk,deGiorgi:2026jqn, Batell:2026avi} for recent studies compatible with  leptogenesis.} The dominant production mechanism depends on whether $U(1)_L$ is broken before or after inflation: in the pre-inflationary scenario the relic abundance is generated primarily through vacuum misalignment~\cite{Akita:2026gzk},
\begin{equation}
\Omega_J^{({\rm mis})}h^2
\simeq
0.12\,\theta_i^2
\left(
\frac{f_M}{10^{12}\ {\rm GeV}}
\right)^2
\left(
\frac{\mu_J}{25.6\ {\rm meV}}
\right)^{1/2},
\end{equation}
with $\theta_i$ denoting the initial value of $\theta_i = J(t_{\rm osc})/f_M$ at the onset of oscillations $t_{\rm osc}$ analogously to the axion misalignment, whereas in the post-inflationary scenario cosmic-string decays provide an additional, comparable contribution \cite{Batell:2026avi},
\begin{equation}
\Omega_J^{({\rm strings})}h^2
\simeq
0.12
\left(
\frac{f_M}{10^{12}\ {\rm GeV}}
\right)^2
\left(
\frac{\mu_J}{0.73\ {\rm meV}}
\right)^{1/2}
\left(
\frac{g_{\ast,s}(t_{\rm osc})}{106.75}
\right)^{-1/4}
\left(
\frac{\xi(t_{\rm osc})}{10}
\right)^{1/2},
\end{equation} with $\xi(t_{\rm osc})$ denoting the string number density per Hubble volume at the onset of oscillations $t_{\rm osc}$ (see \cite{Batell:2026avi} for more details).
For the parameter space considered here, the viable Majoron DM region extends to masses $\mu_J \lesssim 1\,\mathrm{meV}$. In this regime, the observed DM abundance is supplied by the non-thermal Majoron population generated through vacuum misalignment and, in the post-inflationary scenario, by radiation from cosmic strings and the subsequent collapse of the string-domain wall network. Consequently, the pseudo-Majoron provides a well-motivated complement to axion DM in the region $f_a \lesssim 10^{11}\,\mathrm{GeV}$, where the GW detection prospects for the PQ PT are at their most promising, while axion misalignment alone underproduces the observed relic abundance.

\subsection{Two-axion system}

Additional heavy states generating the dimension-six operator in Eq.~\eqref{eq:Vtree_beforeSSb} could naturally originate from an extended PQ sector containing multiple axions. Although multi-axion models were initially motivated by the string axiverse \cite{Svrcek:2006yi,Arvanitaki:2009fg}, they have since been studied extensively because of their rich cosmological phenomenology, independent of any particular UV completion \cite{Kitajima:2014xla,Daido:2015bva,Daido:2015cba,Ho:2018qur,Cyncynates:2023esj,Li:2024okl,Li:2025cep,Dunsky:2025sgz,Muursepp:2024mbb,Murai:2024nsp}. In particular, mixing between axions enables energy transfer between the different mass eigenstates, thereby modifying the conventional single-axion misalignment mechanism.

A concrete example is the resonant level crossing between axion mass eigenstates \cite{Kitajima:2014xla,Daido:2015bva,Daido:2015cba}. As the temperature of the ambient SM plasma approaches the QCD scale, the mass of the QCD axion starts to grow with temperature, while the dark axion associated with an already-confined hidden sector maintains a fixed mass. When these masses cross during the thermal evolution, the off-diagonal terms in the mass matrix result in energy transfer between the eigenstates, altering the DM abundance obtained from the QCD axion. This mechanism enables a heavier QCD axion to account for the observed DM while remaining consistent with cosmological constraints.

The two axion system can be easily realised by considering two PQ fields $\Phi$ and $\chi$ instead of a single scalar $\Phi$ as in the typical KSVZ construction; and by extending the relevant gauge group from the $SU(3)_c$ of QCD to $SU(3)_c \times SU(3)_D$, where the hidden gauge group $SU(3)_D$ confines at much higher energies than $SU(3)_c$. The fermion content is modified from the minimal scenario by including two vector-like $\Psi_c, \Psi_m$ fermions in the fundamental of $SU(3)_c$ and one vector-like fermion $\Psi_d$ under the fundamental of $SU(3)_D$. Finally the Yukawa interactions, as well as the terms in the scalar potential are constrained by two independent global $U(1)_q$ and $U(1)_p$ axial symmetries. The charges of the fields under those symmetries are displayed in Table~\ref{tab:twoaxionKSZV}.  
\begin{table}[h]
\centering
\begin{tabular}{c|ccccc}
\hline
 & $\Phi$ & $\chi$ & $\Psi_c$ & $\Psi_m$ & $\Psi_d$\\
\hline
$U(1)_q$ & $-q$ &  0 & $q/2$ & 0 & 0\\
$U(1)_p$ & 0 & $-p$& 0 & $p/2$ & $p/2$ \\
$SU(3)_c$ & $\mathbf{1}$ & $\mathbf{1}$ & $\mathbf{3}$ & $\mathbf{3}$ & $\mathbf{1}$ \\
$SU(3)_D$ & $\mathbf{1}$ & $\mathbf{1}$ & $\mathbf{1}$ & $\mathbf{1}$ & $\mathbf{3}$ \\
\hline
\end{tabular}
\caption{Field content of the two-axion KSVZ model, representations of the fields under $SU(3)_c \times SU(3)_D$ and charges under $U(1)_q \times U(1)_p$.}
\label{tab:twoaxionKSZV}
\end{table}
The shape of the scalar potential is then fixed by the aforementioned symmetries as
\begin{equation}
    V(\Phi, \chi) = \mu_\Phi^2 \abs{\Phi}^2 + \lambda_\Phi \abs{\Phi}^4 + \mu_{\chi}^2 \abs{\chi}^2 + \lambda_{\chi} \abs{\chi}^4 + \lambda_{\Phi \chi} \abs{\Phi}^2 \abs{\chi}^2.
\end{equation}

Parametrising the PQ scalars as
\begin{equation}
    \Phi(x) = \frac{1}{\sqrt2}\,\phi\, e^{ia/f_a}, \qquad
    \chi(x) = \frac{\xi + f_b}{\sqrt{2}}\, e^{i b/f_b},
\end{equation}
where $\expval{\phi} = f_a$, and $\xi$ denotes the radial fluctuation of $\chi$
about its vev $\expval{\chi} = f_b/\sqrt{2}$, we then see that after choosing $p=q = 1$, the allowed Yukawa terms
\begin{equation}
    \mathcal{L} \supset
y_c \Phi\,\bar{\Psi}_{c,L}\Psi_{c,R}
+y_m \chi\,\bar{\Psi}_{m,L}\Psi_{m,R}
+y_d \chi\,\bar{\Psi}_{d,L}\Psi_{d,R}
+\text{h.c.}
\end{equation}
generate the anomalous couplings of the Goldstone bosons $a$ and $b$ to the gauge bosons $G$ and $G_D$ of $SU(3)_c$ and $SU(3)_D$ respectively:
\begin{equation}
    \mathcal{L} \supset \frac{g_D^2}{32\pi^2} \frac{b}{f_b} \Tilde{G}_D G_D + \frac{g_s^2}{32 \pi^2} \left(\frac{a}{f_a} + \frac{b}{f_b} \right) \Tilde{G} G,
\end{equation} with $g_D$ denoting the gauge coupling of $SU(3)_D$, which, at temperatures below the confinement scale of $SU(3)_D$, i.e at $T \ll \Lambda_D$, generates a potential of the form\footnote{Note that instead of introducing three new fermions, one can also introduce only two fermions with one of the two in the bifundamental representation of $SU(3)_c \times SU(3)_D$~\cite{Muursepp:2024mbb}. However, these models are less attractive because they are not able to produce a unit domain wall number.} \cite{Cyncynates:2023esj}
\begin{equation}
    V_{\rm axions} = m_a^2(T) f_a^2 \left[1 - \cos (\frac{a}{f_a} + \frac{b}{f_b})\right] + m_{b}^2 f_b^2 \left[ 1 - \cos (\frac{b}{f_b}) \right].
    \label{eq:twoaxionpotential}
\end{equation}

Expanding Eq.~\eqref{eq:twoaxionpotential}, one clearly sees that the mass matrix in  $(a, b)^T$ basis contains an off-diagonal term, which allows energy to be transferred between the states. Here we assume that the gauge group $SU(3)_D$ confines early enough such that we can assume $m_b$ having a constant value. Instead, the QCD axion mass grows with temperature as the Universe cools, $m_a^2(T) \sim T^{-n}$ for some $n \in (6,8)$ until $T \simeq \Lambda_{\rm QCD} \simeq 160 \text{ MeV}.$ Furthermore, assuming that $m_{b} > m_a(T)$ for $T\gg \Lambda_{\rm QCD}$ and $m_{b} < m_a(0) \equiv \sqrt{\chi_{\rm QCD}}/f_a$, we see that there exists a range of $T$ around $T_{\ast}$ such that $m_{a}(T_\ast) = m_{b}$, where the energy can be transferred between the two mass eigenstates. If this transfer is adiabatic,\footnote{For more details about the conditions which need to be satisfied for an adiabatic transfer, see \cite{Murai:2024nsp, Muursepp:2024mbb, Dunsky:2025sgz}.} the DM abundance today is given by \cite{Cyncynates:2023esj}
\begin{equation}
    \Omega^a_{\rm DM} h^2 = \frac{22.81 f_b^2 T_0^3 \sqrt{\chi_{\rm QCD}} }{f_a H_0^2 M_{\rm Pl}^{7/2}\sqrt{m_{b}}}\,,
\end{equation}
where $T_0 \simeq 10^{-4}$ eV is the present temperature of the Universe, $H_0 \simeq 10^{-33}$ eV denotes the present value of the Hubble constant and $M_{\rm Pl}$ is the Planck mass. 

Notice that in this case, the breaking of PQ, even at the lowest scales allowed by experimental constraints $f_a \simeq 10^{9}$ GeV, can be compatible with DM abundance. For instance, choosing $f_a = 10^9$ GeV, and $m_b = 10^{-3}$ eV, we would need $f_b \simeq 2.4 \times 10^{12}$ GeV. This further motivates our study of PTs around $f_a \simeq 10^{9}$ GeV, where the detection prospects of the associated GW signals are the most promising.

The generation of the effective dimension-six operator in this setup proceeds
exactly as in Section~\ref{sec62:seesaw}. Fixing
$\mu_\chi^2 = -\lambda_\chi f_b^2$, the equation of motion for $\xi$ takes the
same form as Eq.~\eqref{eq:phiM-eom} under the identifications
\begin{equation}
  \lambda_p \to \lambda_{\Phi\chi}\,,\qquad
  \lambda_M \to \lambda_\chi\,,\qquad
  f_M \to f_b\,,
  \label{eq:twoaxion-dictionary}
\end{equation}
so that integrating out the heavy radial mode, of mass
$m_\xi = \sqrt{2\lambda_\chi}\,f_b$, yields
\begin{equation}
  \tilde\lambda = \lambda_\Phi - \frac{\lambda_{\Phi\chi}^2}{4\lambda_\chi}\,,
  \qquad
  \tilde c_{\Phi\Box} = \frac{\lambda_{\Phi\chi}^2}{4\lambda_\chi}\,,
  \qquad
  \tilde c_{\Phi\delta} = \tilde c_{\Phi} = 0\,.
\end{equation}
As in the seesaw case, the sextic operator is generated entirely by the field
redefinition to canonical kinetic form, giving
\begin{equation}
  m^2 = \mu_\Phi^2 + \frac{\lambda_{\Phi\chi}f_b^2}{2}\,,\qquad
  \lambda = \tilde\lambda
          - \frac{\lambda_{\Phi\chi}^2\,m^2}{6\lambda_\chi^2 f_b^2}\,,\qquad
  \frac{1}{\Lambda^2}
    = -\frac{\lambda_{\Phi\chi}^2\,\tilde\lambda}{3\lambda_\chi^2 f_b^2}\,,
  \label{eq:twoaxion-final}
\end{equation}
so that a strong PQ transition again requires $\tilde\lambda < 0$. The cut-off
is therefore fixed by the heavy axion sector,
\begin{equation}
  \frac{\Lambda}{f_b}
  = \frac{\sqrt{3}\,\lambda_\chi}{\lambda_{\Phi\chi}\sqrt{|\tilde\lambda|}}\,,
  \label{eq:twoaxion-Lambda}
\end{equation}
and the hierarchy $\Lambda\sim f_a \ll f_b$, required by the level-crossing
mechanism, is realised for $\lambda_\chi \ll \lambda_{\Phi\chi}$.

\section{Conclusions}

In this work, we have studied the cosmological PT dynamics and GW phenomenology of the minimal KSVZ axion model, extended by the addition of a Peccei–Quinn invariant dimension-6 operator in the scalar potential.

Unlike the minimal KSVZ scenario, where the absence of a thermal barrier typically forces the PQ PT to be second-order or a smooth crossover, we have shown that the inclusion of a single higher-dimensional operator can generate a barrier between the symmetric ($\langle\phi\rangle=0$) and broken ($\langle\phi\rangle=f_a$) phases at $T=0$, thus enabling strong FOPTs and potentially observable stochastic GW backgrounds.

After briefly reviewing the minimal KSVZ model in Section~\ref{sec:KSZV}, we have derived the effective potential in its dimension-6 operator extension, by considering both quantum and thermal corrections in Section~\ref{sec:quantum_corrections} and~\ref{sec:thermal_corrections} respectively. In the rest of Section~\ref{sec3:PT_background}, we have then defined the key quantities to determine in order to analyse the PT dynamics and compute the power spectrum of the subsequent stochastic GW background.

In Section~\ref{sec:tree-level} we have presented a tree-level analysis with two main constraints: Eq.~\eqref{eq:existence of barrier} about the existence of a tree-level barrier between the true and false vacua and Eq.~\eqref{eq:deeperminima} about the requirement for the broken minimum at $f_a$ to be the global one. We have then derived the one-loop version of these two constraints in Section~\ref{sec:one-loop_constraint}. They respectively resulted in an upper and lower bound on the Yukawa coupling $y_F$, which were depicted in Figure~\ref{fig:yukawa_bound}. Next, in Section~\ref{sec:snr}, we mentioned the possibility of having the non-restoration of the PQ symmetry as the temperature increases. In Figure~\ref{fig:snr}, we showed two scenarios: one in which there is no symmetry restoration with no PT, and one in which the PT first occurs from the broken phase due to SNR, then a subsequent PT between the symmetric and the broken phase takes place. 

This gives rise to a potential two-step PT, in which the Universe first transitions from the symmetry-broken to the symmetric phase and subsequently back to the symmetry-broken phase at lower temperatures. If both transitions are first order and complete, they could produce a distinctive double-peaked GW signal, potentially allowing this scenario to be discriminated from other KSVZ constructions.

Finally, in Section \ref{sec:O4}, we formulated analytical conditions for percolation in the case of $O(4)$ tunneling.

After scanning the allowed parameter space, shown in Figure~\ref{fig:scan}, we have first determined in Figure~\ref{fig:parameters} the region of the parameter space allowing FOPTs, explicitly showing that the PT strength $\alpha$ is the strongest around the diagonal $\lambda=-2m/\Lambda$. Then, through a refined scan around this diagonal, we have computed the GW signal for different values of the cut-off scale $\Lambda$, corresponding to the window $10^9<f_a<10^{12}$ GeV and shown in Figure~\ref{fig:gw_all_LAM} that it was easier to obtain a GW signal reaching the sensitivity of the current interferometers aLIGO, aVirgo and KAGRA for lower values of $\Lambda$. In Figure~\ref{fig:gw_common_BP}, we have analysed the impact of varying the scale $\Lambda$ for ten benchmark points. We have shown that increasing $\Lambda$ generally increases the peak amplitude, while the peak frequency initially increases but, for sufficiently strong FOPTs, can reach a maximum and subsequently decrease due to the competition between the increasing $T_{\rm reh}$ and decreasing $\beta/H(T_p)$. This, in turn, implies that the GW signal, which was too weak for 
low $\Lambda$, could actually be within the sensitivity range of the GW detectors for higher values of $\Lambda$.

We have furthermore shown that for every parameter point at which thermal percolation fails, zero-temperature $O(4)$ tunnelling also fails to complete. Among these points, the bounce action satisfies $S_4\gtrsim296$, well above the threshold $S_4^{\rm crit}\simeq70-110$ required for the false vacuum to decay. Such parameter points are therefore cosmologically excluded, since the Universe remains trapped in the symmetric false vacuum.

We also investigated multimessenger scenarios in which the QCD axion originating from PQ breaking can account for the observed DM relic abundance. Neglecting the additional axion population from topological defects, we find that the observed abundance is reproduced for $\Lambda\simeq(3-5)\times10^{11}$ GeV, where the associated GW signals can potentially be probed by future GW interferometers. Including axion production from the cosmic-string network would shift the scale at which the relic abundance is saturated towards smaller values of $\Lambda$.

Finally, in Section~\ref{sec6:UVcompletions}, we have presented concrete UV completions capable of generating the effective dimension-6 operator at lower energies. These include extensions with a heavy real scalar singlet, a Type-I seesaw scenario involving a pseudo-Majoron, and a two-axion system, where the latter two models can reproduce the observed DM abundance for values of $f_a$ compatible with the range considered in the GW analysis.

There are several interesting directions in which our work could be extended. The residual scale dependence in our treatment of the effective potential could be reduced by using RG-improvement techniques or dimensional reduction. The effective potential could also be extended with additional non-renormalisable terms of higher dimension as well as dimension-6 non-canonical kinetic terms, which would allow to connect our formalism to a larger set of UV theories. For those transitions that result in runaway walls, one could also connect the PQ transition to the production of ultraheavy particles from the PT bubble collisions --- which could play a role in DM physics or baryogenesis. Finally, quantifying the GW signals of multi-step transitions in the SNR scenarios, as well as possible GW signals from $O(4)$ symmetric vacuum transitions from other regions of parameter space, provides an additional compelling avenue for more detailed studies. We plan to return to some of these topics in the future.

In conclusion, higher-dimensional operators within the framework of the KSVZ axion model provide a viable pathway to strong FOPTs. This leads to a promising multimessenger perspective, allowing GW detectors to complement traditional axion search experiments in probing the underlying physics connected to the QCD axion and the spontaneous breaking of the PQ symmetry.

\label{sec7:concl}

\acknowledgments 
It is a pleasure to thank Carlo Tasillo for very interesting and instructive  correspondence on the usage of the TransitionListener code, Ville Vaskonen for very helpful discussion on supercooled first-order phase transitions, Kristjan Kannike on interesting discussion on Majoron models and Enrico Nardi for interesting discussions at the initial stage of this project, for reading the manuscript and suggesting insightful improvements. This work was supported by the Centre of Excellence program TK202 ``Fundamental Universe''. KM was supported by the Estonian Research Council personal grant PUTJD1256.

\appendix

\section{Derivative corrections}
\label{sec:Appendix}

In this section we clarify the assumptions that we make in order to neglect the dimension-6 derivative corrections. We start from the independent PQ-preserving operators involving only the scalar field up to dimension six
\begin{equation}
    \mathcal{L}_{\Phi} \supset  \abs{\partial_\mu \Phi}^2-\Tilde{\mu}^2\abs{\Phi}^2 - \Tilde{\lambda} \abs{\Phi}^4 -  \frac{\Tilde{c}_{\Phi \Box}}{\Tilde{\Lambda}^2} \abs{\Phi}^2 \Box \abs{\Phi}^2 - \frac{\Tilde{c}_{\Phi \delta}}{\Tilde{\Lambda}^2} \abs{\Phi \partial_\mu \Phi }^2 - \frac{\Tilde{c}_{\Phi}}{\Tilde{\Lambda}^2} \abs{\Phi}^6.
    \label{eq:dercorr_lag}
\end{equation}
Using $\Phi = \frac{1}{\sqrt{2}}\Tilde{\phi}(x)\, e^{i a(x)/f_a}$, integrating the fourth term by parts and dropping the $\partial_\mu a$ terms, which do not contribute to the potential, we obtain for the dimension-6 terms\footnote{In general the angular mode $a$ does not contribute to the background dynamics of
the radial mode: its shift symmetry forbids any non-derivative dependence on $a$,
so it cannot enter the effective potential, and $\langle\partial_\mu a\rangle = 0$
for a homogeneous background. The $\partial_\mu a$ terms therefore only renormalise
the axion kinetic term. In any case, in the specific scenarios of Section~\ref{sec6:UVcompletions}
we have $\tilde c_{\Phi\delta} = 0$, so no such term arises in Eq.~\eqref{eq:App_dimsixcorr} at all. }
\begin{equation}
    \mathcal{L}^{D=6}_{\Phi} = -\frac{\Tilde{c}_{\rm k}}{\Tilde{\Lambda}^2} \Tilde{\phi}^2 (\partial_{\mu} \Tilde{\phi)}^2 - \frac{\Tilde{c}_{\Phi}}{8 \Tilde{\Lambda}^2}\Tilde{\phi}^6,
    \label{eq:App_dimsixcorr}
\end{equation}
with $\Tilde{c}_{\rm k} = \frac{1}{4} \Tilde{c}_{\Phi \delta} - \Tilde{c}_{\Phi \Box} $. Thus the Lagrangian takes the form 
\begin{equation}
    \mathcal{L} = \frac{1}{2}Z(\Tilde{\phi}) (\partial \Tilde{\phi})^2 - V(\Tilde{\phi})
\end{equation}
with
\begin{equation}
    Z(\Tilde{\phi}) = 1 - 2 \frac{\Tilde{c}_k}{\Tilde{\Lambda^2}}\Tilde{\phi}^2.
\end{equation}
We can then bring the kinetic term into canonical form by a field redefinition $\Tilde{\phi} \rightarrow \phi$ such that
\begin{equation}
    \frac{d \phi}{d \Tilde{\phi}} = \sqrt{Z(\Tilde{\phi})}.
\end{equation}
Expanding $\sqrt{Z(\Tilde{\phi})}$,
we obtain
\begin{equation}
    \frac{d\phi}{d \Tilde{\phi}} = 1 - \frac{\Tilde{c}_k}{\Tilde{\Lambda}^2}\Tilde{\phi^2} + \order{\Tilde{c}_k^2\frac{\Tilde{\phi^4}}{\Tilde{\Lambda}^4}}.
\end{equation}
Integrating this equation and then inverting the result yields
\begin{equation}
    \Tilde{\phi} = \phi + \frac{\Tilde{c}_k}{3\Tilde{\Lambda}^2}\phi^3 + \order{\Tilde{c}_k^2\frac{\phi^5}{\Tilde{\Lambda}^4}}.
    \label{eq:TildePhi_def}
\end{equation}
The Lagrangian is thus given by (up to canonical field dimension six)
\begin{equation}
    \mathcal{L} \supset \frac{1}{2}\left( \partial_\mu \phi\right)^2 - \left(\frac{m^2}{2} \phi^2 + \frac{\lambda}{4}\phi^4 + \frac{c_\Phi}{8\Tilde{\Lambda}^2} \phi^6 \right),
\end{equation}
with the new parameters $m^2, \lambda, c_{\Phi}$ related to those appearing in Eq.~\eqref{eq:dercorr_lag} by
\begin{equation}
    m^2 = \Tilde{\mu}^2, \quad \lambda = \Tilde{\lambda} + \frac{4}{3}\, \Tilde{c}_k \frac{\Tilde{\mu}^2}{\Tilde{\Lambda}^2}, \quad c_{\Phi} =  \Tilde{c}_{\Phi} + \frac{8}{3} \Tilde{c}_k \Tilde{\lambda}.
    \label{eq:redef:match}
\end{equation}
Lastly, by defining $\Lambda = \frac{\Tilde{\Lambda}}{\sqrt{c_{\Phi}}}$, we obtain the potential in Eq.~\eqref{eq:treelevelpot:main} of the main text.

Finally we note that in principle, the field redefinition in Eq.~\eqref{eq:TildePhi_def} also generates dimension-six interactions involving the
other fields of the KSVZ theory, such as corrections to the Yukawa interaction.
Since these operators do not contribute to the scalar potential at tree level,
their effects on the PT dynamics arise only through loop and
thermal corrections. For this reason we neglect these UV-dependent subleading effects in the
present analysis.

\bibliographystyle{JHEP}
\bibliography{biblio.bib}

@article{Peccei:1977hh,
    author = "Peccei, R. D. and Quinn, Helen R.",
    title = "{CP Conservation in the Presence of Instantons}",
    reportNumber = "ITP-568-STANFORD",
    doi = "10.1103/PhysRevLett.38.1440",
    journal = "Phys. Rev. Lett.",
    volume = "38",
    pages = "1440--1443",
    year = "1977"
}

@article{PhysRevLett.40.223,
  title = {A New Light Boson?},
  author = {Weinberg, Steven},
  journal = {Phys. Rev. Lett.},
  volume = {40},
  issue = {4},
  pages = {223--226},
  numpages = {0},
  year = {1978},
  month = {Jan},
  publisher = {American Physical Society},
  doi = {10.1103/PhysRevLett.40.223},
  url = {https://link.aps.org/doi/10.1103/PhysRevLett.40.223}
}

@article{PhysRevLett.40.279,
  title = {Problem of Strong $P$ and $T$ Invariance in the Presence of Instantons},
  author = {Wilczek, F.},
  journal = {Phys. Rev. Lett.},
  volume = {40},
  issue = {5},
  pages = {279--282},
  numpages = {0},
  year = {1978},
  month = {Jan},
  publisher = {American Physical Society},
  doi = {10.1103/PhysRevLett.40.279},
  url = {https://link.aps.org/doi/10.1103/PhysRevLett.40.279}
}

@article{Preskill:1982cy,
    author = "Preskill, John and Wise, Mark B. and Wilczek, Frank",
    editor = "Srednicki, M. A.",
    title = "{Cosmology of the Invisible Axion}",
    reportNumber = "HUTP-82-A048, NSF-ITP-82-103",
    doi = "10.1016/0370-2693(83)90637-8",
    journal = "Phys. Lett. B",
    volume = "120",
    pages = "127--132",
    year = "1983"
}

@article{Abbott:1982af,
    author = "Abbott, L. F. and Sikivie, P.",
    editor = "Srednicki, M. A.",
    title = "{A Cosmological Bound on the Invisible Axion}",
    reportNumber = "PRINT-82-0695 (BRANDEIS)",
    doi = "10.1016/0370-2693(83)90638-X",
    journal = "Phys. Lett. B",
    volume = "120",
    pages = "133--136",
    year = "1983"
}

@article{Dine:1982ah,
    author = "Dine, Michael and Fischler, Willy",
    editor = "Srednicki, M. A.",
    title = "{The Not So Harmless Axion}",
    reportNumber = "UPR-0201T",
    doi = "10.1016/0370-2693(83)90639-1",
    journal = "Phys. Lett. B",
    volume = "120",
    pages = "137--141",
    year = "1983"
}

@article{Marsh:2015xka,
    author = "Marsh, David J. E.",
    title = "{Axion Cosmology}",
    eprint = "1510.07633",
    archivePrefix = "arXiv",
    primaryClass = "astro-ph.CO",
    reportNumber = "KCL-PH-TH-2015-50",
    doi = "10.1016/j.physrep.2016.06.005",
    journal = "Phys. Rept.",
    volume = "643",
    pages = "1--79",
    year = "2016"
}

@article{DiLuzio:2020wdo,
    author = "Di Luzio, Luca and Giannotti, Maurizio and Nardi, Enrico and Visinelli, Luca",
    title = "{The landscape of QCD axion models}",
    eprint = "2003.01100",
    archivePrefix = "arXiv",
    primaryClass = "hep-ph",
    reportNumber = "DESY 20-036, DESY-20-036",
    doi = "10.1016/j.physrep.2020.06.002",
    journal = "Phys. Rept.",
    volume = "870",
    pages = "1--117",
    year = "2020"
}

@article{Choi:2020rgn,
    author = "Choi, Kiwoon and Im, Sang Hui and Sub Shin, Chang",
    title = "{Recent Progress in the Physics of Axions and Axion-Like Particles}",
    eprint = "2012.05029",
    archivePrefix = "arXiv",
    primaryClass = "hep-ph",
    reportNumber = "CTPU-PTC-20-28",
    doi = "10.1146/annurev-nucl-120720-031147",
    journal = "Ann. Rev. Nucl. Part. Sci.",
    volume = "71",
    pages = "225--252",
    year = "2021"
}

@article{DelleRose:2019pgi,
    author = "Delle Rose, Luigi and Panico, Giuliano and Redi, Michele and Tesi, Andrea",
    title = "{Gravitational Waves from Supercool Axions}",
    eprint = "1912.06139",
    archivePrefix = "arXiv",
    primaryClass = "hep-ph",
    doi = "10.1007/JHEP04(2020)025",
    journal = "JHEP",
    volume = "04",
    pages = "025",
    year = "2020"
}

@article{PhysRevLett.43.103,
  title = {Weak-Interaction Singlet and Strong $\mathrm{CP}$ Invariance},
  author = {Kim, Jihn E.},
  journal = {Phys. Rev. Lett.},
  volume = {43},
  issue = {2},
  pages = {103--107},
  numpages = {0},
  year = {1979},
  month = {Jul},
  publisher = {American Physical Society},
  doi = {10.1103/PhysRevLett.43.103},
  url = {https://link.aps.org/doi/10.1103/PhysRevLett.43.103}
}

@article{SHIFMAN1980493,
title = {Can confinement ensure natural CP invariance of strong interactions?},
journal = {Nuclear Physics B},
volume = {166},
number = {3},
pages = {493-506},
year = {1980},
issn = {0550-3213},
doi = {https://doi.org/10.1016/0550-3213(80)90209-6},
url = {https://www.sciencedirect.com/science/article/pii/0550321380902096},
author = {M.A. Shifman and A.I. Vainshtein and V.I. Zakharov}
}

@article{Zhitnitsky:1980tq,
    author = "Zhitnitsky, A. R.",
    title = "{On Possible Suppression of the Axion Hadron Interactions. (In Russian)}",
    journal = "Sov. J. Nucl. Phys.",
    volume = "31",
    pages = "260",
    year = "1980"
}

@article{DINE1981199,
title = {A simple solution to the strong CP problem with a harmless axion},
journal = {Physics Letters B},
volume = {104},
number = {3},
pages = {199-202},
year = {1981},
issn = {0370-2693},
doi = {https://doi.org/10.1016/0370-2693(81)90590-6},
url = {https://www.sciencedirect.com/science/article/pii/0370269381905906},
author = {Michael Dine and Willy Fischler and Mark Srednicki}
}

@article{Yang:2024npd,
    author = "Yang, Aidi and Huang, Fa Peng",
    title = "{Detectability of the phase transition gravitational waves in the DFSZ axion model}",
    eprint = "2404.18703",
    archivePrefix = "arXiv",
    primaryClass = "hep-ph",
    doi = "10.1088/1475-7516/2025/05/028",
    journal = "JCAP",
    volume = "05",
    pages = "028",
    year = "2025"
}

@article{Ahmadvand:2021vxs,
    author = "Ahmadvand, M.",
    title = "{Filtered asymmetric dark matter during the Peccei-Quinn phase transition}",
    eprint = "2108.00958",
    archivePrefix = "arXiv",
    primaryClass = "hep-ph",
    doi = "10.1007/JHEP10(2021)109",
    journal = "JHEP",
    volume = "10",
    pages = "109",
    year = "2021"
}

@article{Chiang:2020aui,
    author = "Chiang, Cheng-Wei and Lu, Bo-Qiang",
    title = "{Testing clockwork axion with gravitational waves}",
    eprint = "2012.14071",
    archivePrefix = "arXiv",
    primaryClass = "hep-ph",
    doi = "10.1088/1475-7516/2021/05/049",
    journal = "JCAP",
    volume = "05",
    pages = "049",
    year = "2021"
}

@article{Ghoshal:2020vud,
    author = "Ghoshal, Anish and Salvio, Alberto",
    title = "{Gravitational waves from fundamental axion dynamics}",
    eprint = "2007.00005",
    archivePrefix = "arXiv",
    primaryClass = "hep-ph",
    doi = "10.1007/JHEP12(2020)049",
    journal = "JHEP",
    volume = "12",
    pages = "049",
    year = "2020"
}

@article{VonHarling:2019rgb,
    author = "Von Harling, Benedict and Pomarol, Alex and Pujol{\`a}s, Oriol and Rompineve, Fabrizio",
    title = "{Peccei-Quinn Phase Transition at LIGO}",
    eprint = "1912.07587",
    archivePrefix = "arXiv",
    primaryClass = "hep-ph",
    doi = "10.1007/JHEP04(2020)195",
    journal = "JHEP",
    volume = "04",
    pages = "195",
    year = "2020"
}

@article{Athron:2023xlk,
    author = "Athron, Peter and Bal{\'a}zs, Csaba and Fowlie, Andrew and Morris, Lachlan and Wu, Lei",
    title = "{Cosmological phase transitions: From perturbative particle physics to gravitational waves}",
    eprint = "2305.02357",
    archivePrefix = "arXiv",
    primaryClass = "hep-ph",
    doi = "10.1016/j.ppnp.2023.104094",
    journal = "Prog. Part. Nucl. Phys.",
    volume = "135",
    pages = "104094",
    year = "2024"
}

@article{Hindmarsh:2020hop,
    author = {Hindmarsh, Mark B. and L{\"u}ben, Marvin and Lumma, Johannes and Pauly, Martin},
    title = "{Phase transitions in the early universe}",
    eprint = "2008.09136",
    archivePrefix = "arXiv",
    primaryClass = "astro-ph.CO",
    reportNumber = "MPP-2020-163, HIP-2020-27/TH",
    doi = "10.21468/SciPostPhysLectNotes.24",
    journal = "SciPost Phys. Lect. Notes",
    volume = "24",
    pages = "1",
    year = "2021"
}

@article{Croon:2023zay,
    author = "Croon, Djuna",
    title = "{TASI lectures on Phase Transitions, Baryogenesis, and Gravitational Waves}",
    eprint = "2307.00068",
    archivePrefix = "arXiv",
    primaryClass = "hep-ph",
    doi = "10.22323/1.439.0003",
    journal = "PoS",
    volume = "TASI2022",
    pages = "003",
    year = "2024"
}

@article{Mazumdar:2018dfl,
    author = "Mazumdar, Anupam and White, Graham",
    title = "{Review of cosmic phase transitions: their significance and experimental signatures}",
    eprint = "1811.01948",
    archivePrefix = "arXiv",
    primaryClass = "hep-ph",
    doi = "10.1088/1361-6633/ab1f55",
    journal = "Rept. Prog. Phys.",
    volume = "82",
    number = "7",
    pages = "076901",
    year = "2019"
}

@inproceedings{Quiros:1999jp,
    author = "Quiros, Mariano",
    title = "{Finite temperature field theory and phase transitions}",
    booktitle = "{ICTP Summer School in High-Energy Physics and Cosmology}",
    eprint = "hep-ph/9901312",
    archivePrefix = "arXiv",
    reportNumber = "IEM-FT-187-99",
    pages = "187--259",
    month = "1",
    year = "1999"
}

@article{Caprini:2024ofd,
    author = "Caprini, Chiara and Pujol{\`a}s, Oriol and Quelquejay-Leclere, Hippolyte and Rompineve, Fabrizio and Steer, Dani{\`e}le A.",
    title = "{Primordial gravitational wave backgrounds from phase transitions with next generation ground based detectors}",
    eprint = "2406.02359",
    archivePrefix = "arXiv",
    primaryClass = "astro-ph.CO",
    reportNumber = "CERN-TH-2024-065",
    doi = "10.1088/1361-6382/ad9a48",
    journal = "Class. Quant. Grav.",
    volume = "42",
    number = "4",
    pages = "045015",
    year = "2025"
}

@article{Coleman:1973jx,
    author = "Coleman, Sidney R. and Weinberg, Erick J.",
    title = "{Radiative Corrections as the Origin of Spontaneous Symmetry Breaking}",
    doi = "10.1103/PhysRevD.7.1888",
    journal = "Phys. Rev. D",
    volume = "7",
    pages = "1888--1910",
    year = "1973"
}

@article{Wainwright:2011kj,
    author = "Wainwright, Carroll L.",
    title = "{CosmoTransitions: Computing Cosmological Phase Transition Temperatures and Bubble Profiles with Multiple Fields}",
    eprint = "1109.4189",
    archivePrefix = "arXiv",
    primaryClass = "hep-ph",
    doi = "10.1016/j.cpc.2012.04.004",
    journal = "Comput. Phys. Commun.",
    volume = "183",
    pages = "2006--2013",
    year = "2012"
}

@article{Parwani:1991gq,
    author = "Parwani, Rajesh R.",
    title = "{Resummation in a hot scalar field theory}",
    eprint = "hep-ph/9204216",
    archivePrefix = "arXiv",
    reportNumber = "ITP-SB-91-64",
    doi = "10.1103/PhysRevD.45.4695",
    journal = "Phys. Rev. D",
    volume = "45",
    pages = "4695",
    year = "1992",
    note = "[Erratum: Phys.Rev.D 48, 5965 (1993)]"
}

@article{10.1063/1.1338506,
    author = {Lorenz, Christian D. and Ziff, Robert M.},
    title = {Precise determination of the critical percolation threshold for the three-dimensional “Swiss cheese” model using a growth algorithm},
    journal = {The Journal of Chemical Physics},
    volume = {114},
    number = {8},
    pages = {3659-3661},
    year = {2001},
    month = {02}}

@article{LIN2018299,
title = {Continuum percolation of porous media via random packing of overlapping cube-like particles},
journal = {Theoretical and Applied Mechanics Letters},
volume = {8},
number = {5},
pages = {299-303},
year = {2018},
issn = {2095-0349},
doi = {https://doi.org/10.1016/j.taml.2018.05.007},
url = {https://www.sciencedirect.com/science/article/pii/S209503491830196X},
author = {Jianjun Lin and Huisu Chen}
}

@article{LI2020112815,
title = {Numerical study for the percolation threshold and transport properties of porous composites comprising non-centrosymmetrical superovoidal pores},
journal = {Computer Methods in Applied Mechanics and Engineering},
volume = {361},
pages = {112815},
year = {2020},
issn = {0045-7825},
doi = {https://doi.org/10.1016/j.cma.2019.112815},
url = {https://www.sciencedirect.com/science/article/pii/S0045782519307078},
author = {Mingqi Li and Huisu Chen and Jianjun Lin}
}

@article{PhysRevD.23.876,
  title = {Cosmological consequences of a first-order phase transition in the S${\mathrm{U}}_{5}$ grand unified model},
  author = {Guth, Alan H. and Weinberg, Erick J.},
  journal = {Phys. Rev. D},
  volume = {23},
  issue = {4},
  pages = {876--885},
  numpages = {0},
  year = {1981},
  month = {Feb},
  publisher = {American Physical Society},
  doi = {10.1103/PhysRevD.23.876},
  url = {https://link.aps.org/doi/10.1103/PhysRevD.23.876}
}

@article{Guth:1982pn,
    author = "Guth, Alan H. and Weinberg, Erick J.",
    title = "{Could the Universe Have Recovered from a Slow First Order Phase Transition?}",
    reportNumber = "MIT-CTP-950",
    doi = "10.1016/0550-3213(83)90307-3",
    journal = "Nucl. Phys. B",
    volume = "212",
    pages = "321--364",
    year = "1983"
}

@article{PhysRevLett.44.631,
  title = {Phase Transitions and Magnetic Monopole Production in the Very Early Universe},
  author = {Guth, Alan H. and Tye, S. -H. H.},
  journal = {Phys. Rev. Lett.},
  volume = {44},
  issue = {10},
  pages = {631--635},
  numpages = {0},
  year = {1980},
  month = {Mar},
  publisher = {American Physical Society},
  doi = {10.1103/PhysRevLett.44.631},
  url = {https://link.aps.org/doi/10.1103/PhysRevLett.44.631}
}

@article{Ellis:2018mja,
    author = "Ellis, John and Lewicki, Marek and No, Jos{\'e} Miguel",
    title = "{On the Maximal Strength of a First-Order Electroweak Phase Transition and its Gravitational Wave Signal}",
    eprint = "1809.08242",
    archivePrefix = "arXiv",
    primaryClass = "hep-ph",
    reportNumber = "KCL-PH-TH/2018-46, CERN-TH/2018-197, IFT-UAM/CSIC-18-94, CERN-TH-2018-197",
    doi = "10.1088/1475-7516/2019/04/003",
    journal = "JCAP",
    volume = "04",
    pages = "003",
    year = "2019"
}

@article{PhysRevD.46.2384,
  title = {Bubble nucleation in first-order inflation and other cosmological phase transitions},
  author = {Turner, Michael S. and Weinberg, Erick J. and Widrow, Lawrence M.},
  journal = {Phys. Rev. D},
  volume = {46},
  issue = {6},
  pages = {2384--2403},
  numpages = {0},
  year = {1992},
  month = {Sep},
  publisher = {American Physical Society},
  doi = {10.1103/PhysRevD.46.2384},
  url = {https://link.aps.org/doi/10.1103/PhysRevD.46.2384}
}

@article{Henning:2014wua,
    author = "Henning, Brian and Lu, Xiaochuan and Murayama, Hitoshi",
    title = "{How to use the Standard Model effective field theory}",
    eprint = "1412.1837",
    archivePrefix = "arXiv",
    primaryClass = "hep-ph",
    reportNumber = "UCB-PTH-14-40, IPMU14-0353",
    doi = "10.1007/JHEP01(2016)023",
    journal = "JHEP",
    volume = "01",
    pages = "023",
    year = "2016"
}

@article{Co:2019jts,
    author = "Co, Raymond T. and Hall, Lawrence J. and Harigaya, Keisuke",
    title = "{Axion Kinetic Misalignment Mechanism}",
    eprint = "1910.14152",
    archivePrefix = "arXiv",
    primaryClass = "hep-ph",
    reportNumber = "LCTP-19-28",
    doi = "10.1103/PhysRevLett.124.251802",
    journal = "Phys. Rev. Lett.",
    volume = "124",
    number = "25",
    pages = "251802",
    year = "2020"
}

@article{Co:2020dya,
    author = "Co, Raymond T. and Hall, Lawrence J. and Harigaya, Keisuke and Olive, Keith A. and Verner, Sarunas",
    title = "{Axion Kinetic Misalignment and Parametric Resonance from Inflation}",
    eprint = "2004.00629",
    archivePrefix = "arXiv",
    primaryClass = "hep-ph",
    reportNumber = "LCTP-20-06, UMN-TH-3912/20, FTPI-MINN-20/02",
    doi = "10.1088/1475-7516/2020/08/036",
    journal = "JCAP",
    volume = "08",
    pages = "036",
    year = "2020"
}

@article{Eroncel:2024rpe,
    author = {Er{\"o}ncel, Cem and Sato, Ryosuke and Servant, G{\'e}raldine and S{\o}rensen, Philip},
    title = "{Model implementations of axion dark matter from kinetic misalignment}",
    eprint = "2408.08355",
    archivePrefix = "arXiv",
    primaryClass = "hep-ph",
    reportNumber = "DESY 23-194, OU-HET-1152, CERN-TH-2024-093",
    doi = "10.1088/1475-7516/2025/08/087",
    journal = "JCAP",
    volume = "08",
    pages = "087",
    year = "2025"
}

@article{Morgante:2025lav,
    author = "Morgante, Enrico and Natale, Riccardo",
    title = "{A post-inflationary kinetic axion}",
    eprint = "2512.13633",
    archivePrefix = "arXiv",
    primaryClass = "hep-ph",
    reportNumber = "DESY-25-177",
    month = "12",
    year = "2025"
}

@article{GrillidiCortona:2015jxo,
    author = "Grilli di Cortona, Giovanni and Hardy, Edward and Pardo Vega, Javier and Villadoro, Giovanni",
    title = "{The QCD axion, precisely}",
    eprint = "1511.02867",
    archivePrefix = "arXiv",
    primaryClass = "hep-ph",
    doi = "10.1007/JHEP01(2016)034",
    journal = "JHEP",
    volume = "01",
    pages = "034",
    year = "2016"
}

@article{Ballesteros:2016xej,
    author = "Ballesteros, Guillermo and Redondo, Javier and Ringwald, Andreas and Tamarit, Carlos",
    title = "{Standard Model{\textemdash}axion{\textemdash}seesaw{\textemdash}Higgs portal inflation. Five problems of particle physics and cosmology solved in one stroke}",
    eprint = "1610.01639",
    archivePrefix = "arXiv",
    primaryClass = "hep-ph",
    reportNumber = "DESY-16-184, IPPP-16-79, CERN-TH-2016-055",
    doi = "10.1088/1475-7516/2017/08/001",
    journal = "JCAP",
    volume = "08",
    pages = "001",
    year = "2017"
}

@mastersthesis{Turbang:2020,
  author = {Turbang, Kevin},
  title = {The Strong CP Problem and Gravitational Waves},
  school = {Universit\'e Libre de Bruxelles},
  year = {2020},
  type = {Master's thesis},
  url = {https://iihe.ac.be/sites/default/files/thesis-kevin-turbang-pheno-master-2020pdf/thesis-kevin-turbang-pheno-master-2020.pdf}
}

@article{Matuszak:2026xsz,
    author = "Matuszak, Jonas and Tasillo, Carlo",
    title = "{TransitionListener v2.0 -- Robust gravitational wave predictions for cosmological phase transitions}",
    eprint = "2605.15259",
    archivePrefix = "arXiv",
    primaryClass = "hep-ph",
    month = "5",
    year = "2026"
}

@article{Caprini:2024hue,
    author = "Caprini, Chiara and Jinno, Ryusuke and Lewicki, Marek and Madge, Eric and Merchand, Marco and Nardini, Germano and Pieroni, Mauro and Roper Pol, Alberto and Vaskonen, Ville",
    collaboration = "LISA Cosmology Working Group",
    title = "{Gravitational waves from first-order phase transitions in LISA: reconstruction pipeline and physics interpretation}",
    eprint = "2403.03723",
    archivePrefix = "arXiv",
    primaryClass = "astro-ph.CO",
    reportNumber = "LISA-COSWG-24-01, CERN-TH-2024-029",
    doi = "10.1088/1475-7516/2024/10/020",
    journal = "JCAP",
    volume = "10",
    pages = "020",
    year = "2024"
}

@misc{AxionLimits,
  author       = {Ciaran O'Hare},
  title        = {cajohare/AxionLimits: AxionLimits},
  month        = jul,
  year         = 2020,
  publisher    = {Zenodo},
  version      = {v1.0},
  doi          = {10.5281/zenodo.3932430},
  howpublished = {\url{https://cajohare.github.io/AxionLimits/}}
}

@article{Caputo:2024oqc,
    author = "Caputo, Andrea and Raffelt, Georg",
    title = "{Astrophysical Axion Bounds: The 2024 Edition}",
    eprint = "2401.13728",
    archivePrefix = "arXiv",
    primaryClass = "hep-ph",
    reportNumber = "MPP-2024-13, CERN-TH-2024-013",
    doi = "10.22323/1.454.0041",
    journal = "PoS",
    volume = "COSMICWISPers",
    pages = "041",
    year = "2024"
}

@article{Breitbach:2018ddu,
    author = "Breitbach, Moritz and Kopp, Joachim and Madge, Eric and Opferkuch, Toby and Schwaller, Pedro",
    title = "{Dark, Cold, and Noisy: Constraining Secluded Hidden Sectors with Gravitational Waves}",
    eprint = "1811.11175",
    archivePrefix = "arXiv",
    primaryClass = "hep-ph",
    reportNumber = "CERN-TH-2018-255, MITP/18-115",
    doi = "10.1088/1475-7516/2019/07/007",
    journal = "JCAP",
    volume = "07",
    pages = "007",
    year = "2019"
}

@article{Figueroa:2023zhu,
    author = "Figueroa, Daniel G. and Pieroni, Mauro and Ricciardone, Angelo and Simakachorn, Peera",
    title = "{Cosmological Background Interpretation of Pulsar Timing Array Data}",
    eprint = "2307.02399",
    archivePrefix = "arXiv",
    primaryClass = "astro-ph.CO",
    reportNumber = "CERN-TH-2023-132",
    doi = "10.1103/PhysRevLett.132.171002",
    journal = "Phys. Rev. Lett.",
    volume = "132",
    number = "17",
    pages = "171002",
    year = "2024"
}

@article{vonHarling:2017yew,
    author = "von Harling, Benedict and Servant, Geraldine",
    title = "{QCD-induced Electroweak Phase Transition}",
    eprint = "1711.11554",
    archivePrefix = "arXiv",
    primaryClass = "hep-ph",
    reportNumber = "DESY-17-056",
    doi = "10.1007/JHEP01(2018)159",
    journal = "JHEP",
    volume = "01",
    pages = "159",
    year = "2018"
}

@article{Madge:2023dxc,
    author = "Madge, Eric and Morgante, Enrico and Puchades-Ib{\'a}{\~n}ez, Cristina and Ramberg, Nicklas and Ratzinger, Wolfram and Schenk, Sebastian and Schwaller, Pedro",
    title = "{Primordial gravitational waves in the nano-Hertz regime and PTA data {\textemdash} towards solving the GW inverse problem}",
    eprint = "2306.14856",
    archivePrefix = "arXiv",
    primaryClass = "hep-ph",
    reportNumber = "MITP-23-029",
    doi = "10.1007/JHEP10(2023)171",
    journal = "JHEP",
    volume = "10",
    pages = "171",
    year = "2023"
}

@article{Fairbairn:2019xog,
    author = "Fairbairn, Malcolm and Hardy, Edward and Wickens, Alastair",
    title = "{Hearing without seeing: gravitational waves from hot and cold hidden sectors}",
    eprint = "1901.11038",
    archivePrefix = "arXiv",
    primaryClass = "hep-ph",
    reportNumber = "KCL-PH-TH/2019-12",
    doi = "10.1007/JHEP07(2019)044",
    journal = "JHEP",
    volume = "07",
    pages = "044",
    year = "2019"
}

@article{Addazi:2023jvg,
    author = "Addazi, Andrea and Cai, Yi-Fu and Marciano, Antonino and Visinelli, Luca",
    title = "{Have pulsar timing array methods detected a cosmological phase transition?}",
    eprint = "2306.17205",
    archivePrefix = "arXiv",
    primaryClass = "astro-ph.CO",
    reportNumber = "CA21106; CA21136",
    doi = "10.1103/PhysRevD.109.015028",
    journal = "Phys. Rev. D",
    volume = "109",
    number = "1",
    pages = "015028",
    year = "2024"
}

@article{Bringmann:2023opz,
    author = "Bringmann, Torsten and Depta, Paul Frederik and Konstandin, Thomas and Schmidt-Hoberg, Kai and Tasillo, Carlo",
    title = "{Does NANOGrav observe a dark sector phase transition?}",
    eprint = "2306.09411",
    archivePrefix = "arXiv",
    primaryClass = "astro-ph.CO",
    reportNumber = "DESY-23-077",
    doi = "10.1088/1475-7516/2023/11/053",
    journal = "JCAP",
    volume = "11",
    pages = "053",
    year = "2023"
}

@article{Carena:2019une,
    author = "Carena, Marcela and Liu, Zhen and Wang, Yikun",
    title = "{Electroweak phase transition with spontaneous Z$_{2}$-breaking}",
    eprint = "1911.10206",
    archivePrefix = "arXiv",
    primaryClass = "hep-ph",
    reportNumber = "FERMILAB-PUB-19-667-T, FERMILAB-PUB-19-602-T",
    doi = "10.1007/JHEP08(2020)107",
    journal = "JHEP",
    volume = "08",
    pages = "107",
    year = "2020"
}

@article{Xue:2021gyq,
    author = "Xue, Xiao and others",
    title = "{Constraining Cosmological Phase Transitions with the Parkes Pulsar Timing Array}",
    eprint = "2110.03096",
    archivePrefix = "arXiv",
    primaryClass = "astro-ph.CO",
    doi = "10.1103/PhysRevLett.127.251303",
    journal = "Phys. Rev. Lett.",
    volume = "127",
    number = "25",
    pages = "251303",
    year = "2021"
}

@article{Salvio:2023ynn,
    author = "Salvio, Alberto",
    title = "{Supercooling in Radiative Symmetry Breaking: Theory Extensions, Gravitational Wave Detection and Primordial Black Holes}",
    eprint = "2307.04694",
    archivePrefix = "arXiv",
    primaryClass = "hep-ph",
    doi = "10.1088/1475-7516/2023/12/046",
    journal = "JCAP",
    volume = "12",
    pages = "046",
    year = "2023"
}

@article{Megias:2018sxv,
    author = "Meg{\'\i}as, Eugenio and Nardini, Germano and Quir{\'o}s, Mariano",
    title = "{Cosmological Phase Transitions in Warped Space: Gravitational Waves and Collider Signatures}",
    eprint = "1806.04877",
    archivePrefix = "arXiv",
    primaryClass = "hep-ph",
    reportNumber = "UAB-FT-776",
    doi = "10.1007/JHEP09(2018)095",
    journal = "JHEP",
    volume = "09",
    pages = "095",
    year = "2018"
}

@article{Ghosh:2023aum,
    author = "Ghosh, Tathagata and Ghoshal, Anish and Guo, Huai-Ke and Hajkarim, Fazlollah and King, Stephen F. and Sinha, Kuver and Wang, Xin and White, Graham",
    title = "{Did we hear the sound of the Universe boiling? Analysis using the full fluid velocity profiles and NANOGrav 15-year data}",
    eprint = "2307.02259",
    archivePrefix = "arXiv",
    primaryClass = "astro-ph.HE",
    doi = "10.1088/1475-7516/2024/05/100",
    journal = "JCAP",
    volume = "05",
    pages = "100",
    year = "2024"
}

@article{Wu:2023hsa,
    author = "Wu, Yu-Mei and Chen, Zu-Cheng and Huang, Qing-Guo",
    title = "{Cosmological interpretation for the stochastic signal in pulsar timing arrays}",
    eprint = "2307.03141",
    archivePrefix = "arXiv",
    primaryClass = "astro-ph.CO",
    doi = "10.1007/s11433-023-2298-7",
    journal = "Sci. China Phys. Mech. Astron.",
    volume = "67",
    number = "4",
    pages = "240412",
    year = "2024"
}

@article{Ellis:2022lft,
    author = "Ellis, John and Lewicki, Marek and Merchand, Marco and No, Jos{\'e} Miguel and Zych, Mateusz",
    title = "{The scalar singlet extension of the Standard Model: gravitational waves versus baryogenesis}",
    eprint = "2210.16305",
    archivePrefix = "arXiv",
    primaryClass = "hep-ph",
    reportNumber = "CERN-TH-2022-150, KCL-PH-TH/2022-48, IFT{\textendash}UAM/CSIC{\textendash}22-133",
    doi = "10.1007/JHEP01(2023)093",
    journal = "JHEP",
    volume = "01",
    pages = "093",
    year = "2023"
}

@article{Azatov:2022tii,
    author = "Azatov, Aleksandr and Barni, Giulio and Chakraborty, Sabyasachi and Vanvlasselaer, Miguel and Yin, Wen",
    title = "{Ultra-relativistic bubbles from the simplest Higgs portal and their cosmological consequences}",
    eprint = "2207.02230",
    archivePrefix = "arXiv",
    primaryClass = "hep-ph",
    reportNumber = "SISSA 12/2022/FISI TU-1157",
    doi = "10.1007/JHEP10(2022)017",
    journal = "JHEP",
    volume = "10",
    pages = "017",
    year = "2022"
}

@article{Agashe:2019lhy,
    author = "Agashe, Kaustubh and Du, Peizhi and Ekhterachian, Majid and Kumar, Soubhik and Sundrum, Raman",
    title = "{Cosmological Phase Transition of Spontaneous Confinement}",
    eprint = "1910.06238",
    archivePrefix = "arXiv",
    primaryClass = "hep-ph",
    reportNumber = "UMD-PP-019-05, YITP-SB-19-32",
    doi = "10.1007/JHEP05(2020)086",
    journal = "JHEP",
    volume = "05",
    pages = "086",
    year = "2020"
}

@article{Romero:2021kby,
    author = "Romero, Alba and Martinovic, Katarina and Callister, Thomas A. and Guo, Huai-Ke and Mart{\'\i}nez, Mario and Sakellariadou, Mairi and Yang, Feng-Wei and Zhao, Yue",
    title = "{Implications for First-Order Cosmological Phase Transitions from the Third LIGO-Virgo Observing Run}",
    eprint = "2102.01714",
    archivePrefix = "arXiv",
    primaryClass = "hep-ph",
    doi = "10.1103/PhysRevLett.126.151301",
    journal = "Phys. Rev. Lett.",
    volume = "126",
    number = "15",
    pages = "151301",
    year = "2021"
}

@article{An:2023jxf,
    author = "An, Haipeng and Su, Boye and Tai, Hanwen and Wang, Lian-Tao and Yang, Chen",
    title = "{Phase transition during inflation and the gravitational wave signal at pulsar timing arrays}",
    eprint = "2308.00070",
    archivePrefix = "arXiv",
    primaryClass = "astro-ph.CO",
    doi = "10.1103/PhysRevD.109.L121304",
    journal = "Phys. Rev. D",
    volume = "109",
    number = "12",
    pages = "L121304",
    year = "2024"
}

@article{DiBari:2021dri,
    author = "Di Bari, Pasquale and Marfatia, Danny and Zhou, Ye-Ling",
    title = "{Gravitational waves from first-order phase transitions in Majoron models of neutrino mass}",
    eprint = "2106.00025",
    archivePrefix = "arXiv",
    primaryClass = "hep-ph",
    doi = "10.1007/JHEP10(2021)193",
    journal = "JHEP",
    volume = "10",
    pages = "193",
    year = "2021"
}

@article{Brandenburg:2021tmp,
    author = "Brandenburg, Axel and Clarke, Emma and He, Yutong and Kahniashvili, Tina",
    title = "{Can we observe the QCD phase transition-generated gravitational waves through pulsar timing arrays?}",
    eprint = "2102.12428",
    archivePrefix = "arXiv",
    primaryClass = "astro-ph.CO",
    reportNumber = "NORDITA-2021-016",
    doi = "10.1103/PhysRevD.104.043513",
    journal = "Phys. Rev. D",
    volume = "104",
    number = "4",
    pages = "043513",
    year = "2021"
}

@article{Greljo:2019xan,
    author = "Greljo, Admir and Opferkuch, Toby and Stefanek, Ben A.",
    title = "{Gravitational Imprints of Flavor Hierarchies}",
    eprint = "1910.02014",
    archivePrefix = "arXiv",
    primaryClass = "hep-ph",
    reportNumber = "CERN-TH-2019-162",
    doi = "10.1103/PhysRevLett.124.171802",
    journal = "Phys. Rev. Lett.",
    volume = "124",
    number = "17",
    pages = "171802",
    year = "2020"
}

@article{Ertas:2021xeh,
    author = "Ertas, Fatih and Kahlhoefer, Felix and Tasillo, Carlo",
    title = "{Turn up the volume: listening to phase transitions in hot dark sectors}",
    eprint = "2109.06208",
    archivePrefix = "arXiv",
    primaryClass = "astro-ph.CO",
    reportNumber = "TTK-21-36, DESY-22-014",
    doi = "10.1088/1475-7516/2022/02/014",
    journal = "JCAP",
    volume = "02",
    number = "02",
    pages = "014",
    year = "2022"
}

@article{Agashe:2020lfz,
    author = "Agashe, Kaustubh and Du, Peizhi and Ekhterachian, Majid and Kumar, Soubhik and Sundrum, Raman",
    title = "{Phase Transitions from the Fifth Dimension}",
    eprint = "2010.04083",
    archivePrefix = "arXiv",
    primaryClass = "hep-th",
    reportNumber = "UMD-PP-020-5, YITP-SB-2020-29",
    doi = "10.1007/JHEP02(2021)051",
    journal = "JHEP",
    volume = "02",
    pages = "051",
    year = "2021"
}

@article{Harigaya:2019shz,
    author = "Harigaya, Keisuke and Mcgehee, Robert and Murayama, Hitoshi and Schutz, Katelin",
    title = "{A predictive mirror twin Higgs with small Z$_{2}$ breaking}",
    eprint = "1905.08798",
    archivePrefix = "arXiv",
    primaryClass = "hep-ph",
    doi = "10.1007/JHEP05(2020)155",
    journal = "JHEP",
    volume = "05",
    pages = "155",
    year = "2020"
}

@article{Baldes:2023cih,
    author = "Baldes, Iason and Dichtl, Maximilian and Gouttenoire, Yann and Sala, Filippo",
    title = "{Ultrahigh-Energy Particle Collisions and Heavy Dark Matter at Phase Transitions}",
    eprint = "2306.15555",
    archivePrefix = "arXiv",
    primaryClass = "hep-ph",
    doi = "10.1103/PhysRevLett.134.061001",
    journal = "Phys. Rev. Lett.",
    volume = "134",
    number = "6",
    pages = "061001",
    year = "2025"
}

@article{Blasi:2023rqi,
    author = "Blasi, Simone and Jinno, Ryusuke and Konstandin, Thomas and Rubira, Henrique and Stomberg, Isak",
    title = "{Gravitational waves from defect-driven phase transitions: domain walls}",
    eprint = "2302.06952",
    archivePrefix = "arXiv",
    primaryClass = "astro-ph.CO",
    doi = "10.1088/1475-7516/2023/10/051",
    journal = "JCAP",
    volume = "10",
    pages = "051",
    year = "2023"
}

@article{Athron:2023rfq,
    author = "Athron, Peter and Morris, Lachlan and Xu, Zhongxiu",
    title = "{How robust are gravitational wave predictions from cosmological phase transitions?}",
    eprint = "2309.05474",
    archivePrefix = "arXiv",
    primaryClass = "hep-ph",
    doi = "10.1088/1475-7516/2024/05/075",
    journal = "JCAP",
    volume = "05",
    pages = "075",
    year = "2024"
}

@article{Cataldi:2024pgt,
    author = "Cataldi, Martina and Shakya, Bibhushan",
    title = "{Leptogenesis via bubble collisions}",
    eprint = "2407.16747",
    archivePrefix = "arXiv",
    primaryClass = "hep-ph",
    reportNumber = "DESY-24-110",
    doi = "10.1088/1475-7516/2024/11/047",
    journal = "JCAP",
    volume = "11",
    pages = "047",
    year = "2024"
}

@article{Kumar:2021ffi,
    author = "Kumar, Soubhik and Sundrum, Raman and Tsai, Yuhsin",
    title = "{Non-Gaussian stochastic gravitational waves from phase transitions}",
    eprint = "2102.05665",
    archivePrefix = "arXiv",
    primaryClass = "astro-ph.CO",
    doi = "10.1007/JHEP11(2021)107",
    journal = "JHEP",
    volume = "11",
    pages = "107",
    year = "2021"
}

@article{Conaci:2024tlc,
    author = "Conaci, Angela and Delle Rose, Luigi and Dev, P. S. Bhupal and Ghoshal, Anish",
    title = "{Slaying axion-like particles via gravitational waves and primordial black holes from supercooled phase transition}",
    eprint = "2401.09411",
    archivePrefix = "arXiv",
    primaryClass = "astro-ph.CO",
    doi = "10.1007/JHEP12(2024)196",
    journal = "JHEP",
    volume = "12",
    pages = "196",
    year = "2024"
}

@article{Kobakhidze:2017mru,
    author = "Kobakhidze, Archil and Lagger, Cyril and Manning, Adrian and Yue, Jason",
    title = "{Gravitational waves from a supercooled electroweak phase transition and their detection with pulsar timing arrays}",
    eprint = "1703.06552",
    archivePrefix = "arXiv",
    primaryClass = "hep-ph",
    doi = "10.1140/epjc/s10052-017-5132-y",
    journal = "Eur. Phys. J. C",
    volume = "77",
    number = "8",
    pages = "570",
    year = "2017"
}

@article{Cataldi:2025nac,
    author = {Cataldi, Martina and M{\"u}{\"u}rsepp, Kristjan and Vanvlasselaer, Miguel},
    title = "{CP-violation in production of heavy neutrinos from bubble collisions}",
    eprint = "2506.12123",
    archivePrefix = "arXiv",
    primaryClass = "hep-ph",
    reportNumber = "DESY-25-082",
    doi = "10.1007/JHEP01(2026)058",
    journal = "JHEP",
    volume = "01",
    pages = "058",
    year = "2026"
}

@article{Chun:2023ezg,
    author = "Chun, Eung Jin and Dutka, Tomasz P. and Jung, Tae Hyun and Nagels, Xander and Vanvlasselaer, Miguel",
    title = "{Bubble-assisted leptogenesis}",
    eprint = "2305.10759",
    archivePrefix = "arXiv",
    primaryClass = "hep-ph",
    reportNumber = "CTPU-PTC-23-17",
    doi = "10.1007/JHEP09(2023)164",
    journal = "JHEP",
    volume = "09",
    pages = "164",
    year = "2023"
}

@article{Azatov:2021ifm,
    author = "Azatov, Aleksandr and Vanvlasselaer, Miguel and Yin, Wen",
    title = "{Dark Matter production from relativistic bubble walls}",
    eprint = "2101.05721",
    archivePrefix = "arXiv",
    primaryClass = "hep-ph",
    reportNumber = "SISSA 03/2021/FISI",
    doi = "10.1007/JHEP03(2021)288",
    journal = "JHEP",
    volume = "03",
    pages = "288",
    year = "2021"
}

@article{Azatov:2021irb,
    author = "Azatov, Aleksandr and Vanvlasselaer, Miguel and Yin, Wen",
    title = "{Baryogenesis via relativistic bubble walls}",
    eprint = "2106.14913",
    archivePrefix = "arXiv",
    primaryClass = "hep-ph",
    reportNumber = "SISSA 13/2021/FISI TU-1127",
    doi = "10.1007/JHEP10(2021)043",
    journal = "JHEP",
    volume = "10",
    pages = "043",
    year = "2021"
}

@article{Baldes:2021vyz,
    author = "Baldes, Iason and Blasi, Simone and Mariotti, Alberto and Sevrin, Alexander and Turbang, Kevin",
    title = "{Baryogenesis via relativistic bubble expansion}",
    eprint = "2106.15602",
    archivePrefix = "arXiv",
    primaryClass = "hep-ph",
    reportNumber = "ULB-TH/21-09",
    doi = "10.1103/PhysRevD.104.115029",
    journal = "Phys. Rev. D",
    volume = "104",
    number = "11",
    pages = "115029",
    year = "2021"
}

@article{Vaskonen:2016yiu,
    author = "Vaskonen, Ville",
    title = "{Electroweak baryogenesis and gravitational waves from a real scalar singlet}",
    eprint = "1611.02073",
    archivePrefix = "arXiv",
    primaryClass = "hep-ph",
    doi = "10.1103/PhysRevD.95.123515",
    journal = "Phys. Rev. D",
    volume = "95",
    number = "12",
    pages = "123515",
    year = "2017"
}

@article{Marzola:2017jzl,
    author = "Marzola, Luca and Racioppi, Antonio and Vaskonen, Ville",
    title = "{Phase transition and gravitational wave phenomenology of scalar conformal extensions of the Standard Model}",
    eprint = "1704.01034",
    archivePrefix = "arXiv",
    primaryClass = "hep-ph",
    doi = "10.1140/epjc/s10052-017-4996-1",
    journal = "Eur. Phys. J. C",
    volume = "77",
    number = "7",
    pages = "484",
    year = "2017"
}

@article{Marzo:2018nov,
    author = "Marzo, Carlo and Marzola, Luca and Vaskonen, Ville",
    title = "{Phase transition and vacuum stability in the classically conformal B{\textendash}L model}",
    eprint = "1811.11169",
    archivePrefix = "arXiv",
    primaryClass = "hep-ph",
    reportNumber = "KCL-PH-TH/2018-68",
    doi = "10.1140/epjc/s10052-019-7076-x",
    journal = "Eur. Phys. J. C",
    volume = "79",
    number = "7",
    pages = "601",
    year = "2019"
}

@article{Lewicki:2021xku,
    author = "Lewicki, Marek and Pujol{\`a}s, Oriol and Vaskonen, Ville",
    title = "{Escape from supercooling with or without bubbles: gravitational wave signatures}",
    eprint = "2106.09706",
    archivePrefix = "arXiv",
    primaryClass = "astro-ph.CO",
    doi = "10.1140/epjc/s10052-021-09669-6",
    journal = "Eur. Phys. J. C",
    volume = "81",
    number = "9",
    pages = "857",
    year = "2021"
}

@article{Lewicki:2024ghw,
    author = "Lewicki, Marek and Toczek, Piotr and Vaskonen, Ville",
    title = "{Black Holes and Gravitational Waves from Slow First-Order Phase Transitions}",
    eprint = "2402.04158",
    archivePrefix = "arXiv",
    primaryClass = "astro-ph.CO",
    doi = "10.1103/PhysRevLett.133.221003",
    journal = "Phys. Rev. Lett.",
    volume = "133",
    number = "22",
    pages = "221003",
    year = "2024"
}

@article{Lewicki:2024sfw,
    author = "Lewicki, Marek and Toczek, Piotr and Vaskonen, Ville",
    title = "{Black holes and gravitational waves from phase transitions in realistic models}",
    eprint = "2412.10366",
    archivePrefix = "arXiv",
    primaryClass = "astro-ph.CO",
    doi = "10.1016/j.dark.2025.102075",
    journal = "Phys. Dark Univ.",
    volume = "50",
    pages = "102075",
    year = "2025"
}

@article{Bringmann:2026xcx,
    author = "Bringmann, Torsten and Konstandin, Thomas and Matuszak, Jonas and Schmidt-Hoberg, Kai and Tasillo, Carlo",
    title = "{Tuning the violins: dark sector phase transition models for the PTA signal}",
    eprint = "2602.09092",
    archivePrefix = "arXiv",
    primaryClass = "hep-ph",
    reportNumber = "DESY-26-014, P3H-26-009, TTP26-002",
    month = "2",
    year = "2026"
}

@article{Balan:2025uke,
    author = "Balan, Sowmiya and Bringmann, Torsten and Kahlhoefer, Felix and Matuszak, Jonas and Tasillo, Carlo",
    title = "{Sub-GeV dark matter and nano-Hertz gravitational waves from a classically conformal dark sector}",
    eprint = "2502.19478",
    archivePrefix = "arXiv",
    primaryClass = "hep-ph",
    doi = "10.1088/1475-7516/2025/08/062",
    journal = "JCAP",
    volume = "08",
    pages = "062",
    year = "2025"
}

@article{Bringmann:2023iuz,
    author = "Bringmann, Torsten and Gonzalo, Tom{\'a}s E. and Kahlhoefer, Felix and Matuszak, Jonas and Tasillo, Carlo",
    title = "{Hunting WIMPs with LISA: correlating dark matter and gravitational wave signals}",
    eprint = "2311.06346",
    archivePrefix = "arXiv",
    primaryClass = "astro-ph.CO",
    doi = "10.1088/1475-7516/2024/05/065",
    journal = "JCAP",
    volume = "05",
    pages = "065",
    year = "2024"
}

@article{Benincasa:2026dhg,
    author = "Benincasa, Nico and Li, Ji-Wei and Pu, Hanxiao and Mann, Robert B. and Shokrollahic, Vahid and Steele, T. G. and Wang, Zhi-Wei",
    title = "{Conformal versus non-conformal two-Higgs-doublet model: phase transitions and gravitational waves}",
    eprint = "2603.07189",
    archivePrefix = "arXiv",
    primaryClass = "hep-ph",
    month = "3",
    year = "2026"
}

@article{Benincasa:2025tdr,
    author = "Benincasa, Nico and Delle Rose, Luigi and Panizzi, Luca and Razzaq, Maimoona and Urzetta, Savio",
    title = "{Phase transitions and gravitational waves in a non-Abelian vector dark matter scenario}",
    eprint = "2506.22248",
    archivePrefix = "arXiv",
    primaryClass = "hep-ph",
    doi = "10.1103/xd9f-bq4p",
    journal = "Phys. Rev. D",
    volume = "112",
    number = "9",
    pages = "095004",
    year = "2025"
}

@article{Benincasa:2023vyp,
    author = "Benincasa, Nico and Hryczuk, Andrzej and Kannike, Kristjan and Laletin, Maxim",
    title = "{Phase transitions and gravitational waves in a model of~$\mathbb{Z}_{3}$ scalar dark matter}",
    eprint = "2312.04627",
    archivePrefix = "arXiv",
    primaryClass = "hep-ph",
    doi = "10.1007/JHEP02(2024)207",
    journal = "JHEP",
    volume = "02",
    pages = "207",
    year = "2024"
}

@article{Benincasa:2022elt,
    author = "Benincasa, Nico and Delle Rose, Luigi and Kannike, Kristjan and Marzola, Luca",
    title = "{Multi-step phase transitions and gravitational waves in the inert doublet model}",
    eprint = "2205.06669",
    archivePrefix = "arXiv",
    primaryClass = "hep-ph",
    doi = "10.1088/1475-7516/2022/12/025",
    journal = "JCAP",
    volume = "12",
    pages = "025",
    year = "2022"
}

@article{Alanne:2020jwx,
    author = "Alanne, Tommi and Benincasa, Nico and Heikinheimo, Matti and Kannike, Kristjan and Keus, Venus and Koivunen, Niko and Tuominen, Kimmo",
    title = "{Pseudo-Goldstone dark matter: gravitational waves and direct-detection blind spots}",
    eprint = "2008.09605",
    archivePrefix = "arXiv",
    primaryClass = "hep-ph",
    reportNumber = "HIP-2020-24/TH",
    doi = "10.1007/JHEP10(2020)080",
    journal = "JHEP",
    volume = "10",
    pages = "080",
    year = "2020"
}

@article{Kannike:2019wsn,
    author = "Kannike, Kristjan and Raidal, Martti",
    title = "{Phase Transitions and Gravitational Wave Tests of Pseudo-Goldstone Dark Matter in the Softly Broken U(1) Scalar Singlet Model}",
    eprint = "1901.03333",
    archivePrefix = "arXiv",
    primaryClass = "hep-ph",
    doi = "10.1103/PhysRevD.99.115010",
    journal = "Phys. Rev. D",
    volume = "99",
    number = "11",
    pages = "115010",
    year = "2019"
}

@article{Kannike:2019mzk,
    author = "Kannike, Kristjan and Loos, Kaius and Raidal, Martti",
    title = "{Gravitational wave signals of pseudo-Goldstone dark matter in the $\mathbb{Z}_{3}$ complex singlet model}",
    eprint = "1907.13136",
    archivePrefix = "arXiv",
    primaryClass = "hep-ph",
    doi = "10.1103/PhysRevD.101.035001",
    journal = "Phys. Rev. D",
    volume = "101",
    number = "3",
    pages = "035001",
    year = "2020"
}

@article{Jinno:2022fom,
    author = "Jinno, Ryusuke and Shakya, Bibhushan and van de Vis, Jorinde",
    title = "{Gravitational Waves from Feebly Interacting Particles in a First Order Phase Transition}",
    eprint = "2211.06405",
    archivePrefix = "arXiv",
    primaryClass = "gr-qc",
    reportNumber = "DESY-22-172, IFT-UAM/CSIC-22-140, MITP-22-095, RESCEU-22/22",
    doi = "10.1103/phwp-jsvq",
    journal = "Phys. Rev. Lett.",
    volume = "136",
    number = "13",
    pages = "131002",
    year = "2026"
}

@article{Giudice:2024tcp,
    author = "Giudice, Gian F. and Lee, Hyun Min and Pomarol, Alex and Shakya, Bibhushan",
    title = "{Nonthermal heavy dark matter from a first-order phase transition}",
    eprint = "2403.03252",
    archivePrefix = "arXiv",
    primaryClass = "hep-ph",
    reportNumber = "CERN-TH-2024-031, DESY-24-031",
    doi = "10.1007/JHEP12(2024)190",
    journal = "JHEP",
    volume = "12",
    pages = "190",
    year = "2024"
}

@article{Inomata:2024rkt,
    author = "Inomata, Keisuke and Kamionkowski, Marc and Kasai, Kentaro and Shakya, Bibhushan",
    title = "{Gravitational waves from particles produced from bubble collisions in first-order phase transitions}",
    eprint = "2412.17912",
    archivePrefix = "arXiv",
    primaryClass = "astro-ph.CO",
    doi = "10.1103/k4s5-8zqy",
    journal = "Phys. Rev. D",
    volume = "112",
    number = "8",
    pages = "083523",
    year = "2025"
}

@article{Schwaller:2015tja,
    author = "Schwaller, Pedro",
    title = "{Gravitational Waves from a Dark Phase Transition}",
    eprint = "1504.07263",
    archivePrefix = "arXiv",
    primaryClass = "hep-ph",
    reportNumber = "CERN-PH-TH-2015-093",
    doi = "10.1103/PhysRevLett.115.181101",
    journal = "Phys. Rev. Lett.",
    volume = "115",
    number = "18",
    pages = "181101",
    year = "2015"
}

@article{Morgante:2022zvc,
    author = "Morgante, Enrico and Ramberg, Nicklas and Schwaller, Pedro",
    title = "{Gravitational waves from dark SU(3) Yang-Mills theory}",
    eprint = "2210.11821",
    archivePrefix = "arXiv",
    primaryClass = "hep-ph",
    doi = "10.1103/PhysRevD.107.036010",
    journal = "Phys. Rev. D",
    volume = "107",
    number = "3",
    pages = "036010",
    year = "2023"
}

@article{Pascoli:2026tuu,
    author = "Pascoli, S. and Rosauro-Alcaraz, S. and Zandi, M.",
    title = "{Cosmological phase transitions: from particle physics to gravitational waves, semi-analytically}",
    eprint = "2602.02829",
    archivePrefix = "arXiv",
    primaryClass = "hep-ph",
    month = "2",
    year = "2026"
}

@article{Levi:2022bzt,
    author = "Levi, Noam and Opferkuch, Toby and Redigolo, Diego",
    title = "{The supercooling window at weak and strong coupling}",
    eprint = "2212.08085",
    archivePrefix = "arXiv",
    primaryClass = "hep-ph",
    doi = "10.1007/JHEP02(2023)125",
    journal = "JHEP",
    volume = "02",
    pages = "125",
    year = "2023"
}

@article{Kierkla:2022odc,
    author = "Kierkla, Maciej and Karam, Alexandros and Swiezewska, Bogumila",
    title = "{Conformal model for gravitational waves and dark matter: a status update}",
    eprint = "2210.07075",
    archivePrefix = "arXiv",
    primaryClass = "astro-ph.CO",
    doi = "10.1007/JHEP03(2023)007",
    journal = "JHEP",
    volume = "03",
    pages = "007",
    year = "2023"
}

@article{Kierkla:2023von,
    author = "Kierkla, Maciej and Swiezewska, Bogumila and Tenkanen, Tuomas V. I. and van de Vis, Jorinde",
    title = "{Gravitational waves from supercooled phase transitions: dimensional transmutation meets dimensional reduction}",
    eprint = "2312.12413",
    archivePrefix = "arXiv",
    primaryClass = "hep-ph",
    doi = "10.1007/JHEP02(2024)234",
    journal = "JHEP",
    volume = "02",
    pages = "234",
    year = "2024"
}

@article{Prokopec:2018tnq,
    author = "Prokopec, Tomislav and Rezacek, Jonas and {\'S}wie{\.z}ewska, Bogumi{\l}a",
    title = "{Gravitational waves from conformal symmetry breaking}",
    eprint = "1809.11129",
    archivePrefix = "arXiv",
    primaryClass = "hep-ph",
    doi = "10.1088/1475-7516/2019/02/009",
    journal = "JCAP",
    volume = "02",
    pages = "009",
    year = "2019"
}

@article{Croon:2020cgk,
    author = "Croon, Djuna and Gould, Oliver and Schicho, Philipp and Tenkanen, Tuomas V. I. and White, Graham",
    title = "{Theoretical uncertainties for cosmological first-order phase transitions}",
    eprint = "2009.10080",
    archivePrefix = "arXiv",
    primaryClass = "hep-ph",
    reportNumber = "HIP-2020-26/TH",
    doi = "10.1007/JHEP04(2021)055",
    journal = "JHEP",
    volume = "04",
    pages = "055",
    year = "2021"
}

@article{Guo:2020grp,
    author = "Guo, Huai-Ke and Sinha, Kuver and Vagie, Daniel and White, Graham",
    title = "{Phase Transitions in an Expanding Universe: Stochastic Gravitational Waves in Standard and Non-Standard Histories}",
    eprint = "2007.08537",
    archivePrefix = "arXiv",
    primaryClass = "hep-ph",
    doi = "10.1088/1475-7516/2021/01/001",
    journal = "JCAP",
    volume = "01",
    pages = "001",
    year = "2021"
}

@article{Dror:2019syi,
    author = "Dror, Jeff A. and Hiramatsu, Takashi and Kohri, Kazunori and Murayama, Hitoshi and White, Graham",
    title = "{Testing the Seesaw Mechanism and Leptogenesis with Gravitational Waves}",
    eprint = "1908.03227",
    archivePrefix = "arXiv",
    primaryClass = "hep-ph",
    reportNumber = "IPMU19-0108, DESY-19-138, DESY 19-138, KEK-TH-2147, KEK-Cosmo-241",
    doi = "10.1103/PhysRevLett.124.041804",
    journal = "Phys. Rev. Lett.",
    volume = "124",
    number = "4",
    pages = "041804",
    year = "2020"
}

@article{Croon:2018erz,
    author = "Croon, Djuna and Sanz, Ver{\'o}nica and White, Graham",
    title = "{Model Discrimination in Gravitational Wave spectra from Dark Phase Transitions}",
    eprint = "1806.02332",
    archivePrefix = "arXiv",
    primaryClass = "hep-ph",
    doi = "10.1007/JHEP08(2018)203",
    journal = "JHEP",
    volume = "08",
    pages = "203",
    year = "2018"
}

@article{Croon:2018kqn,
    author = "Croon, Djuna and Gonzalo, Tom{\'a}s E. and White, Graham",
    title = "{Gravitational Waves from a Pati-Salam Phase Transition}",
    eprint = "1812.02747",
    archivePrefix = "arXiv",
    primaryClass = "hep-ph",
    reportNumber = "CoEPP-MN-18-11, CoEPP-MN-18-10",
    doi = "10.1007/JHEP02(2019)083",
    journal = "JHEP",
    volume = "02",
    pages = "083",
    year = "2019"
}

@article{Balazs:2016tbi,
    author = "Balazs, Csaba and Fowlie, Andrew and Mazumdar, Anupam and White, Graham",
    title = "{Gravitational waves at aLIGO and vacuum stability with a scalar singlet extension of the Standard Model}",
    eprint = "1611.01617",
    archivePrefix = "arXiv",
    primaryClass = "hep-ph",
    reportNumber = "COEPP-MN-16-27",
    doi = "10.1103/PhysRevD.95.043505",
    journal = "Phys. Rev. D",
    volume = "95",
    number = "4",
    pages = "043505",
    year = "2017"
}

@article{Guo:2021qcq,
    author = "Guo, Huai-Ke and Sinha, Kuver and Vagie, Daniel and White, Graham",
    title = "{The benefits of diligence: how precise are predicted gravitational wave spectra in models with phase transitions?}",
    eprint = "2103.06933",
    archivePrefix = "arXiv",
    primaryClass = "hep-ph",
    doi = "10.1007/JHEP06(2021)164",
    journal = "JHEP",
    volume = "06",
    pages = "164",
    year = "2021"
}

@article{Croon:2018new,
    author = "Croon, Djuna and White, Graham",
    title = "{Exotic Gravitational Wave Signatures from Simultaneous Phase Transitions}",
    eprint = "1803.05438",
    archivePrefix = "arXiv",
    primaryClass = "hep-ph",
    doi = "10.1007/JHEP05(2018)210",
    journal = "JHEP",
    volume = "05",
    pages = "210",
    year = "2018"
}

@article{Addazi:2016fbj,
    author = "Addazi, Andrea",
    title = "{Limiting First Order Phase Transitions in Dark Gauge Sectors from Gravitational Waves experiments}",
    eprint = "1607.08057",
    archivePrefix = "arXiv",
    primaryClass = "hep-ph",
    doi = "10.1142/S0217732317500493",
    journal = "Mod. Phys. Lett. A",
    volume = "32",
    number = "08",
    pages = "1750049",
    year = "2017"
}

@article{Pasechnik:2023hwv,
    author = "Pasechnik, Roman and Reichert, Manuel and Sannino, Francesco and Wang, Zhi-Wei",
    title = "{Gravitational waves from composite dark sectors}",
    eprint = "2309.16755",
    archivePrefix = "arXiv",
    primaryClass = "hep-ph",
    doi = "10.1007/JHEP02(2024)159",
    journal = "JHEP",
    volume = "02",
    pages = "159",
    year = "2024"
}

@article{Reichert:2021cvs,
    author = "Reichert, Manuel and Sannino, Francesco and Wang, Zhi-Wei and Zhang, Chen",
    title = "{Dark confinement and chiral phase transitions: gravitational waves vs matter representations}",
    eprint = "2109.11552",
    archivePrefix = "arXiv",
    primaryClass = "hep-ph",
    doi = "10.1007/JHEP01(2022)003",
    journal = "JHEP",
    volume = "01",
    pages = "003",
    year = "2022"
}

@article{Huang:2020bbe,
    author = "Huang, Wei-Chih and Sannino, Francesco and Wang, Zhi-Wei",
    title = "{Gravitational Waves from Pati-Salam Dynamics}",
    eprint = "2004.02332",
    archivePrefix = "arXiv",
    primaryClass = "hep-ph",
    reportNumber = "CP3-Origins-2020-05 DNRF90",
    doi = "10.1103/PhysRevD.102.095025",
    journal = "Phys. Rev. D",
    volume = "102",
    number = "9",
    pages = "095025",
    year = "2020"
}

@article{Caprini:2009fx,
    author = "Caprini, Chiara and Durrer, Ruth and Konstandin, Thomas and Servant, Geraldine",
    title = "{General Properties of the Gravitational Wave Spectrum from Phase Transitions}",
    eprint = "0901.1661",
    archivePrefix = "arXiv",
    primaryClass = "astro-ph.CO",
    doi = "10.1103/PhysRevD.79.083519",
    journal = "Phys. Rev. D",
    volume = "79",
    pages = "083519",
    year = "2009"
}

@article{Grojean:2006bp,
    author = "Grojean, Christophe and Servant, Geraldine",
    title = "{Gravitational Waves from Phase Transitions at the Electroweak Scale and Beyond}",
    eprint = "hep-ph/0607107",
    archivePrefix = "arXiv",
    reportNumber = "CERN-PH-TH-2006-125",
    doi = "10.1103/PhysRevD.75.043507",
    journal = "Phys. Rev. D",
    volume = "75",
    pages = "043507",
    year = "2007"
}

@article{Grojean:2004xa,
    author = "Grojean, Christophe and Servant, Geraldine and Wells, James D.",
    title = "{First-order electroweak phase transition in the standard model with a low cutoff}",
    eprint = "hep-ph/0407019",
    archivePrefix = "arXiv",
    reportNumber = "SACLAY-T04-084, MCTP-04-37, ANL-HEP-PR-04-63, EFI-04-23",
    doi = "10.1103/PhysRevD.71.036001",
    journal = "Phys. Rev. D",
    volume = "71",
    pages = "036001",
    year = "2005"
}

@article{Jinno:2016knw,
    author = "Jinno, Ryusuke and Takimoto, Masahiro",
    title = "{Probing a classically conformal B-L model with gravitational waves}",
    eprint = "1604.05035",
    archivePrefix = "arXiv",
    primaryClass = "hep-ph",
    reportNumber = "KEK-TH-1896",
    doi = "10.1103/PhysRevD.95.015020",
    journal = "Phys. Rev. D",
    volume = "95",
    number = "1",
    pages = "015020",
    year = "2017"
}

@article{Hashino:2018wee,
    author = "Hashino, Katsuya and Jinno, Ryusuke and Kakizaki, Mitsuru and Kanemura, Shinya and Takahashi, Tomo and Takimoto, Masahiro",
    title = "{Selecting models of first-order phase transitions using the synergy between collider and gravitational-wave experiments}",
    eprint = "1809.04994",
    archivePrefix = "arXiv",
    primaryClass = "hep-ph",
    reportNumber = "CTPU-PTC-18-26, KEK-TH-2073, OU-HET-977, UT-HET-128",
    doi = "10.1103/PhysRevD.99.075011",
    journal = "Phys. Rev. D",
    volume = "99",
    number = "7",
    pages = "075011",
    year = "2019"
}

@article{Espinosa:2011ax,
    author = "Espinosa, Jose R. and Konstandin, Thomas and Riva, Francesco",
    title = "{Strong Electroweak Phase Transitions in the Standard Model with a Singlet}",
    eprint = "1107.5441",
    archivePrefix = "arXiv",
    primaryClass = "hep-ph",
    reportNumber = "CERN-PH-TH-2011-171",
    doi = "10.1016/j.nuclphysb.2011.09.010",
    journal = "Nucl. Phys. B",
    volume = "854",
    pages = "592--630",
    year = "2012"
}

@article{Dolan:1973qd,
    author = "Dolan, L. and Jackiw, R.",
    title = "{Symmetry Behavior at Finite Temperature}",
    reportNumber = "MIT-CTP-406",
    doi = "10.1103/PhysRevD.9.3320",
    journal = "Phys. Rev. D",
    volume = "9",
    pages = "3320--3341",
    year = "1974"
}

@article{Weinberg:1974hy,
    author = "Weinberg, Steven",
    title = "{Gauge and Global Symmetries at High Temperature}",
    reportNumber = "PRINT-74-0689 (HARVARD)",
    doi = "10.1103/PhysRevD.9.3357",
    journal = "Phys. Rev. D",
    volume = "9",
    pages = "3357--3378",
    year = "1974"
}

@article{Ford:1992mv,
    author = "Ford, C. and Jones, D. R. T. and Stephenson, P. W. and Einhorn, M. B.",
    title = "{The Effective potential and the renormalization group}",
    eprint = "hep-lat/9210033",
    archivePrefix = "arXiv",
    reportNumber = "LTH-288, UM-TH-92-21",
    doi = "10.1016/0550-3213(93)90206-5",
    journal = "Nucl. Phys. B",
    volume = "395",
    pages = "17--34",
    year = "1993"
}

@article{Bando:1992wy,
    author = "Bando, Masako and Kugo, Taichiro and Maekawa, Nobuhiro and Nakano, Hiroaki",
    title = "{Improving the effective potential: Multimass scale case}",
    eprint = "hep-ph/9210229",
    archivePrefix = "arXiv",
    reportNumber = "KUNS-1162, KUNS1162",
    doi = "10.1143/PTP.90.405",
    journal = "Prog. Theor. Phys.",
    volume = "90",
    pages = "405--418",
    year = "1993"
}

@article{Ginsparg:1980ef,
    author = "Ginsparg, Paul H.",
    title = "{First Order and Second Order Phase Transitions in Gauge Theories at Finite Temperature}",
    reportNumber = "SACLAY-DPh-T 80/27",
    doi = "10.1016/0550-3213(80)90418-6",
    journal = "Nucl. Phys. B",
    volume = "170",
    pages = "388--408",
    year = "1980"
}

@article{Appelquist:1981vg,
    author = "Appelquist, Thomas and Pisarski, Robert D.",
    title = "{High-Temperature Yang-Mills Theories and Three-Dimensional Quantum Chromodynamics}",
    reportNumber = "Print-81-0020 (YALE), YTP-81-01, COO-3075-203",
    doi = "10.1103/PhysRevD.23.2305",
    journal = "Phys. Rev. D",
    volume = "23",
    pages = "2305",
    year = "1981"
}

@article{Coleman:1977py,
    author = "Coleman, Sidney R.",
    title = "{The Fate of the False Vacuum. 1. Semiclassical Theory}",
    reportNumber = "HUTP-77-A004",
    doi = "10.1103/PhysRevD.15.2929",
    journal = "Phys. Rev. D",
    volume = "15",
    pages = "2929--2936",
    year = "1977",
    note = "[Erratum: Phys.Rev.D 16, 1248 (1977)]"
}

@article{Callan:1977pt,
    author = "Callan, Jr., Curtis G. and Coleman, Sidney R.",
    title = "{The Fate of the False Vacuum. 2. First Quantum Corrections}",
    reportNumber = "HUTP-77-A032",
    doi = "10.1103/PhysRevD.16.1762",
    journal = "Phys. Rev. D",
    volume = "16",
    pages = "1762--1768",
    year = "1977"
}

@article{Linde:1981zj,
    author = "Linde, Andrei D.",
    title = "{Decay of the False Vacuum at Finite Temperature}",
    reportNumber = "LEBEDEV-81-265",
    doi = "10.1016/0550-3213(83)90072-X",
    journal = "Nucl. Phys. B",
    volume = "216",
    pages = "421",
    year = "1983",
    note = "[Erratum: Nucl.Phys.B 223, 544 (1983)]"
}

@article{Chikashige:1980ui,
    author = "Chikashige, Y. and Mohapatra, Rabindra N. and Peccei, R. D.",
    title = "{Are There Real Goldstone Bosons Associated with Broken Lepton Number?}",
    reportNumber = "MPI-PAE-PTH-36-80",
    doi = "10.1016/0370-2693(81)90011-3",
    journal = "Phys. Lett. B",
    volume = "98",
    pages = "265--268",
    year = "1981"
}

@article{Brune:2018sab,
    author = {Brune, Tim and P{\"a}s, Heinrich},
    title = "{Massive Majorons and constraints on the Majoron-neutrino coupling}",
    eprint = "1808.08158",
    archivePrefix = "arXiv",
    primaryClass = "hep-ph",
    reportNumber = "DO-TH 18/23",
    doi = "10.1103/PhysRevD.99.096005",
    journal = "Phys. Rev. D",
    volume = "99",
    number = "9",
    pages = "096005"}

@article{deGiorgi:2026jqn,
    author = "de Giorgi, Arturo and Naredo-Tuero, Daniel and Ponce D{\'\i}az, Xavier",
    title = "{The Majoron Cosmological Window: Dark Matter and Thermal Leptogenesis}",
    eprint = "2605.18944",
    archivePrefix = "arXiv",
    primaryClass = "hep-ph",
    reportNumber = "IPPP/26/39",
    month = "5",
    year = "2026"
}

@article{Batell:2026avi,
    author = "Batell, Brian and Dasgupta, Arnab and Dutta, Swapnil and Ghalsasi, Akshay",
    title = "{Majoron Dark Matter, High-Scale Seesaw, and Leptogenesis}",
    eprint = "2606.02706",
    archivePrefix = "arXiv",
    primaryClass = "hep-ph",
    reportNumber = "PITT-PACC-2606",
    month = "6",
    year = "2026"
}

@article{Akita:2026gzk,
    author = "Akita, Kensuke and Hamaguchi, Koichi and Kitagawa, Haruto and Yokoyama, Tatsuya",
    title = "{Minimal Majoron Dark Matter}",
    eprint = "2605.12946",
    archivePrefix = "arXiv",
    primaryClass = "hep-ph",
    month = "5",
    year = "2026"
}

@article{Heeck:2019guh,
    author = "Heeck, Julian and Patel, Hiren H.",
    title = "{Majoron at two loops}",
    eprint = "1909.02029",
    archivePrefix = "arXiv",
    primaryClass = "hep-ph",
    reportNumber = "UCI-TR-2019-23",
    doi = "10.1103/PhysRevD.100.095015",
    journal = "Phys. Rev. D",
    volume = "100",
    number = "9",
    pages = "095015",
    year = "2019"
}

@article{Frigerio:2011in,
    author = "Frigerio, Michele and Hambye, Thomas and Masso, Eduard",
    title = "{Sub-GeV dark matter as pseudo-Goldstone from the seesaw scale}",
    eprint = "1107.4564",
    archivePrefix = "arXiv",
    primaryClass = "hep-ph",
    reportNumber = "ULB-TH-11-17",
    doi = "10.1103/PhysRevX.1.021026",
    journal = "Phys. Rev. X",
    volume = "1",
    pages = "021026",
    year = "2011"
}

@article{Ringwald:2020vei,
    author = "Ringwald, Andreas and Saikawa, Ken'ichi and Tamarit, Carlos",
    title = "{Primordial gravitational waves in a minimal model of particle physics and cosmology}",
    eprint = "2009.02050",
    archivePrefix = "arXiv",
    primaryClass = "hep-ph",
    reportNumber = "DESY 20-135, DESY-20-135, KANAZAWA-20-06, TUM-HEP-1279-20",
    doi = "10.1088/1475-7516/2021/02/046",
    journal = "JCAP",
    volume = "02",
    pages = "046",
    year = "2021"
}

@article{Svrcek:2006yi,
    author = "Svrcek, Peter and Witten, Edward",
    title = "{Axions In String Theory}",
    eprint = "hep-th/0605206",
    archivePrefix = "arXiv",
    reportNumber = "SLAC-PUB-11894",
    doi = "10.1088/1126-6708/2006/06/051",
    journal = "JHEP",
    volume = "06",
    pages = "051",
    year = "2006"
}

@article{Arvanitaki:2009fg,
    author = "Arvanitaki, Asimina and Dimopoulos, Savas and Dubovsky, Sergei and Kaloper, Nemanja and March-Russell, John",
    title = "{String Axiverse}",
    eprint = "0905.4720",
    archivePrefix = "arXiv",
    primaryClass = "hep-th",
    doi = "10.1103/PhysRevD.81.123530",
    journal = "Phys. Rev. D",
    volume = "81",
    pages = "123530",
    year = "2010"
}

@article{Kitajima:2014xla,
    author = "Kitajima, Naoya and Takahashi, Fuminobu",
    title = "{Resonant conversions of QCD axions into hidden axions and suppressed isocurvature perturbations}",
    eprint = "1411.2011",
    archivePrefix = "arXiv",
    primaryClass = "hep-ph",
    reportNumber = "TU-985, IPMU14-0334",
    doi = "10.1088/1475-7516/2015/01/032",
    journal = "JCAP",
    volume = "01",
    pages = "032",
    year = "2015"
}

@article{Daido:2015bva,
    author = "Daido, Ryuji and Kitajima, Naoya and Takahashi, Fuminobu",
    title = "{Domain Wall Formation from Level Crossing in the Axiverse}",
    eprint = "1505.07670",
    archivePrefix = "arXiv",
    primaryClass = "hep-ph",
    reportNumber = "TU-995, IPMU15-0076",
    doi = "10.1103/PhysRevD.92.063512",
    journal = "Phys. Rev. D",
    volume = "92",
    number = "6",
    pages = "063512",
    year = "2015"
}

@article{Daido:2015cba,
    author = "Daido, Ryuji and Kitajima, Naoya and Takahashi, Fuminobu",
    title = "{Level crossing between the QCD axion and an axionlike particle}",
    eprint = "1510.06675",
    archivePrefix = "arXiv",
    primaryClass = "hep-ph",
    reportNumber = "TU-1007, APCTP-PRE2015-026, IPMU15-0181",
    doi = "10.1103/PhysRevD.93.075027",
    journal = "Phys. Rev. D",
    volume = "93",
    number = "7",
    pages = "075027",
    year = "2016"
}

@article{Ho:2018qur,
    author = "Ho, Shu-Yu and Saikawa, Ken'ichi and Takahashi, Fuminobu",
    title = "{Enhanced photon coupling of ALP dark matter adiabatically converted from the QCD axion}",
    eprint = "1806.09551",
    archivePrefix = "arXiv",
    primaryClass = "hep-ph",
    reportNumber = "IPMU18-0110, MPP-2018-140, TU-1065, MIT-CTP/5026",
    doi = "10.1088/1475-7516/2018/10/042",
    journal = "JCAP",
    volume = "10",
    pages = "042",
    year = "2018"
}

@article{Cyncynates:2023esj,
    author = "Cyncynates, David and Thompson, Jedidiah O.",
    title = "{Heavy QCD axion dark matter from avoided level crossing}",
    eprint = "2306.04678",
    archivePrefix = "arXiv",
    primaryClass = "hep-ph",
    doi = "10.1103/PhysRevD.108.L091703",
    journal = "Phys. Rev. D",
    volume = "108",
    number = "9",
    pages = "L091703",
    year = "2023"
}

@article{Li:2024okl,
    author = "Li, Hai-Jun and Zhou, Yu-Feng",
    title = "{Mass mixing between QCD axions*}",
    eprint = "2408.00267",
    archivePrefix = "arXiv",
    primaryClass = "hep-ph",
    reportNumber = "ITP-CAS-24-166, ITP-24-166",
    doi = "10.1088/1674-1137/aded01",
    journal = "Chin. Phys. C",
    volume = "49",
    number = "11",
    pages = "115101",
    year = "2025"
}

@article{Li:2025cep,
    author = "Li, Hai-Jun and Zhou, Yu-Feng",
    title = "{Axion Mixing in the String Axiverse}",
    eprint = "2504.10170",
    archivePrefix = "arXiv",
    primaryClass = "hep-th",
    reportNumber = "ITP-CAS-25-088",
    month = "4",
    year = "2025"
}

@article{Murai:2024nsp,
    author = "Murai, Kai and Narita, Yuma and Takahashi, Fuminobu and Yin, Wen",
    title = "{QCD axion dark matter from level crossing with refined adiabatic condition}",
    eprint = "2412.10232",
    archivePrefix = "arXiv",
    primaryClass = "hep-ph",
    reportNumber = "TU-1252",
    doi = "10.1007/JHEP04(2025)124",
    journal = "JHEP",
    volume = "04",
    pages = "124",
    year = "2025"
}

@article{Dunsky:2025sgz,
    author = "Dunsky, David I. and Manzari, Claudio Andrea and Qu{\'\i}lez, Pablo and Ramos, Maria and S{\o}rensen, Philip",
    title = "{Resonant Landau-Zener conversion in multi-axion systems}",
    eprint = "2507.06287",
    archivePrefix = "arXiv",
    primaryClass = "hep-ph",
    reportNumber = "CERN-TH-2025-131",
    doi = "10.1007/JHEP01(2026)077",
    journal = "JHEP",
    volume = "01",
    pages = "077",
    year = "2026"
}

@article{Muursepp:2024mbb,
    author = {M{\"u}{\"u}rsepp, Kristjan and Nardi, Enrico and Smarra, Clemente},
    title = "{Accelerated cosmic expansion, mass creation, and the QCD axion}",
    eprint = "2405.00090",
    archivePrefix = "arXiv",
    primaryClass = "hep-ph",
    month = "4",
    year = "2024"
}

@article{Minkowski:1977sc,
    author = "Minkowski, Peter",
    title = "{$\mu \to e\gamma$ at a Rate of One Out of $10^{9}$ Muon Decays?}",
    reportNumber = "Print-77-0182 (BERN)",
    doi = "10.1016/0370-2693(77)90435-X",
    journal = "Phys. Lett. B",
    volume = "67",
    pages = "421--428",
    year = "1977"
}

@article{Yanagida:1979as,
    author = "Yanagida, Tsutomu",
    editor = "Sawada, Osamu and Sugamoto, Akio",
    title = "{Horizontal gauge symmetry and masses of neutrinos}",
    reportNumber = "KEK-79-18-95",
    journal = "Conf. Proc. C",
    volume = "7902131",
    pages = "95--99",
    year = "1979"
}

@article{Giese:2020rtr,
    author = "Giese, Felix and Konstandin, Thomas and van de Vis, Jorinde",
    title = "{Model-independent energy budget of cosmological first-order phase transitions{\textemdash}A sound argument to go beyond the bag model}",
    eprint = "2004.06995",
    archivePrefix = "arXiv",
    primaryClass = "astro-ph.CO",
    reportNumber = "DESY-20-064",
    doi = "10.1088/1475-7516/2020/07/057",
    journal = "JCAP",
    volume = "07",
    number = "07",
    pages = "057",
    year = "2020"
}

@article{Giese:2020znk,
    author = "Giese, Felix and Konstandin, Thomas and Schmitz, Kai and van de Vis, Jorinde",
    title = "{Model-independent energy budget for LISA}",
    eprint = "2010.09744",
    archivePrefix = "arXiv",
    primaryClass = "astro-ph.CO",
    reportNumber = "DESY-20-173, DESY 20-173, CERN-TH-2020-170",
    doi = "10.1088/1475-7516/2021/01/072",
    journal = "JCAP",
    volume = "01",
    pages = "072",
    year = "2021"
}

@book{Maggiore:2018sht,
    author = "Maggiore, Michele",
    title = "{Gravitational Waves. Vol. 2: Astrophysics and Cosmology}",
    isbn = "978-0-19-857089-9",
    publisher = "Oxford University Press",
    month = "3",
    year = "2018"
}

@article{Servant:2023tua,
    author = "Servant, G{\'e}raldine and Simakachorn, Peera",
    title = "{Ultrahigh frequency primordial gravitational waves beyond the kHz: The case of cosmic strings}",
    eprint = "2312.09281",
    archivePrefix = "arXiv",
    primaryClass = "hep-ph",
    reportNumber = "DESY-23-202, CERN-TH-2023-226",
    doi = "10.1103/PhysRevD.109.103538",
    journal = "Phys. Rev. D",
    volume = "109",
    number = "10",
    pages = "103538",
    year = "2024"
}

@article{Goldstein:2026iuu,
    author = "Goldstein, Samuel and Hill, J. Colin",
    title = "{A 2{\%} determination of $N_{\rm eff}$ from primordial element abundance, cosmic microwave background, and baryon acoustic oscillation measurements}",
    eprint = "2603.13226",
    archivePrefix = "arXiv",
    primaryClass = "astro-ph.CO",
    month = "3",
    year = "2026"
}

@article{Thrane:2013oya,
    author = "Thrane, Eric and Romano, Joseph D.",
    title = "{Sensitivity curves for searches for gravitational-wave backgrounds}",
    eprint = "1310.5300",
    archivePrefix = "arXiv",
    primaryClass = "astro-ph.IM",
    doi = "10.1103/PhysRevD.88.124032",
    journal = "Phys. Rev. D",
    volume = "88",
    number = "12",
    pages = "124032",
    year = "2013"
}

@article{LISA:2017pwj,
    author = "Amaro-Seoane, Pau and others",
    collaboration = "LISA",
    title = "{Laser Interferometer Space Antenna}",
    eprint = "1702.00786",
    archivePrefix = "arXiv",
    primaryClass = "astro-ph.IM",
    month = "2",
    year = "2017"
}

@article{Corbin:2005ny,
    author = "Corbin, Vincent and Cornish, Neil J.",
    title = "{Detecting the cosmic gravitational wave background with the big bang observer}",
    eprint = "gr-qc/0512039",
    archivePrefix = "arXiv",
    doi = "10.1088/0264-9381/23/7/014",
    journal = "Class. Quant. Grav.",
    volume = "23",
    pages = "2435--2446",
    year = "2006"
}

@article{Kawamura:2011zz,
    author = "Kawamura, Seiji and others",
    editor = "Buchman, Sasha and Sun, Ke-Xun",
    title = "{The Japanese space gravitational wave antenna: DECIGO}",
    doi = "10.1088/0264-9381/28/9/094011",
    journal = "Class. Quant. Grav.",
    volume = "28",
    pages = "094011",
    year = "2011"
}

@article{TianQin:2015yph,
    author = "Luo, Jun and others",
    collaboration = "TianQin",
    title = "{TianQin: a space-borne gravitational wave detector}",
    eprint = "1512.02076",
    archivePrefix = "arXiv",
    primaryClass = "astro-ph.IM",
    doi = "10.1088/0264-9381/33/3/035010",
    journal = "Class. Quant. Grav.",
    volume = "33",
    number = "3",
    pages = "035010",
    year = "2016"
}

@article{Hu:2017mde,
    author = "Hu, Wen-Rui and Wu, Yue-Liang",
    title = "{The Taiji Program in Space for gravitational wave physics and the nature of gravity}",
    doi = "10.1093/nsr/nwx116",
    journal = "Natl. Sci. Rev.",
    volume = "4",
    number = "5",
    pages = "685--686",
    year = "2017"
}

@article{Punturo:2010zz,
    author = "Punturo, M. and others",
    editor = "Ricci, Fulvio",
    title = "{The Einstein Telescope: A third-generation gravitational wave observatory}",
    doi = "10.1088/0264-9381/27/19/194002",
    journal = "Class. Quant. Grav.",
    volume = "27",
    pages = "194002",
    year = "2010"
}

@article{Reitze:2019iox,
    author = "Reitze, David and others",
    title = "{Cosmic Explorer: The U.S. Contribution to Gravitational-Wave Astronomy beyond LIGO}",
    eprint = "1907.04833",
    archivePrefix = "arXiv",
    primaryClass = "astro-ph.IM",
    reportNumber = "LIGO-P1900316",
    journal = "Bull. Am. Astron. Soc.",
    volume = "51",
    number = "7",
    pages = "035",
    year = "2019"
}

@article{LIGOScientific:2014pky,
    author = "Aasi, J. and others",
    collaboration = "LIGO Scientific",
    title = "{Advanced LIGO}",
    eprint = "1411.4547",
    archivePrefix = "arXiv",
    primaryClass = "gr-qc",
    doi = "10.1088/0264-9381/32/7/074001",
    journal = "Class. Quant. Grav.",
    volume = "32",
    pages = "074001",
    year = "2015"
}

@article{Cooper:2022jfr,
    author = "Cooper, S. J. and Mow-Lowry, C. M. and Hoyland, D. and Bryant, J. and Ubhi, A. and O'Dell, J. and Huddart, A. and Aston, S. and Vecchio, A.",
    title = "{Sensors and actuators for the advanced LIGO A+ upgrade}",
    eprint = "2208.00798",
    archivePrefix = "arXiv",
    primaryClass = "astro-ph.IM",
    doi = "10.1063/5.0117605",
    journal = "Rev. Sci. Instrum.",
    volume = "94",
    number = "1",
    pages = "014502",
    year = "2023"
}

@article{LIGO:2020xsf,
    author = "Adhikari, R. X. and others",
    collaboration = "LIGO",
    title = "{A cryogenic silicon interferometer for gravitational-wave detection}",
    eprint = "2001.11173",
    archivePrefix = "arXiv",
    primaryClass = "astro-ph.IM",
    reportNumber = "LIGO-P1800072",
    doi = "10.1088/1361-6382/ab9143",
    journal = "Class. Quant. Grav.",
    volume = "37",
    number = "16",
    pages = "165003",
    year = "2020"
}

@article{VIRGO:2014yos,
    author = "Acernese, F. and others",
    collaboration = "VIRGO",
    title = "{Advanced Virgo: a second-generation interferometric gravitational wave detector}",
    eprint = "1408.3978",
    archivePrefix = "arXiv",
    primaryClass = "gr-qc",
    doi = "10.1088/0264-9381/32/2/024001",
    journal = "Class. Quant. Grav.",
    volume = "32",
    number = "2",
    pages = "024001",
    year = "2015"
}

@article{Somiya:2011np,
    author = "Somiya, Kentaro",
    editor = "Hannam, Mark and Sutton, Patrick and Hild, Stefan and van den Broeck, Chris",
    collaboration = "KAGRA",
    title = "{Detector configuration of KAGRA: The Japanese cryogenic gravitational-wave detector}",
    eprint = "1111.7185",
    archivePrefix = "arXiv",
    primaryClass = "gr-qc",
    doi = "10.1088/0264-9381/29/12/124007",
    journal = "Class. Quant. Grav.",
    volume = "29",
    pages = "124007",
    year = "2012"
}

@article{Ai:2023see,
    author = "Ai, Wen-Yuan and Laurent, Benoit and van de Vis, Jorinde",
    title = "{Model-independent bubble wall velocities in local thermal equilibrium}",
    eprint = "2303.10171",
    archivePrefix = "arXiv",
    primaryClass = "astro-ph.CO",
    reportNumber = "KCL-PH-TH/2023-19",
    doi = "10.1088/1475-7516/2023/07/002",
    journal = "JCAP",
    volume = "07",
    pages = "002",
    year = "2023"
}

@article{Bodeker:2009qy,
    author = "Bodeker, Dietrich and Moore, Guy D.",
    title = "{Can electroweak bubble walls run away?}",
    eprint = "0903.4099",
    archivePrefix = "arXiv",
    primaryClass = "hep-ph",
    doi = "10.1088/1475-7516/2009/05/009",
    journal = "JCAP",
    volume = "05",
    pages = "009",
    year = "2009"
}

@article{Espinosa:2010hh,
    author = "Espinosa, Jose R. and Konstandin, Thomas and No, Jose M. and Servant, Geraldine",
    title = "{Energy Budget of Cosmological First-order Phase Transitions}",
    eprint = "1004.4187",
    archivePrefix = "arXiv",
    primaryClass = "hep-ph",
    reportNumber = "CERN-PH-TH-2010-027",
    doi = "10.1088/1475-7516/2010/06/028",
    journal = "JCAP",
    volume = "06",
    pages = "028",
    year = "2010"
}

@article{Kannike:2025ykx,
    author = {Kannike, Kristjan and Marzola, Luca and M{\"u}{\"u}rsepp, Kristjan},
    title = "{An EFT approach to the study of multi-phase criticality scenarios}",
    eprint = "2511.11367",
    archivePrefix = "arXiv",
    primaryClass = "hep-ph",
    month = "11",
    year = "2025"
}

@article{Manohar:2020nzp,
    author = "Manohar, Aneesh V. and Nardoni, Emily",
    title = "{Renormalization Group Improvement of the Effective Potential: an EFT Approach}",
    eprint = "2010.15806",
    archivePrefix = "arXiv",
    primaryClass = "hep-ph",
    doi = "10.1007/JHEP04(2021)093",
    journal = "JHEP",
    volume = "04",
    pages = "093",
    year = "2021"
}

@article{Cline:1996mga,
    author = "Cline, James M. and Lemieux, Pierre-Anthony",
    title = "{Electroweak phase transition in two Higgs doublet models}",
    eprint = "hep-ph/9609240",
    archivePrefix = "arXiv",
    reportNumber = "MCGILL-96-16",
    doi = "10.1103/PhysRevD.55.3873",
    journal = "Phys. Rev. D",
    volume = "55",
    pages = "3873--3881",
    year = "1997"
}

@article{Cline:2011mm,
    author = "Cline, James M. and Kainulainen, Kimmo and Trott, Michael",
    title = "{Electroweak Baryogenesis in Two Higgs Doublet Models and B meson anomalies}",
    eprint = "1107.3559",
    archivePrefix = "arXiv",
    primaryClass = "hep-ph",
    doi = "10.1007/JHEP11(2011)089",
    journal = "JHEP",
    volume = "11",
    pages = "089",
    year = "2011"
}

@article{Martin:2013gka,
    author = "Martin, Stephen P.",
    title = "{Three-Loop Standard Model Effective Potential at Leading Order in Strong and Top Yukawa Couplings}",
    eprint = "1310.7553",
    archivePrefix = "arXiv",
    primaryClass = "hep-ph",
    reportNumber = "NSF-KITP-13-248, FERMILAB-PUB-13-676-T",
    doi = "10.1103/PhysRevD.89.013003",
    journal = "Phys. Rev. D",
    volume = "89",
    number = "1",
    pages = "013003",
    year = "2014"
}

@article{Martin:2014bca,
    author = "Martin, Stephen P.",
    title = "{Taming the Goldstone contributions to the effective potential}",
    eprint = "1406.2355",
    archivePrefix = "arXiv",
    primaryClass = "hep-ph",
    reportNumber = "FERMILAB-PUB-14-361-T",
    doi = "10.1103/PhysRevD.90.016013",
    journal = "Phys. Rev. D",
    volume = "90",
    number = "1",
    pages = "016013",
    year = "2014"
}

@article{Elias-Miro:2014pca,
    author = "Elias-Miro, J. and Espinosa, J. R. and Konstandin, T.",
    title = "{Taming Infrared Divergences in the Effective Potential}",
    eprint = "1406.2652",
    archivePrefix = "arXiv",
    primaryClass = "hep-ph",
    reportNumber = "DESY-14-093",
    doi = "10.1007/JHEP08(2014)034",
    journal = "JHEP",
    volume = "08",
    pages = "034",
    year = "2014"
}

@article{Gorghetto:2018myk,
    author = "Gorghetto, Marco and Hardy, Edward and Villadoro, Giovanni",
    title = "{Axions from Strings: the Attractive Solution}",
    eprint = "1806.04677",
    archivePrefix = "arXiv",
    primaryClass = "hep-ph",
    doi = "10.1007/JHEP07(2018)151",
    journal = "JHEP",
    volume = "07",
    pages = "151",
    year = "2018"
}

@article{Gorghetto:2020qws,
    author = "Gorghetto, Marco and Hardy, Edward and Villadoro, Giovanni",
    title = "{More axions from strings}",
    eprint = "2007.04990",
    archivePrefix = "arXiv",
    primaryClass = "hep-ph",
    doi = "10.21468/SciPostPhys.10.2.050",
    journal = "SciPost Phys.",
    volume = "10",
    number = "2",
    pages = "050",
    year = "2021"
}

@article{Buschmann:2019icd,
    author = "Buschmann, Malte and Foster, Joshua W. and Safdi, Benjamin R.",
    title = "{Early-Universe Simulations of the Cosmological Axion}",
    eprint = "1906.00967",
    archivePrefix = "arXiv",
    primaryClass = "astro-ph.CO",
    reportNumber = "LCTP-19-08",
    doi = "10.1103/PhysRevLett.124.161103",
    journal = "Phys. Rev. Lett.",
    volume = "124",
    number = "16",
    pages = "161103",
    year = "2020"
}

@article{Buschmann:2021sdq,
    author = "Buschmann, Malte and Foster, Joshua W. and Hook, Anson and Peterson, Adam and Willcox, Don E. and Zhang, Weiqun and Safdi, Benjamin R.",
    title = "{Dark matter from axion strings with adaptive mesh refinement}",
    eprint = "2108.05368",
    archivePrefix = "arXiv",
    primaryClass = "hep-ph",
    doi = "10.1038/s41467-022-28669-y",
    journal = "Nature Commun.",
    volume = "13",
    number = "1",
    pages = "1049",
    year = "2022"
}

@article{Klaer:2017ond,
    author = "Klaer, Vincent B. . and Moore, Guy D.",
    title = "{The dark-matter axion mass}",
    eprint = "1708.07521",
    archivePrefix = "arXiv",
    primaryClass = "hep-ph",
    doi = "10.1088/1475-7516/2017/11/049",
    journal = "JCAP",
    volume = "11",
    pages = "049",
    year = "2017"
}

@article{Hiramatsu:2012gg,
    author = "Hiramatsu, Takashi and Kawasaki, Masahiro and Saikawa, Ken'ichi and Sekiguchi, Toyokazu",
    title = "{Production of dark matter axions from collapse of string-wall systems}",
    eprint = "1202.5851",
    archivePrefix = "arXiv",
    primaryClass = "hep-ph",
    reportNumber = "ICRR-REPORT-608-2011-25, IPMU12-0025, YITP-12-9",
    doi = "10.1103/PhysRevD.85.105020",
    journal = "Phys. Rev. D",
    volume = "85",
    pages = "105020",
    year = "2012",
    note = "[Erratum: Phys.Rev.D 86, 089902 (2012)]"
}

@article{Davis:1986xc,
    author = "Davis, Richard Lynn",
    title = "{Cosmic Axions from Cosmic Strings}",
    reportNumber = "SLAC-PUB-3895",
    doi = "10.1016/0370-2693(86)90300-X",
    journal = "Phys. Lett. B",
    volume = "180",
    pages = "225--230",
    year = "1986"
}

@article{Vilenkin:1984ib,
    author = "Vilenkin, Alexander",
    title = "{Cosmic Strings and Domain Walls}",
    reportNumber = "PRINT-84-0840 (TUFTS)",
    doi = "10.1016/0370-1573(85)90033-X",
    journal = "Phys. Rept.",
    volume = "121",
    pages = "263--315",
    year = "1985"
}

@article{Chang:2019mza,
    author = "Chang, Chia-Feng and Cui, Yanou",
    title = "{Stochastic Gravitational Wave Background from Global Cosmic Strings}",
    eprint = "1910.04781",
    archivePrefix = "arXiv",
    primaryClass = "hep-ph",
    doi = "10.1016/j.dark.2020.100604",
    journal = "Phys. Dark Univ.",
    volume = "29",
    pages = "100604",
    year = "2020"
}

@article{Chang:2021afa,
    author = "Chang, Chia-Feng and Cui, Yanou",
    title = "{Gravitational waves from global cosmic strings and cosmic archaeology}",
    eprint = "2106.09746",
    archivePrefix = "arXiv",
    primaryClass = "hep-ph",
    doi = "10.1007/JHEP03(2022)114",
    journal = "JHEP",
    volume = "03",
    pages = "114",
    year = "2022"
}

@article{Delaunay:2007wb,
    author = "Delaunay, Cedric and Grojean, Christophe and Wells, James D.",
    title = "{Dynamics of Non-renormalizable Electroweak Symmetry Breaking}",
    eprint = "0711.2511",
    archivePrefix = "arXiv",
    primaryClass = "hep-ph",
    reportNumber = "CERN-PH-TH-2007-219, MCTP-07-31, SACLAY-T07-141",
    doi = "10.1088/1126-6708/2008/04/029",
    journal = "JHEP",
    volume = "04",
    pages = "029",
    year = "2008"
}

@article{Chala:2018ari,
    author = "Chala, Mikael and Krause, Claudius and Nardini, Germano",
    title = "{Signals of the electroweak phase transition at colliders and gravitational wave observatories}",
    eprint = "1802.02168",
    archivePrefix = "arXiv",
    primaryClass = "hep-ph",
    reportNumber = "FERMILAB-PUB-18-241-T",
    doi = "10.1007/JHEP07(2018)062",
    journal = "JHEP",
    volume = "07",
    pages = "062",
    year = "2018"
}

@article{Postma:2020toi,
    author = "Postma, Marieke and White, Graham",
    title = "{Cosmological phase transitions: is effective field theory just a toy?}",
    eprint = "2012.03953",
    archivePrefix = "arXiv",
    primaryClass = "hep-ph",
    doi = "10.1007/JHEP03(2021)280",
    journal = "JHEP",
    volume = "03",
    pages = "280",
    year = "2021"
}

@article{Bigazzi:2020phm,
    author = "Bigazzi, Francesco and Caddeo, Alessio and Cotrone, Aldo L. and Paredes, Angel",
    title = "{Fate of false vacua in holographic first-order phase transitions}",
    eprint = "2008.02579",
    archivePrefix = "arXiv",
    primaryClass = "hep-th",
    doi = "10.1007/JHEP12(2020)200",
    journal = "JHEP",
    volume = "12",
    pages = "200",
    year = "2020"
}

@article{Bodeker:2017cim,
    author = "Bodeker, Dietrich and Moore, Guy D.",
    title = "{Electroweak Bubble Wall Speed Limit}",
    eprint = "1703.08215",
    archivePrefix = "arXiv",
    primaryClass = "hep-ph",
    doi = "10.1088/1475-7516/2017/05/025",
    journal = "JCAP",
    volume = "05",
    pages = "025",
    year = "2017"
}

@article{Fujikawa:1979ay,
    author = "Fujikawa, Kazuo",
    title = "{Path Integral Measure for Gauge Invariant Fermion Theories}",
    reportNumber = "INS-328",
    doi = "10.1103/PhysRevLett.42.1195",
    journal = "Phys. Rev. Lett.",
    volume = "42",
    pages = "1195--1198",
    year = "1979"
}






\end{document}